\documentclass[useAMS,usenatbib]{mn2e}
\usepackage{times}
\usepackage{rotating}
\usepackage{subfigure}

\newcommand{\Ha}{H$\alpha$}
\newcommand{\Hb}{H$\beta$}
\newcommand{\SII}{[S {\sc ii}]}

\newcommand{\NII}{[N {\sc ii}]}
\newcommand{\OI}{[O {\sc i}]}
\newcommand{\OII}{[O {\sc ii}]}
\newcommand{\OIII}{[O {\sc iii}]}
\newcommand{\HeI}{[He {\sc i}]}
\newcommand{\HeII}{[He {\sc ii}]}
\newcommand{\ArIII}{[Ar {\sc iii}]}

\newcommand{\HII}{H{\sc ii}}

\title[New optically selected SNRs in nearby galaxies]{A multiwavelength study of Supernova Remnants in six nearby galaxies. 
II. New optically selected Supernova Remnants}
\author[Leonidaki et al.]{I.Leonidaki,$^{1,2}$\thanks{E-mail:
ileonid@astro.noa.gr; ptb@astro.noa.gr; azezas@physics.uoc.gr} P. Boumis,$^{1}$\footnotemark[1] A.Zezas$^{3,4,5}$\footnotemark[1] \\
$^{1}$Institute of Astronomy, Astrophysics, Space Applications $\&$ Remote Sensing, National Observatory of Athens, I.Metaxa and V.Pavlou,\\ Lofos Koufou, Penteli, 15236, Athens, Greece\\
$^{2}$Astronomical Laboratory, Physics Department, University of Patras, 26500, Rio-Patra, Greece\\
$^{3}$Physics Department, University of Crete, P.O. Box 2208, GR-710 03, Heraklion, Crete, Greece\\
$^{4}$Harvard-Smithsonian Center for Astrophysics, 60 Garden Street, Cambridge, MA 02138, USA\\
$^{5}$IESL/Foundation for Research and Technology-Hellas, P.O. Box 1527, GR-711 10, Heraklion,\\ Crete, Greece}

\begin{document}

\date{}

\pagerange{\pageref{firstpage}--\pageref{lastpage}} \pubyear{2012}

\maketitle

\label{firstpage}

\begin{abstract}

We present results from a study of optically emitting Supernova Remnants (SNRs) in six nearby galaxies (NGC\,2403, NGC\,3077, NGC\,4214, NGC\,4395, NGC\,4449 and NGC\,5204) based on deep narrow band \Ha\ and \SII\ images as well as spectroscopic observations. The SNR classification was based on the detected sources that fulfill the well-established emission line flux criterion of \SII/\Ha $>$ 0.4. This study revealed $\sim$400 photometric SNRs down to a limiting \Ha\ flux of 10$^{-15}$ erg sec$^{-1}$ cm$^{-2}$. Spectroscopic observations confirmed the shock-excited nature of 56 out of the 96 sources with (\SII/\Ha)$_{phot}>$ 0.3 (our limit for an SNR classification) for which we obtained spectra. 11 more sources were spectroscopically identified as SNRs although their photometric \SII/\Ha\ ratio was below 0.3. We discuss the properties of the optically-detected SNRs in our sample for different types of galaxies and hence different environments, in order to address their connection with the surrounding interstellar medium. We find that there is a difference in \NII/\Ha\ line ratios of the SNR populations between different types of galaxies which indicates that this happens due to metallicity. We cross-correlate parameters of the optically detected SNRs (\SII/\Ha\ ratio, luminosity) with parameters of coincident X- ray emitting SNRs, resulted from our previous studies in the same sample of galaxies, in order to understand their evolution and investigate possible selection effects. We do not find a correlation between their \Ha\ and X-ray luminosities, which we attribute to the presence of material in a wide range of temperatures. We also find evidence for a linear relation between the number of luminous optical SNRs (10$^{37}$ erg sec$^{-1}$) and SFR in our sample of galaxies.

\end{abstract}


\section{Introduction}

Supernova Remnants (SNRs) inject the predominant fraction of mechanical energy that heats and shapes the Interstellar Medium (ISM) since the generated shock waves are responsible for the compression, acceleration and excitation of their surrounding medium. At the same time they replenish the ISM with heavy elements formed during the evolution of massive stars while when immersed in molecular clouds the compression may trigger formation of the next generation of stars. SNRs can yield significant information on the global properties of a galaxy's ISM such as its density, temperature or composition \citep[e.g.][]{Blair04}. Furthermore, being the endpoints of core-collapse massive stars ($M > 8M_{\sun}$) they can be used as proxies for measurements of massive star formation rate (SFR) and studies of stellar evolution \citep{Condon90}.

Detecting large samples of SNRs in a multi-wavelength context can provide several key aspects of the physical processes taking place during their evolution. For example, the blast waves of newly formed SNRs can heat the material behind the shock front to temperatures up to $10^{8}$ K producing thermal X-rays. Synchrotron radiation in radio wavelengths is produced at the vicinity of the shock as well as from the cooling regions behind the shock front, owing to relativistic electrons gyrating in the magnetic field of the SNRs \citep[e.g.][]{Dickel99, CS95}. Optical filaments are signs of older SNRs since they form in the cooling regions behind the shock \citep[i.e.][]{Stupar09} producing shock-heated collisionally ionized species (such as \SII, \OIII\ or \Ha\ recombination lines). Therefore, multi-wavelength studies (optical, radio, X-rays, infrared) can surmount possible selection effects inherent in 'monochromatic' samples of SNRs and provide a more complete picture of their nature and evolution as well as their interplay with the ISM and their correlation with star forming activity.

About 274 SNRs detected in different wavebands are known in our Galaxy, a comprehensive catalogue of which is presented in \citet{Green09}. A large number of them has been studied in detail in various wavebands (e.g. radio: \citealt{Green09}; optical: \citealt[]{Boumis02, Boumis05, Boumis09, Fesen10}; X-rays: \citealt[]{Reynolds09, Slane02}; infrared: \citealt{Reach06}) providing significant information on the properties of individual objects and SNR physics. However these studies are impeded by distance uncertainties and Galactic absorption, hampering the investigation of SNRs in a wide variety of environments. On the other hand extragalactic studies of SNRs offer several advantages: they can be achieved in determined distances with much fewer observations while they cover a broader range of metallicities and ISM parameters than our Galaxy, giving us a more complete picture of the SNR population parameters.

Numerous studies of extragalactic SNRs have been conducted since the pioneering work of \citet{MC73} on the Magellanic Clouds. The availability of sensitive, high-resolution observations in the radio and X-ray bands have allowed the systematic investigation of extragalactic SNRs in these wavebands \citep[e.g.][]{Leonidaki10, Long10, Pannuti07, Ghavamian05}. However, since the first studies in a small sample of nearby galaxies \citep[e.g.][]{MF97, MFBL97} there have not been any systematic pursuits of the optical populations of extragalactic SNRs.

The availability of sensitive wide-field imagers and spectrographs allow us to greatly extend these initial efforts to a larger set of galaxies, while probing fainter SNRs populations. We have embarked in an extensive multi-wavelength investigation of the SNR populations in six nearby galaxies (NGC\,2403, NGC\,3077, NGC\,4214, NGC\,4395, NGC\,4449 and NGC\,5204), involving optical and X-ray data. These galaxies are selected from the Third Catalog of Bright Galaxies (RC3; \citealt{de Vaucouleurs95}) to be (a) late type (T $>$ 4; Hubble type), (b) close ($\leq$5 Mpc) in order to minimize source confusion, (c) at low inclination $\leq 60\degr$) in order to minimize internal extinction and projection effects, and (d) be above the Galactic plane ($|b| > 20\degr$). The properties of the galaxies in our sample are presented in Table 1 while a list of previous multi-wavelength SNR surveys are presented in \S2 of \citet{Leonidaki10}.

From the pool of objects drawn from these selection criteria, we selected galaxies that have Chandra archival data with exposure times long enough to achieve a uniform detection limit of $10^{36}$ erg s$^{-1}$. We opted to focus on Chandra data owing to its superior spatial resolution which allows the detection of faint sources in crowded environments and therefore can provide the most unbiased sample to be correlated with the optical data. This X-ray investigation revealed 37 thermal X-ray SNRs (based on their spectra or hardness ratio colours), 30 of which were new discoveries. In many cases, the X-ray classification was confirmed based on counterparts with SNRs identified in other wavelengths. We found that X-ray selected SNRs in irregular galaxies appear to be more luminous than those in spirals due to the lower metallicities and therefore more massive progenitor stars of irregular galaxies or the higher local densities of the interstellar medium. A comparison of the numbers of observed luminous X-ray-selected SNRs with those expected based on the luminosity functions of X-ray SNRs in the Magellanic Clouds and M33 suggested different luminosity distributions between the SNRs in spiral and irregular galaxies with the latter tending to have flatter distributions. These results are presented in the first paper of this series (\citealp{Leonidaki10}, hereafter Paper I).

In this paper we present a detailed optical spectro-photometric study of the SNR populations in this sample of galaxies. The optical identification of SNRs is based on the elevated \SII/\Ha\ ratio ($\geq$ 0.4), pioneered by \citet{MC73}. The outline of this paper is as follows: In \S2, we describe the observations, including data reduction and techniques used for source detection and photometry. In \S3, we describe the long-slit and multi-slit spectroscopic observations, while in \S4 we give the SNR classification criteria as well as aggregate results of the detected SNRs in our sample of galaxies. In \S5 we discuss the results of our spectro-photometric investigation. Finally, in \S6 we present the conclusions of this work.

\section{Imaging}

\subsection{Observations}

Optical images were obtained with the 1.3m (f/7.7) Ritchey--Chr\'{e}tien telescope at Skinakas Observatory on June 6-12, 2008 and Nov 16-18, 2009. A 2048 $\times$ 2048 ANDOR Tech CCD was used which has a $9.6\arcmin \times 9.6\arcmin$ field of view and an image scale of 0.28$\arcsec$ per pixel. Apart from Nov 18 2009, all the other observing nights were photometric with seeing conditions ranging between 1.3$\arcsec$-2.5$\arcsec$. The observations were performed with the narrow band \Ha\ + \NII, \SII\, and \OIII\ filters.  Broadband continuum filters in red and blue were used to subtract the continuum from the \Ha + \NII, \SII\, and \OIII\ images respectively. The continuum filters used for the observations are centered on line-free regions of the spectra in order to avoid strong SNR emission lines to pass. The interference filter characteristics are listed in Table 2.

The exposure time was 3600 sec for each \Ha\ + \NII\ filter exposure, 7200 sec for each \SII\ filter exposure and 300 sec for the exposures through the continuum filters. The airmass of the galaxies during observations ranged between 1.06 and 1.87. In the case of NGC\,2403, the $9.6\arcmin$ CCD field of view did not cover the whole D$_{25}$\footnote{The D$_{25}$ area is defined as the optical isophote at the B-band surface brightness of 25 mag arcsec$^{-2}$} area of the galaxy, therefore we obtained a 2$\times$2 mosaic. We did not observe NGC\,2403 and NGC\,4395 through \OIII\ or continuum blue filters owing to weather conditions. Bias frames and well-exposed twilight flats were taken on each run as well as spectro-photometric standard stars from the list of \citet{Hamuy92}.

\subsection{Data Reduction}

The data reduction was performed using the IRAF V2.14 package\footnote{http://iraf.net/irafdocs/ccduser3/}. All images were bias-subtracted and flat-field corrected while the data sets of each filter for each galaxy were median-combined in order to reject the cosmic rays. Star-free areas outside the body of the galaxies were selected in \Ha + \NII, \SII, \OIII\ and continuum images in order to subtract the sky background and obtain just the light from each galaxy.

The sky-background subtracted images were aligned to a reference image for each galaxy (e.g. continuum red for \Ha + \NII\ and \SII, continuum blue for \OIII) and astrometrically calibrated using the USNO-B1.0 Catalog or SDSS Data Release 7. For each galaxy's \Ha + \NII, \SII\ and continuum red images, we selected an adequate number (8-10) of the same faint stars ($\sim$15-20 mag) for which we calculated their (\Ha + \NII/cont red) and (\SII/cont red) ratios. The mean value of those ratios were used to create normalized-continuum red images for \Ha + \NII\ and \SII\ images, respectively. We then subtracted the corresponding normalized-continuum red images from the \Ha + \NII, \SII\ images in order to eliminate the star-light continuum. 
We did not follow the same procedure for the \OIII\ images since they were used only for visual examination of each source's \OIII\ emission. The continuum-subtracted \Ha\ + \NII\ and \SII\ images were flux calibrated, using several spectrophotometric standard stars observed each night and reduced the same way as the galaxy images. We note that the used intereference \Ha\ + \NII\ filter includes the \NII\ 6548 \AA\ and 6584 \AA\ lines. In order to estimate the net \Ha\ flux-calibrated images of our galaxies, we corrected for the \NII\ contamination using the \NII($\lambda\lambda$ 6548, 6584)/\Ha\ ratios from integrated spectroscopy of the galaxies from the work of \citet{Kennicutt08}. 


\subsection{Source Detection}

Sources present considerably higher S/N in the \Ha\ than in the \SII\ images therefore we searched for sources in the continuum-subtracted, flux calibrated \Ha\ data sets of each galaxy, using the Sextractor V2.5.0 package \citep{Bertin96}. Our goal is to identify faint nebulae in relatively isolated regions as well as to separate possible SNRs from \HII\ or diffuse emission regions. Since Sextractor was used only for detection, the main parameters we adjusted are the following: a) detection threshold set to 1.3-3.5 $\sigma$ above background, b) minimum number of pixels for a detection to be triggered between 3 to 7 for different galaxies, c) a background mesh size of 6-10 pixels in order to detect faint sources and account for local variations of the background within the galaxies. All the above parameters were adjusted for each galaxy depending on its background and the detection efficiency of faint sources. 

The non-uniform \Ha\ background and diffuse emission within the galaxies did not allow us to base the source detection solely on the Sextractor output. For that reason, the results of the Sextractor run for each of the continuum-subtracted, flux-calibrated \Ha\ images were visually inspected in order to discard local maxima of \Ha\ background or spurious sources associated with bad pixels, and then were used to create source lists. In the case of NGC\,2403, the fourth frame of the mosaic was observed at a non-photometric night (November 18, 2009) therefore we excluded it from the data analysis procedure. The source detection for NGC\,2403 was performed individually on each of the three frames of the galaxy mosaic. The detection results from each mosaic were then combined in order to form the final list of sources in NGC\,2403. Each source list was overlaid on the relevant continuum red image of each galaxy in order to eliminate any obvious star-like objects. Since SNRs are identified on the basis of strong \SII\ emission, we visually inspected the significance of the sources on the relevant continuum-subtracted, flux-calibrated \SII\ images. The final source list was defined on the basis of clear detection of sources in both \Ha\ and \SII\ images. We note that we opted to use the individual \Ha\ and \SII\ images (rather than the \SII/\Ha\ ratio images) for source detection since they tend to be less noisy.

\subsection{Photometry}

We used the {\it apphot} package in IRAF in order to perform photometry of the sources identified on the continuum-subtracted, flux-calibrated \Ha\ and \SII\ images. In the case of NGC\,2403, we created a final list from each mosaic frame and performed photometry on each source in every frame that it was observed. The final photometric parameters for each source in NGC\,2403 were derived from the mean value of the parameters in each frame. We used source apertures with diameter set to 8-10 pixels (which corresponds to $\sim$2$\arcsec$-3$\arcsec$) and physical scales of $\sim$32 to $\sim$66 pc for the closest and most distant galaxy, respectively. These apertures were chosen to cover most of the source's flux in both the \Ha\ and \SII\ images (taking into account the seeing conditions of 1.3$\arcsec$ - 2.5$\arcsec$) while taking care not to encompass other neighbouring sources or diffuse emission. The local background for each source was measured from appropriate annuli of typical sizes of 10 pixels ($\sim$3$\arcsec$). In some special cases with highly non-uniform background, the local background was measured from a neighbouring region. We note that accurate photometry depends strongly on the selected background area, especially in cases of sources embedded in large filaments or regions with enhanced diffuse emission. From measurements for different background regions we find a typical uncertainty on the flux of 40\% stemming from the background selection. This uncertainty is minimized in the case of \SII\ images due to the very low background and the point-like nature of most sources. Extinction correction was not applied on the \Ha, \SII\ fluxes since no \Hb\ observations were obtained for the galaxies in our sample.

We calculated variance maps by applying error propagation to the error map of the initial bias-subtracted, flat-fielded combined images (including readout noise, gain etc). In this calculation we accounted for all the analysis steps taken from the derivation of the fluxed images (continuum-subtraction, flux calibration). The \Ha\ and \SII\ flux errors were estimated by calculating the square root of the measured sum of the variance map within the aperture of each source. The \SII/\Ha\ ratio errors were calculated through standard error propagation. 

Based on the photometric properties of the detected sources, we calculated the \SII/\Ha\ flux ratio of the final source list in each galaxy. In order to further examine whether the correction we applied for the \NII\ contamination (based on the integrated \NII/\Ha\ ratios of \citealt{Kennicutt08}) in the flux-calibrated \Ha\ images is appropriate and thus inspect the validity of the measured \Ha\ fluxes and \SII/\Ha\ ratios, we used the spectroscopic \NII/\Ha\ ratios of our spectroscopically-observed SNRs (see \S 3). From the histogram of the spectroscopic \NII/\Ha\ ratios of our SNR sample (Fig. 1) we see that they form two distinct loci: irregular galaxies (apart from NGC\,3077) extend to lower \NII/\Ha\ ratios than spiral galaxies, probably owing to differences in their metallicities (\S5.2). This led us to redefine our \NII/\Ha\ correction factor to the median values of the SNRs in NGC\,2403 + NGC\,3077 and the rest of the irregular galaxies and correct accordingly the \Ha\ fluxes and \SII/\Ha\ ratios of the detected sources.

Distinct sources with strong emission in both continuum-subtracted, flux-calibrated \Ha\ and \SII\ images that present (\SII/\Ha)$_{phot}$ $>$ 0.4 (within their errorbars) are considered photometric SNRs. We also include sources with 0.3 $<$ (\SII/\Ha)$_{phot}$ $<$ 0.4 which are well possible to belong to the SNR regime since photometry can result in many cases to ambiguous \SII/\Ha\ ratios, especially for sources embedded in diffuse emission or near H{\sc ii} regions.

In Tables 3-8 we present the photometric properties of the photometric SNRs in each galaxy while in Table 9 (available at the electronic version) we present the photometric properties of all spectroscopically-observed sources which were not identified as SNRs ((\SII/\Ha)$_{spec}$ $<$ 0.4). In Column 1 we give the source ID, in Columns 2 and 3 the Right Accension and Declination (J2000) of each photometric SNR, in Column 4 the radius in pixels used for the source aperture on which the photometry was performed, in Columns 5 and 6 the inner and outer radius of the annulus used for background subtraction (sources for which the background was measured from a nearby region are indicated as ext), in Columns 7 and 8 their photometric, non-extinction corrected \Ha\ and \SII\ fluxes respectively and in Column 9 their (\SII/\Ha)$_{phot}$ ratio. In Column 10 we indicate whether there is available spectrum ({\it M} and {\it S} for Mayall and Skinakas telescopes respectively) and in Column 11 their classification based on the criteria mentioned in \S4. In some cases of large filaments the imaging resulted in multiple detections. In these cases it is not possible to distinguish between a single or multiple SNRs. These sources are indicated as LBZ\,XX-Y in the relevant tables. Each Table is separated for clarity into three frames: the first frame with spectroscopically-verified SNRs, the second frame with sources with (\SII/\Ha)$_{phot}$ $>$ 0.4 within their errorbars and the third frame with sources presenting 0.3 $<$ (\SII/\Ha)$_{phot}$ $<$ 0.4 (within their errorbars).

\section{Spectroscopy}

Spectroscopic observations are the only way to unambiguously verify the shock-heated nature of these sources and therefore classify them as SNRs ((\SII/\Ha)$_{spec}$ $\geq$ 0.4). They can be used to obtain accurate emission line ratios that provide physical information (e.g. electron density, shock velocities) while inspecting the accuracy of the photometric parameters. 

The spectroscopically-observed sources were selected based on: a) their (\SII/\Ha)$_{phot}$ ratio, b) their strong S/N and c) the physical parameters of the multi-slit masks. We also opted to obtain spectra for a few additional sources in each galaxy with (\SII/\Ha)$_{phot}$ $\leq$ 0.3 in order to investigate any systematic effects in the \SII/\Ha\ photometric ratios. 

The spectroscopic observations were obtained during the course of 2 observing runs: 4 nights (long-slit spectra) at the 1.3m Skinakas telescope in Crete, Greece and 4 nights (multi-slit spectra) at the 4m Mayall telescope, Kitt Peak, Arizona, USA.

\subsection{Multi-slit observations with the 4m Mayall telescope}

Multi-slit spectra were obtained with the 4m Mayall telescope at Kitt Peak on May 3-6, 2010. We used the $5\arcmin\times5\arcmin$ T2KB CCD detector and the BL420 600 lines mm$^{-1}$ grating at the 1st order, centered at 6000 \AA. This setup gives a spectral coverage of 2300 \AA\ with a spectral resolution of 3.8 \AA\ which allows the separation of the \Ha\ from the \NII\ doublet and the measurements of the individual lines of the \SII\ doublet. 

Each slitlet was 2.5$\arcsec$ wide and included most of the source light given the seeing conditions (1.2$\arcsec$-1.5$\arcsec$), without significantly degrading the spectral resolution. The slit length was between 4$\arcsec$-5$\arcsec$ which allowed the subtraction of the local diffuse background. The weather conditions provided photometric nights while the exposure time per mask varied between 2100-3600 sec, depending on the brightness of the targets and the time constraints. Bias-frames, comparison lamp exposures, projector flats and spectrophotometric standard stars were observed each night for CCD, wavelength and flux calibrations. 

\subsection{Long-slit observations with the 1.3m Skinakas telescope}
Long-slit spectra of individual objects were obtained with the 1.3m telescope at Skinakas Observatory on May 25-28, 2009. A 1302 line mm$^{-1}$ grating, blazed at 5500 \AA, was used with the 2000 $\times$ 800 SITe CCD, giving a spectral coverage of 4700-6700 \AA\ (dispersion of $\sim$ 1 \AA/pixel) and a spectral resolution of $\sim$6 \AA\ and $\sim$4 \AA\ (FWHM) in the blue and red wavelenghths respectively. The slit we used has a width of 6.3$\arcsec$, including most of the source light (given the seeing conditions of 1.3$\arcsec$-1.8$\arcsec$), while its length of 7.8$\arcmin$ allowed for background subtraction. In all cases the slit was oriented in the north-south direction. Since the faintness of our target nebulae makes the positioning of the slit a hard task, we positioned each slit on the required targets by offseting from a field star. We opted to observe sources outside regions of strong diffuse emission or crowded fields in order to position them accurately in the slit and be able to subtract their background with better precision and accurately. The slit centres and the exposure times for each slit are presented in Table 3. The nights were all photometric. Spectrophotometric flux standard stars were observed each night as well as calibration frames consisted of biases, twilight flats and comparison lamp exposures.

\subsection{Reduction of spectra}

The IRAF package was used for the standard data reduction as well as for the extraction of flux-calibrated spectra. The slit length in each case allowed the subtraction of the local diffuse background. In cases where the extraction of multiple spectra along the slit was difficult due to extended sources embedded in diffuse regions, we defined the spectrum extraction by comparing the spatial dimension of the spectrum with the photometric images. In the case of long-slit spectroscopy, the centre of the slit for each observation was chosen so that more than one targets to be included in each spectrum. Line measurements were performed by fitting Gaussians to the spectra.

In Table 11 (available at the electronic version) we give the absorbed (F) and extinction-corrected (I) emission line fluxes of all spectroscopically-observed SNRs ((\SII/\Ha)$_{spec}$ $\geq$ 0.4) in our sample of galaxies. The presented emission line fluxes are normalized to F(\Ha)=100 and I(\Ha)=100 respectively. We also give the signal-to-noise (S/N) ratio of the quoted fluxes which was estimated based on the spectral counts of the emission lines and their relevant background. Lines for which no values are given were not detected.

Tables 12-13 (Table 13 is available at the electronic version) present the emission line parameters of the spectroscopically observed sources of our sample with ((\SII/\Ha)$_{spec}$ $\geq$ 0.4 and ((\SII/\Ha)$_{spec}$ $<$ 0.4 respectively. Cols 1 and 2 present the galaxy and the ID of the source respectively. Column 3 gives the absolute, extinction corrected F(\Ha) in units of 10$^{-14}$ erg s$^{-1}$ cm$^{-2}$. In Column 4 we give the extinction c(\Hb), Column 5 gives the colour excess E(B-V) using the ''standard'' reddening law E(B-V)$\approx$0.77c with R=3.1 \citep{Osterbrock06}, while Column 6 presents the unabsorbed \Ha/\Hb\ ratios. The remaining columns present various emission line ratios derived from the relevant extinction-corrected emission line fluxes (when \Hb\ line was detected), otherwise from the absorbed emission line fluxes. The extinction-corrected emission line fluxes were normalized to the \Ha\ emission line and were estimated using the R=3.1 reddening curve \citep{Osterbrock06}. All errors were calculated through standard error propagation.

In Fig. 2 we present the individual, zoomed-in display of the 67 spectroscopically-observed SNRs (SNRs in Tables 3-8) over the \Ha\ image of each galaxy in order to show their distinct morphology where possible. The images cover and area of 30$\arcsec$$\times$30$\arcsec$ while the arrows point at the SNRs. In Fig. 3 we show representative spectra of three spectroscopically-verified SNRs (at low, medium and high resolution) while the electronic version presents the extracted spectra of all spectroscopically-observed SNRs.

\section{RESULTS AND SNR CLASSIFICATION} 

Overall, a large number of sources (269) were detected with (\SII/\Ha)$_{phot}>$0.4 while 138 more present 0.3$<$(\SII/\Ha)$_{phot}<$0.4. 134 sources were spectroscopically observed with the 1.3m Skinakas and 4m Mayall telescopes, resulting in a total of 67 sources identified as SNRs (12 in NGC\,2403, 6 in NGC\,3077, 18 in NGC\,4214, 6 in NGC\,4395, 18 in NGC\,4449 and 7 in NGC\,5204). This number of spectroscopically observed sources does not include only objects with (\SII/\Ha)$_{phot}$$>$0.3 (our limit for an SNR classification) but also sources with ratio below that limit in order to investigate any systematic effects in the \SII/\Ha\ photometric ratios (for an aggregate view, see Table 14).

On the basis of these results, we divide our optically-selected SNRs into the following types: 1) SNRs, 2) candidate SNRs, and 3) probable candidate SNRs. As SNRs we consider all spectroscopically observed sources with (\SII/\Ha)$_{spec}$ flux ratio $\geq$ 0.4 within their error-bars. We consider as candidate SNRs all sources with (\SII/\Ha)$_{phot}$ $\geq$ 0.4 (within their error-bars) but with no available spectra. As probable candidate SNRs we consider sources with 0.3 $<$ (\SII/\Ha)$_{phot}$ $<$ 0.4 within their error-bars (see \S2.4 for details).
This diagnostic tool (\SII/\Ha $>$0.4) has been proven to differentiate shock excited processes occuring in SNRs from photo-ionised nebulae (\HII\ regions or planetary nebulae). This is because in the case of SNRs, most of the sulfur content in the cooling regions behind the shock front are in the form of S$^{+}$ and their collisional excitation yield to enhanced \SII/\Ha\ ratio. In typical \HII\ regions, S$^{++}$ ions are mainly present due to strong photo-ionisation and therefore the \SII/\Ha\ ratio is expected to be generally lower than 0.4. Additional forbidden lines (e.g. \OI\ 6300 \AA\ or \OIII\ 4959, 5007 \AA) or enhanced \NII/\Ha\ ratios with respect to \HII\ regions, can be used as evidence for shock-heating mechanisms and therefore verify the nature of sources as SNRs.

In Table 14 we present the census of the SNRs in our sample of galaxies and the success rates in the photometric SNR classification. In Column 1, we split the (\SII/\Ha)$_{phot}$ ratios into three categories: $>$0.4 (candidate SNRs), 0.3 - 0.4 (probable candidate SNRs) and $<$0.3. In Column 2 we present the number of the photomeric sources that were detected in each category (within the error-bars). In Column 3 we give the number of photometric SNRs presented in Tables 3-8. These numbers result from the detected SNRs (Column 2) if we subtract the spectroscopically observed, non-SNRs (Column 4 minus Column 5). In the case of $<$0.3 for this column, we included only the sources that were spectroscopically verified as SNRs. In Column 4 we give the number of spectroscopically observed sources while in Column 5 we present the number of sources that were spectroscopically verified as SNRs. Finally, in Column 6 we give the percentage of the photometric SNRs that were spectroscopically confirmed as SNRs (success rate in SNRs).

\subsection{Individual objects}

Below we present notable cases of sources for each galaxy:\\
{\it \bf NGC\,2403}: {\it LBZ\,6, LBZ\,95}: These two sources are located within a larger complex of nebulosity (Fig. 4a, small circles) and present (\SII/\Ha)$_{phot}$ 0.63 and 0.42 respectively. LBZ\,6 was also spectroscopically observed to have (\SII/\Ha)$_{spec}$ = 0.61. \citet{MFBL97} photometrically identified the whole region as one SNR (SNR-15). This is most likely due to the fact that their SNR identification is based on the \SII/\Ha\ ratio images which tend to be more noisy than the individual \Ha\ and \SII\ images. For comparison, we performed photometry on the the whole area (large circle in Fig. 4a) and resulted to (\SII/\Ha)$_{phot}\approx$0.41. However, in our images this area is clearly split in several individual sources with large values of (\SII/\Ha)$_{phot}$, suggesting that it is most likely an SNR complex.\\
{\it LBZ\,1}: We performed photometry on the whole region of this arc-like source (circle in Fig. 4b). However, the spectroscopy was performed on the edge of the arc as can be seen from the slit in Fig. 4b. \\
{\it LBZ\,12}: This stellar-like source was photometrically identified as an SNR and stands besides an arc (Fig. 4c). The combination of the two objects was photometrically identified as SNR-32 by \citet{MFBL97}. However, the placemenet of the slit helped us to further investigate the nature of this region (see Fig. 4c). The left edge of the slit falls on our photometrically-detected LBZ\,12 which has (\SII/\Ha)$_{spec}$ = 0.43. The right edge of the slit covers part of the arc with (\SII/\Ha)$_{spec}$ = 0.23 indicating that it is not a shock-excited region. We also calculated the integrated (\SII/\Ha)$_{spec}$ for the entire object in order to compare our results with those of \citet{MFBL97}. We find a ratio of 0.30, possibly consistent with the classification of \citet{MFBL97} for the entire region. Therefore, we suggest that only the stellar-like source (LBZ\,12) is an SNR while the arc is part of an \HII\ region.\\
{\it \bf NGC\,4214}:{\it LBZ\,5}: The photometry of this source was performed on a considerably smaller area than that used to extract the spectrum. However the latter does not show any peaks along the spatial direction that would allow us to extract spectra for individual regions.\\
{\it LBZ\,87}: This source is located in the vicinity of a large \HII\ region. The area presents enhanced diffuse emission over the \Ha\ image, preventing its detection by Sextractor as a discrete source. However, we opted to perform photometry because the \SII\ emission of the particular source is distinct and bright while it is already known X-ray, radio and optical SNR (see Table 18). Its measured (\SII/\Ha)$_{phot}$ ratio (=0.36) allows us to include it in the final list of photometric SNRs (Table 5).\\
{\it \bf NGC\,4449}:{\it LBZ\,6}: The slit was placed along the source and spectroscopy revealed the existence of two peaks in the overall spectrum of the source. We examined the detected source in the \Ha\ image of the galaxy (Fig. 2) and indeed the presence of two lobes is unequivocal. We opted to present the properties and spectra for both regions (LBZ\,6a-LBZ\,6b) but nonetheless we consider it as one source.\\
{\it \bf NGC\,5204}: {\it LBZ\,16}: This source stands beside a very bright \HII\ region and for that reason it was not detected by Sextractor. However, we performed photometry on the source which resulted to (\SII/\Ha)$_{phot}$ = 0.91. This ratio ($>$0.4) as well as the fact it is an already-known optical SNR \citep{MF97} allows us to include the source in the final list of our photometric SNRs.

\subsection{Physical parameters}

The photometric investigation revealed a large number of photometric SNRs (418; see Tables 3-8, Table 14) in our sample of galaxies reaching \Ha\ and \SII\ fluxes as low as $\sim$1.2$\times$10$^{-15}$ and $\sim$7$\times$10$^{-16}$ erg sec$^{-1}$ cm$^{-2}$ respectively. In Fig. 5 we plot the (\SII/\Ha)$_{phot}$ ratio of all photometric SNRs (\SII/\Ha$_{phot} >$0.3 within their error-bars) in our sample of galaxies against their photometric \Ha\ flux. The vast majority of the SNRs, apart from those in NGC\,2403, have fluxes between 3$\times$10$^{-15}$ and 3$\times$10$^{-14}$ erg sec$^{-1}$ cm$^{-2}$. On the other hand, the majority of the SNRs in NGC\,2403 have fluxes between 1$\times$10$^{-14}$ and 6$\times$10$^{-14}$ erg sec$^{-1}$ cm$^{-2}$, almost half order of magnitude brighter than the mean flux value of the SNRs in the other galaxies of our sample. This is consistent, within the photometric errors, with the sensitivity limit of the SNR survey of \citet{MFBL97} performed with a similar telescope. The difference in the sensitivity limits between NGC\,2403 and the other galaxies is most likely due to the stronger and non-unifom diffuse emission in this galaxy. As pointed out by \citet{Pannuti07} optical surveys are not very sensitive in identifying SNRs in these environments.\\

We also derived the electron densities of the 67 spectroscopically observed SNRs based on their \SII(6716)/ \SII(6731) ratios (see Table 12) which is a good indicator of electron density \citep{Osterbrock06}. We used the {\it temden} task of the {\it nebular} package in IRAF software\footnote{http://stsdas.stsci.edu/nebular/temden.html}, assuming a temperature of 10$^{4}$ K. The \SII(6716)/ \SII(6731) ratios of our sample of SNRs indicate electron densities ranging between 170 to 580 cm$^{-3}$ for the sample of our galaxies.\\
In Fig. 6 we plot the number of spectroscopically observed SNRs against their \SII(6716 \AA)/\SII(6731 \AA) ratios. The red histogram corresponds to SNRs in NGC\,2403 (the only spiral galaxy in our sample), the black histogram shows the SNRs in the remaining galaxies of our sample (irregulars) while we have included (magenta) the spectroscopically observed SNRs of four spiral galaxies (NGC\,5585, NGC\,6946, M81, and M101) from the work of \citet{MF97}. One would expect SNRs in irregular galaxies to present lower \SII(6716)/\SII(6731) ratios (higher densities) than those in spirals, since local enhancements of ISM are usually the case in irregular galaxies. However,  there is no trend in the sulfur-line ratios between SNRs in different types of galaxies. This indicates that there are not significant differences in the density of the ejecta or the circumstellar environment between spiral and irregular galaxies. On the other hand, the majority of the SNRs in Fig. 6 have \SII(6716)/\SII(6731) $>$ 1, which following \citet{Stupar09} indicates old SNRs. The trend of detecting preferentially older SNRs in the optical band (e.g. \citealt{Rosado83}), in combination with the age-dependence of their density may explain the fact that we do not see any significant differences between the SNR populations of elliptical and spiral galaxies.

\subsection{Multiwavelength associations}

We have compiled a catalogue of all known optical SNRs in our sample of galaxies from this study and the literature \citep[]{MFBL97, Dopita10, Blair83, MF97} as well as SNRs from X-ray (Paper I and the literature), and radio \citep[]{Eck02, Turner94, Rosa05, Chomiuk09, Vukotic05} observations. We searched for possible associations between these three wavebands by cross-correlating the source catalogue with a search radius of 2$\arcsec$. This search radius was based on the absolute astrometric error of the individual catalogs (which in most cases was very small; e.g. USNO-B1.0 at 0.2$\arcsec$) and the typical error of our optical data ($\sim$1$\arcsec$-1.5$\arcsec$).\\ 
Sources identified as SNRs in the X-ray or radio band but present (\SII/\Ha)$_{phot} < $ 0.3 in this study are denoted as SNR/\HII. For X-ray and radio SNRs not identified as such in our analysis but were correlated with distinct nebular features in the \Ha\ images, we performed photometry at the location of the associated multi-wavelength source for comparison. \\
The results of the cross-correlation are presented in Tables 16-21. Column 1 shows the source identification. Sources with a questionmark have offsets from their multiwavelength associations, somewhat larger than the defined search radius. In most cases however no other sources appear to be encompassed by this search radius, unless otherwise stated. Column 2 gives the source classification based on this study (see \S4). Columns 3 and 4 give the RA and Dec (J2000) of the sources in this study. If the source is detected in this study we report the coordinates of the optical source, otherwise we give the coordinates of the multi-wavelength counterparts. Column 5 shows the optically associated SNR by other studies, while Column 6 gives the coordinate offset of the source between this study and other optical studies. Column 7 shows the X-ray associated SNR while Column 8 gives the coordinate offset of the source between this study and the X-ray associate SNR. Column 9 shows the radio associated SNR while Column 10 gives the coordinate offset of the optical source and the radio asscociated SNR.
 
{\it NGC\,3077}: The X-ray (LZB\,18)/radio SNR is located between two detected SNR/\HII\ sources by this study (LBZ\,299 and LBZ\,300). For a possible interpretation see \S 5.3.3. The offsets of these sources with the X-ray/radio SNR association are similar, therefore we opted to present both of them.\\
{\it NGC\,4449}: The Cas\,A-like, oxygen-rich SNR in NGC\,4449 (e.g. \citealt{Blair83}) presented (\SII/\Ha)$_{phot} < $ 0.4 in this study. However, the identification of this source from previous optical studies was not based on the narrow lines of \Ha, \Hb, \NII, and \SII\ but on broad lines of \OI, \OII\ and \OIII\ which associated it with ejecta of a young, O-rich SNR. We are aware that the photometric method used in this study ((\SII/\Ha)$_{phot} > $ 0.4 criterion for identifying SNRs) is not helpful for identifying young oxygen-rich SNRs since we focus on different strong emission lines of SNRs.\\
{\it NGC\,2403}: We have spectroscopically verified the shock-heated mechanism of three photometric SNRs of \citet{MFBL97} (SNR-3, SNR-15 and SNR-32). On the other hand, two photometric SNR of \citet{MFBL97} (SNR-26, SNR-28) are denoted as SNR/\HII\ ((\SII/\Ha)$_{phot} < $ 0.3) in this study. For these sources, a spectroscopic investigation is necessary in order to verify their nature.

\section{Discussion}

\subsection{Validity of the photometric method}

In order to examine the precision of the photometric \SII/\Ha\ ratios we plot them against the (\SII/\Ha)$_{spec}$ ratios of all spectroscopically observed sources in each galaxy (Fig. 7). The red points denote SNRs (see Tables 3-8, column 5 of Table 14) while the green points indicate sources with (\SII/\Ha)$_{phot}$ $\geq$ 0.3 but were not spectroscopically verified as SNRs ((\SII/\Ha)$_{spec}$ $\leq$ 0.4, see Tables 9, 13, 14). In order to further test the validity of the photometric method, as mentioned in \S 3 we randomly selected sources with (\SII/\Ha)$_{phot} \le$ 0.3 for spectroscopic follow-up. These sources present (\SII/\Ha)$_{spec}$ $\leq$ 0.3 and are denoted as black points in Fig. 7 (see Tables 9, 13, 14). The solid line represents the 1:1 relation between photometric and spectroscopic \SII/\Ha\ ratios while the dashed lines denote the borderline area for SNRs ((\SII/\Ha)$\ge$ 0.4). 

Based on the above plots or the success rates in Table 14, we can estimate the detection rate expected for the candidate SNRs/probable candidate SNRs presented in Tables 3-8. The number of candidate SNRs (\SII/\Ha$_{phot}>$0.4 within their error-bars, without spectra) in all six galaxies of our sample is 229 sources while the probable candidate SNRs (0.3$<$\SII/\Ha$_{phot}<$0.4 within their error-bars, without spectra) are 122 (These numbers result from Table 14, {\it All Galaxies} section if we subtract Column 5 from Column 3). Taking into account the success percentage in SNRs from Table 14 for each galaxy, we expect to have $\sim$ 155 falsely identified SNRs. \\
This is by no means a complete catalogue of SNRs in these galaxies, particularly in the faint end where incompleteness becomes important, and photometric errors dominate in our measurements of the (\SII/\Ha)$_{phot}$ ratio. Another case would be of sources embedded in diffuse emission or near \HII\ regions for which the detection limit is higher, due to the increased background.

\subsection{Line Ratio Diagnostics}

We calculated the emission line ratios log(\Ha/(\NII\ 6548, 6584 \AA)), log(\Ha/(\SII\ 6716, 6731 \AA)) and \SII\ (6716)/\SII\ (6731)\AA\ of the spectroscopically-detected SNRs in order to place them in the diagnostic plots of \citet{Sabbadin77} and \citet{Garcia91} and investigate the region they occupy (Figs. 8-10). The extinction corrected emission lines were used when available, otherwise we used the uncorrected ones (Table 12). The locus of the different types of sources in these diagrams (dashed lines in Figs 8-10) have been created using the emission line ratios of a large number of Galactic SNRs, \HII\ regions and planetary nebulae (PNe) and can help us distinguish the excitation mechanism of the emission lines (photoionization for \HII\ regions and PNe or collisional excitation for SNRs). For comparison, we also included the spectroscopically-observed optical SNRs of four more spiral galaxies (M\,81, M\,101, NGC\,6946 and NGC\,5585) from the work of \citet{MF97}.

In Fig. 8 we plot the log(\Ha/\SII) against the log(\Ha/\NII) emission line ratios of the spectroscopically observed SNRs. All sources are within the range of \SII/\Ha\ = 0.4 - 1 which is typical for SNRs. What is intriguing though is that along the \Ha/\NII\ axis, the vast majority of the SNRs in irregular galaxies extend outside the region of Galactic SNRs in contrast to the SNRs of spiral galaxies which occupy that specific region. The region of SNRs in irregular galaxies is shifted in the direction of higher \Ha/\NII\ ratios, indicating weaker emission in the \NII\ lines. This could be due either to a difference in excitation or difference in metallicity. However, since there is no particular difference between the \SII/\Ha\ ratios (a powerful shock-excitation indicator for SNRs; Fig. 8) for the SNR populations in spiral and irregular galaxies, this suggests a difference in \Ha/\NII\ line ratios of the SNR populations between different types of galaxies due to metallicity. Indeed, irregular galaxies present typically lower metallicities in relation to spiral galaxies (as can be seen from Table 15 or in the work of e.g. \citealt{Pagel81}; \citealt{Garnett02}). Similar behaviour is seen for most of the SNRs in LMC (see \citealt{Meaburn10}) that have significantly higher \Ha/\NII\ ratios than those of Galactic SNRs. The nitrogen abundance of the LMC is lower by a factor of 2 compared with that of our Galaxy \citep{Russell92}. The effect of metallicity on the \Ha/\NII\ line ratios is also observed in other nebulae such as \HII\ regions (e.g. \citealt{Viironen07}).

In Figs 9 and 10 we plot the \SII\ \AA\ (6716)/\SII\ \AA\ (6731) line ratio against log(\Ha/\SII) and log(\Ha/\NII). The \SII\ \AA\ (6716)/\SII\ \AA\ (6731) line ratio is a good indicator of density in interstellar gas therefore it can be used to probe the effects of e.g. non-uniform ISM (which is often the case in irregular galaxies) on the properties of SNR populations between different types of galaxies. In this context, the majority of the SNRs in our sample present \SII\ \AA\ (6716)/\SII\ \AA\ (6731) line ratios between 1.06 - 1.43, indicating electron densities up to $\sim$470 cm$^{-3}$. These SNRs have low densities (\SII\ \AA\ (6716)/\SII\ \AA\ (6731) $>$ 1) and are expected to be old (c.f. \citealt{Stupar09}). Nonetheless, we do not see any trend in electron densities between SNRs in spiral and irregular galaxies. This indicates that there are not significant differences in ejecta or the circumstellar environment between spiral and irregular galaxies. Yet, the SNRs of the irregular galaxies in the \SII\ \AA\ (6716)/\SII\ \AA\ (6731) versus  log(\Ha/(\NII\ 6548 \& 6584 \AA) diagram still extend outside the Galactic region of SNRs. Although they seem to overlap with the Galactic \HII\ region, their SNR nature is beyond doubt based on their high \SII/\Ha\ ratio which appears to be sensitive to metallicity. This is the result to the definition of the \SII/\Ha\ $>$ 0.4 limit which is based on the low metallicity SNRs in the Magellanic Clouds \citep{MC73}. Their higher \Ha/\NII\ ratio could instead be the result of lower metallicity, defining this way an extended region for extragalactic SNRs in low metallicity irregular galaxies. Similar arguments hold for the shifts of our SNRs in the other diagnostic diagrams.\\

The \OIII/\Hb\ ratio is a usefull diagnostic tool for complete/incomplete recombination zones. Theoretical models of \citet{Cox85} and \citet{Hartigan87} suggest that \OIII/\Hb\ ratios below 6 indicate shocks with complete recombination zones while this value is easily exceeded for shocks with incomplete recombination zones \citep{Raymond88}. The measured values from our spectroscopically observed SNRs (Table 12) indicate shocks with complete recombination zones.

The temperature sensitive \OIII\ emission line is a good indicator of shock activity and velocity as the faster the shock propagates the stronger \OIII\ emission is produced. The absence of \OIII\ emission in many of our spectroscopically observed SNRs (Table 12) indicates slow shocks ($<$ 100 km s$^{-1}$; \citealt{Hartigan87}). In an attempt to measure the shock velocities of sources with detectable \OIII\ emission we used the plot of log(\OIII \AA\ 5007/\Hb) versus log(\NII\ \AA\ 6584/\Ha) by \citet{Allen08} (the left panel of their Fig. 21). This plot is based on the commonly used BPT diagram \citep{Baldwin81}, and uses theoretical shock model grids for different values of shock velocity, magnetic field parameters and chemical abundance. \citet{Allen08}, apart from solar abundances, calculate also grids for other chemical abundance sets such as those of LMC and SMC (0.33Z$_{\sun}$ and 0.20Z$_{\sun}$ respectively). They also created grids for shock + precursor theoretical models in various chemical abundances. We note that in both models, the \NII\ \AA\ 6584/\Ha\ ratio changes significantly with abundance in contrast to the \OIII\ \AA\ 5007/\Hb\ since nitrogen shows larger abundance differences mainly due to its secondary nucleosynthesis. 

We calculated the log(\OIII(5007)/\Hb) and log(\NII(6584)/\Ha) ratios of our spectroscopically identified SNRs and placed them on the two diagrams (Figs 11-12). For comparison, we also placed the spectroscopic SNRs of \citet{MF97}. The horizontal lines of each grid denote shock velocities of 200 - 1000  km s$^{-1}$ (from top to bottom for shock only grids and from bottom to top for shock + precursor grids) with a step of 50 km s$^{-1}$. We notice that the SNRs in the irregular galaxies (apart from NGC\,3077) of the present study are located between the LMC and SMC grids. This is not surprising since our sample of galaxies present metallicities between the metallicities of the LMC, SMC (see Table 15). However, in order to obtain accurate shock velocities for these sources new grids should be constructed taking into account the metallicities of each galaxy. Nonetheless, there are galaxies in our sample that present similar metallicities with that of LMC (e.g. NGC\,4395 and NGC\,4449). For SNRs in these galaxies that lie on the LMC grid (and not on its degenerate parts) we can give quite safe shock velocity values. For example, on the shock + precurson grid of LMC, LBZ\,2 in NGC\,4449 presents $>$ 500 km s$^{-1}$ or LBZ\,7 in the same galaxy appears a shock velocity of $\sim$ 330 km s$^{-1}$.

\subsection{Cross-correlating SNRs in multiwavelength bands}

\subsubsection{Venn diagrams}
In Figure 13 we present the overlap between optical, X-ray and radio-selected SNRs (see \S 4.3), in the form of Venn diagrams for NGC\,2403 and for all galaxies in our sample. In the case of all galaxies we include the results of NGC\,5204 even though no X-ray or radio SNRs are detected so far. The optical sources we consider in this comparison are all SNRs identified in the study, and all those previously reported in the literature. For completeness, we included the oxygen-rich SNR in NGC\,4449 as well as sources not detected by this study or classified as SNR/\HII\ but are already known optical SNRs (SNR-21, SNR-25, SNR-26, SNR-27, SNR-28, SNR-34 and SNR-35 in NGC\,2403 from \citealp{MFBL97}, SNR-3 from \citealp{Dopita10}). All multi-wavelength comparisons were performed for the same area of each galaxy. For that reason we excluded the radio SNR in NGC\,4395 \citep{Vukotic05} as it is outside the field of the Chandra data used in Paper I. We also excluded the X-ray selected candidate SNR LZB\,10 in NGC\,4395 as it is outside the field of view used in the present study. In addition, we excluded the radio candidate SNRs $\alpha$ and $\beta$ in NGC\,4214 from the work of \citet{Vukotic05}, the nature of which is debated \citep{Chomiuk09} while from the work of \citet{Chomiuk09} we consider only radio candidate SNRs, excluding SNR/\HII\ composite objects which present spectral index consistent either with an \HII\ region or SNR. \\
From the 427 optically identified SNRs in Fig. 13a (mainly on the basis of narrow-band photometry) 19 possess X-ray counterparts (corresponding to a match rate of 4.4\%), while 7 out of the 20 radio-candidate SNRs have X-ray counterparts (corresponding to a match rate of 35\%). Little overlap appears to be between optical and radio SNRs (2.1\%).

We note that oxygen-rich or Balmer dominated SNRs detected in the X-ray band, they are excluded from our optical sample since their detection method is based on the strong emission of \OIII\ and \Ha, \Hb\ Balmer lines respectively, rather than their enhanced \SII/\Ha\ ratio. However, their detection rate is expected to be low and their fraction hardly affects the match rates in the Venn diagrams. Another case of possible unidentified SNRs is plerion-type SNRs.  The known X-ray SNRs included in the Venn diagrams are selected based on their thermal, soft X-ray spectra (thermal X-ray SNRs). In this manner, plerion-type SNRs, which have hard X-ray spectra (e.g. \citealt[]{Safi-Harb01, Asaoka90}), and which nonetheless present optical properties consistent with those of SNRs (i.e. \SII/\Ha$ >$ 0.4) are excluded from the X-ray SNR sample. In order to investigate to what extent plerion-type SNRs may be missed from our comparison, we used the CSRC\footnote{http://hea-www.cfa.harvard.edu/ChandraSNR/} (Chandra Supernova Remnant Catalog) which is a comprehensive catalog of X-ray emitting SNRs detected in our Galaxy and the MCs . Based on their hard spectra (fitted with a power law model) and/or their compact emission core, we find $\sim$50 Galactic (out of 90), 3-4 LMC (out of 23), and none (out of 6) SMC X-ray SNRs. The intriguing result of higher metallicity galaxies presenting higher fractions of plerions (e.g. $>$50$\%$ for our Galaxy, 17$\%$ for the LMC and 0$\%$ for the SMC), suggests that NGC\,2403 and NGC\,3077 (the galaxies in our sample with higher metallicities, see Table 15) may host more plerion-type SNRs than the rest of the galaxies in our sample, resulting to an increased but not substantially different match rate between X-ray and optical SNRs. Additionally, as discussed in section 5.3.4, a number of wind-blown bubbles may be misclassified as optical SNRs ($\sim$10$\%$). These sources present mainly thermal radio emission instead of non-thermal synchrotron emission which is typically the case in SNRs. Therefore a significant number of wind-blown bubbles may dilute an otherwise close correlation between non-thermal radio sources and shock-excited sources identified in optical wavelengths. However, given the relatively small percentage of wind-blown bubbles in our sample we do not expect a dramatic change in match rates between optical, radio and X-ray SNRs.

 The number of optical SNRs exceed by far the number of X-ray or radio SNRs. Even if many photometric SNRs (mostly probable candidate SNRs) are not spectroscopically verified as such (see \S 5.1), the match rates will still remain low. The poor match rates between optical and X-ray/radio SNRs have also been the case for various other multi-wavelength SNR surveys (e.g. \citealt{Long10} for M33, \citealt{Pannuti07} for five nearby galaxies). This effect could be the result of various factors. The sample of radio SNRs is limited by the lack of deep radio surveys for half of our galaxies (as also discussed by \citealt{Long10} for M33). Sample issues aside, these differences could arise from physical effects since the detection rate of SNRs in different wavebands strongly depends on the properties of the surrounding medium of the source. For example, \citet{Pannuti07} point out that optical searches are more likely to detect SNRs located in regions of low diffuse emission, while radio and X-ray searches are more likely to detect SNRs in regions of high optical confusion. The same is pointed out by \citet{Long10} who find that confused environments in the optical, do not influence the detectability of the SNRs in the X-ray band. Furthermore, the trend of detecting more easily older SNRs in the optical band gives rise to the large difference in the match rate between optical and X-ray SNRs. All these facts could contribute to the large differences in match rates between optical, X-ray and radio SNRs, and highlight the importance of multi-wavelength surveys for the study of extragalactic SNR populations.

\subsubsection{SNRs or X-ray Binaries ?}

Six optically detected candidate SNRs/probable candidate SNRs (LBZ\,6, LBZ\,102, LBZ\,108, LBZ\,127 in NGC\,2403; LBZ\,80 in NGC\,4214 and LBZ\,60 in NGC\,4449) are associated with XRBs (see Tables 16, 18, 20) on the basis of their hard X-ray emission and high X-ray luminosities (Paper I). Although these sources could be considered as plerionic SNRs (due to their hard X-ray spectra and their enhanced \SII/\Ha\ ratio), their X-ray luminosities ($\sim$10$^{38}$) and their variability ($>$15$\%$ between flux observations, see Paper I) place them in the regime of XRBs rather than this of plerions (typical L$_{X}$ $\sim$10$^{35}$; \citealt{Gaensler06}). Therefore, one possible interpretation for these sources is that of an X-ray binary coincident with a SNR, possibly associated with the supernova (SN) that produced the compact object in the binary. In this case the SNR is responsible for the observed optical (and radio emission) while the binary system produces the X-ray emission. The X-ray luminosity of active XRBs (10$^{37}$ erg s$^{-1}$) is higher than that of SNRs (typically 10$^{35}$ - 10$^{37}$ erg s$^{-1}$; \citealt{Mathewson83}) and therefore they can overshadow the latter. The exemplar of this type of objects is the SS 433/W50 SNR/XRB system (e.g. \citealt{Boumis07}, \citealt{Safi99}), while a few other candidates have been identified in other galaxies on the basis of hard and/or variable X-ray sources associated with optically or radio identified SNRs (\citealt{Pannuti07}).

\subsubsection{Young SNRs or SNRs embedded in \HII\ regions ?}

The photometric investigation of the detected sources in our sample of galaxies revealed 20 low-excitation sources ((\SII/\Ha)$_{phot} <$ 0.3) which are not known optical SNRs from other studies and are associated with known X-ray or radio SNRs (see Tables 16-21). There are two possibilities for the nature of these sources: a) the SNR is in its first evolutionary stages where the optical emission is considerably fainter than the X-ray/radio emission. In this case it is also possible that their expanded shock fronts form a precursor \HII\ region \citep{Allen08} which gives the observed optical emission while the SNRs produce the X-ray or radio emission, and b) we observe SNRs embedded in \HII\ regions. In that case the \Ha\ emission we observe comes from the \HII\ region, which is enhanced in relation to that radiated from the SNR, resulting to \SII/\Ha\ ratios below our 0.3 limit for the latter which nonetheless emits in the X-ray or radio band. Follow-up spectroscopic observations will help us clarify the nature of these sources. These sources are considered as SNR/\HII\ composite objects in this study.

 \subsubsection{SNRs or wind blown bubbles?}

Multiple supernovae and/or blown out stellar winds of massive stars in OB associations can create cavities of hot gas in the ISM, known as wind blown bubbles (bubbles or superbubbles). The shock-excited structure of these objects can grant them with moderate \SII/\Ha\ values ($>$0.45; \citealt{Chen00}, \citealt{Lasker77}) since the expansion velocities of their radiative shocks are too low to produce enhanced \SII/\Ha\ ratios (e.g. \citealt{Long10}) or the UV radiation of the OB associations in a superbubble photoionizes sulfur to higher ionization stages which lead to weaker \SII/\Ha\ ratios (e.g. \citealt{Chen99}). Nonetheless there might be some superbubbles that have \SII/\Ha\ ratios within the range of SNRs. The only way to identify these objects is to use their typically larger sizes and lower luminosities compared to SNRs.
 Superbubbles present large sizes ($>$100 pc), which are rare among known SNRs \citep{Williams99}, and slower expansion velocities than those of SNRs ($<$100 km/sec; e.g. \citealt{Franchetti12}). On the other hand, their low-density environment is responsible for their rather faint X-ray emission (below that of SNRs: 10$^{34}$ - 10$^{36}$ erg s$^{-1}$; e.g. \citealt{Chu90}). In cases when a source with the above characteristics exhibits diffuse X-ray emission with luminosities similar to those of SNRs and its \Ha\ expansion velocity is high, then it is most probable that a superbubble encompasses a recently created SNR \citep{Chu90}.\\
Based on the above, we opted to investigate whether some of the SNRs we identify based on their \SII/\Ha\ ratios need to be reclassified as superbubbles. Since our low-resolution images do not allow us to reliably investigate for OB associations, and measurements for \Ha\ expansion velocities of the objects are not yet available (reduction of echelle spectra for $\sim$30 large SNRs with the 2.1m telescope in SPM, Mexico are in progress), we relied solely on identifying SNRs with large diameters ($>$80 pc). The physical size of the sources was estimated by subtracting in quadrature the seeing in each exposure (typically 1.3$\arcsec$-2.5$\arcsec$) from the aperture used to perform the source photometry. The latter was defined including most of a source's flux, while avoiding any neighbouring sources and minimizing the encompassed diffuse emission. Following this approach we set the following limits in order to reject a source as a possible wind-blown bubble: $\ge$8 pixels for NGC\,2403 (which corresponds to $\ga$2.24$\arcsec$ and physical scale of $\ga$75 pc in diameter),  $\ge$6 pixels for NGC\,5204 ($\ga$1.68$\arcsec$, $\ga$90pc diameter), $\ge$12 pixels for NGC\,4395 ($\ga$3.36$\arcsec$, $\ga$90pc diameter), $\ge$7 pixels for NGC\,4449 ($\ga$1.96$\arcsec$, $\ga$90 pc diameter), $\ge$8 pixels for NGC\,3077 ($\ga$2.24$\arcsec$, $\ga$85pc diameter) and $\ge$ 6 pixels for NGC\,4214 ($\ga$1.68$\arcsec$, $\ga$90 pc diameter). We find 51 (39 in NGC\,2403, 2 in NGC\,5204 and 10 in NGC\,4214) that fulfill the above criteria. Four of these sources are also X-ray selected SNRs based on their X-ray colors and/or their soft spectra (see Tables 3-8 and Paper I). The large sizes of these objects in combination with their relatively high X-ray luminosities  which is typical of SNRs, suggest the existence of a SNR inside a superbubble. Three more sources present X-ray properties consistent with those of XRBs, with even larger X-ray luminosities. Therefore, if we remove these seven sources from the 51 initially selected ones, we result to 44 possibly misclassified superbubbles as SNRs which constitute $\sim$10$\%$ of our optical SNR sample. This percentage is expected to be somewhat larger if we take into account the lack of expansion velocity measurements which can give us a reliable discrimination between SNRs and wind-blown bubbles.\\
We note that the estimation of the sizes described above was based on the \Ha\ morphology of the sources, rather than the \SII\ morphology which traces better the higher excitation part of the nebula associated with the shock-blown bubble. This choice was driven by the much higher S/N of the \Ha\ images, but has the tendency to overestimate the size of the SNRs or bubbles. Therefore, the estimated sizes are an upper bound on the true size of the shock-blown bubbles. Furthermore, the majority of these sources present \SII/\Ha\ ratios within the secure range of SNRs ($>$0.5), and X-ray luminosities much higher than the typical luminosities of superbubbles. Based on these two facts we expect that the majority of these 44 objects are bona-fide SNRs and for this reason we do not exclude them from our analysis. However, we do indicate them as possible superbubbles in Tables 3-8.

\subsection{Correlating X-ray selected SNRs with their optical properties}

Mining for optical SNRs within the six nearby galaxies of our sample revealed 18 sources (SNRs/candidate SNRs/probable candidate SNRs) that are associated with X-ray selected SNRs from Paper I. In order to investigate the relation between the optical and X-ray properties of these sources and examine whether the optical properties of SNRs are good predictors of X-ray SNRs, we calculated their \Ha\ luminosity and correlated it with their X-ray luminosity derived in Paper I. The \Ha\ luminosities were calculated based on the non-extinction corrected photometric fluxes F(\Ha) in Tables 3-8 and the distances from Table 1. We used the photometric F(\Ha) instead of the spectroscopic ones since we did not have spectra for all 18 sources. Two X-ray SNRs (candidate SNRs, see Paper I) were excluded from the sample since because of their small number of counts we could not extract spectra and calculate accurate fluxes (their identification was based on their X-ray colours).

In Fig. 14 we plot the \Ha\ luminosity against the non extinction-corrected X-ray luminosity of the 16 optically selected, X-ray emitting SNRs. Different colours in the plot indicate SNRs of different galaxies while the dashed line indicates the 1:1 relation between the two luminosities. The majority of SNRs tend to have higher \Ha\ than X-ray luminosities while the most luminous X-ray SNRs typical present the highest optical luminosities. However, we do not find a correlation at a statistically significant level (linear correlation coefficient:-0.12). The variation of the ratio of X-ray-to-optical luminosities indicates the existence of different materials in a wide range of temperatures: the X-ray emission originates from hot material behind the shock front (plasma temperature of $\sim$10$^{7}$ K) and long cooling timescales (typical values of a few hundreds of kyr) while the \Ha\ emission comes from cooling regions (plasma temperatures of $\sim$10$^{5}$ K) of dense recombining gas around the edges of the remnant and short cooling timescales (up to a few hundred years). With the same rationale we can interpret the lack of a significant correlation between the \SII/\Ha\ ratios of the 16 optically selected, X-ray emitting SNRs and their X-ray luminosities (Fig. 15). In a sample model one would expect that stronger shocks (higher \SII/\Ha\ ratios) would correlate with higher L$_{X}$. However, because of the long-cooling time of the X-ray material the shock-velocity we are measuring does not necessarily correspond to the shock that generated the bulk of the X-ray emission material.

\subsection{SNRs and SFR}

In order to investigate the optical properties of SNRs in different star-forming environments and derive safe conclusions on their connection with SFR, we need primarily to examine to what luminosity limit our sample of SNRs is complete. For that reason we plot the luminosity distributions of photometric SNRs in each galaxy (Fig 16). Previous studies in extragalactic SNRs showed that these populations are distributed in the form of power laws \citep[e.g.][]{Ghavamian05}. Therefore, the turnover in histograms of \Ha\ luminosities indicates the effect of incompleteness for each galaxy's SNR population. Three galaxies in our sample (NGC\,3077, NGC\,4214 and NGC\,5204) present the same limiting luminosity of 1.6$\times$10$^{37}$ erg s$^{-1}$ while for NGC\,2403 is 3.2$\times$10$^{37}$ erg s$^{-1}$, for NGC\,4395 is 4$\times$10$^{36}$ erg s$^{-1}$ and for NGC\,4449 is 2$\times$10$^{37}$ erg s$^{-1}$.

All galaxies in our sample have accurate measurements of their integrated \Ha\ luminosity \citep{Kennicutt08}, so we opted to use this as an SFR proxy. For consistency, we rederived the luminosities from \citet{Kennicutt08} based on the distances in Table 1. No corrections for extinction internal to the galaxies themselves has been applied.

Since core-collapse SNe are the end points of the evolution of the most massive stars, their SNRs are good indicators of the current SFR. This work combining multi-wavelength samples of SNRs which are often not overlaping, provides an excellent census of the SNR populations in diferent galaxies. Therefore, we would expect a linear relation between the number of opically-selected SNRs and SFR (e.g., \citealt{Condon90}). To verify this connection, we plot the number of photometric SNRs (Tables 3-8) above the completeness limit of each galaxy in our sample (see Fig. 16) against the integrated \Ha\ luminosity of each galaxy (Fig. 17, top). We find a linear relation between them but the small number of objects does not allow us to quantify their scaling relation. However, a linear correlation coefficient of 0.87 for the photometric SNRs shows that this is a significant correlation.

The non-thermal radio emission is another indicator of the SN rate and hence high-mass star formation (e.g., \citealt{Condon90}) since it comes from electrons in the magnetic field of the galaxies which are produced and accelerated in SNRs. Therefore, the radio emission should be correlated with the number of SNRs and SFR. We investigate the relation between the 1.4 GHz radio emission of the galaxies in our sample and the detected number of photometric SNRs (Fig. 17, bottom). We use integrated radio fluxes from \citet{Condon87} and we find a correlation coefficient of 0.59. However, if we remove NGC\,4449 which seems to drive out the correlation we find a linear correlation coefficient of 0.86. The weaker correlation between the number of SNRs and the radio 1.4 GHz luminosity could be due to a significant contribution of thermal radio emission to the 1.4 GHz luminosity.

\section{Conclusions}

In this paper we presented a systematic spectrophotometric study of optically emitting SNRs in a sample of six nearby galaxies. The SNRs are initially selected on the basis of their \SII/\Ha$\ge$0.4 ratio revealing a total number of $\sim$400 photometric SNRs (including sources with 0.3$<$(\SII/\Ha)$_{phot}<$0.4). Spectroscopic observations verified the nature of 67 shock-excited sources. 23 optical SNRs in this study are also detected in other wavebands. From the analysis of the sample we find that $\sim$4$\%$ and $\sim$2$\%$ of the optically selected SNRs have X-ray and radio counterparts respectively. The overlap between X-ray and radio classifications is 35$\%$. This little overlap between detection rates in different wavelengths could be either due to environmental effects (e.g. the properties of the surrounding medium which strongly affect the detection rates) or selection effects (such as the poor sensitivity of the existing radio surveys and /or the poor sensitivity of optical surveys in regions with strong star-formation and hence significant \Ha\ emission) and the different evolutionary stages in the life of a remnant.\\ 
Six sources identified as optically-selected SNRs in this study exhibit X-ray properties more consistent with XRBs. We propose that these sources are X-ray binaries coincident with an SNR.\\
The present study revealed 20 SNR/\HII\ sources (based on their narrow-band photometry) that are associated with known X-ray or radio SNRs. Two possible interpretations are of young sources that have not entered the phase of their optical radiation or of sources embedded in \HII\ regions where the SNR produces the X-ray/radio emission and the \HII\ region outshadows the shock-excited optical gas to be detected.\\
There is a trend for irregular galaxies to have lower \NII/\Ha\ ratios. This is due to the lower metallicities of these galaxies since \NII/\Ha\ is a very sensitive metallicity indicator (more than \SII/\Ha) mainly due to its secondary nucleosynthesis. \\
For the optically-emitting SNRs with X-ray counterparts, we do not see a correlation between their \Ha\ and X-ray luminosities, which is due to the presence of material in a wide range of temperatures. Additionally, we do not find any trend between the X-ray luminosity of SNRs and their \SII/\Ha\ ratios.\\
We find evidence for a linear relation between the number of luminous optical SNRs ($\sim$10$^{37}$ erg s$^{-1}$) and SFR in our sample of galaxies.

\section*{Acknowledgments}
The authors would like to thank the referee John Danziger for providing constructive comments that have improved the clarity of this manuscript. We also thank M. Allen for the fruitful discussion and for providing us with different plots on the Mapping III shock model. This program has been supported by NASA grant GO6-7086X and NASA LTSA grant G5-13056. AZ acknowledge support by the  EU IRG grant 224878. Space Astrophysics at the University of Crete is supported by EU FP7-REGPOT grant 206469 (ASTROSPACE). Skinakas Observatory is a collaborative project of the University of Crete, the Foundation for Research and Technology-Hellas and Max-Planck-Institute. The Kitt Peak National Observatory, National Optical Astronomy Observatory, is operated by the Association of Universities for Research in Astronomy (AURA), Inc., under cooperative agreement with the National Science Foundation.

\clearpage

\begin{table*}
 \centering
 \begin{minipage}{180mm}
  \caption{\,\, Properties of Our Sample of Galaxies}
\begin{tabular}{@{}ccccccccc@{}}
  \hline
\footnotetext{Note -- Units of right ascension are hours, minutes and seconds and units of declination are degrees, arcminutes and arcseconds. We note that the SA(s)m galaxies are Magellanic--type which are classified as irregulars with some spiral structure.  Coordinates, diameters, Galactic latitudes and types of galaxies are from {\it NED}.}
Galaxy & RA & DEC & Distance & Major/Minor axis &Inclination\footnotemark[1] &Galactic latitude & Type & Phys. scale\footnotemark[2] \\
 & (J2000) &(J2000) &(Mpc) &(arcmin) &(degrees) &(degrees)  & &(pc)\\ 
\hline
NGC\,2403 &07:36:51.4 &65:36:09 &3.2\footnotemark[3]  &21.9{$\times$}12.3 &62   &29    &SAB(s)cd  &21.7 \\
NGC\,5204 &13:29:36.5 &58:25:07 &4.8\footnotemark[3]  &5.0{$\times$}3.0   &53   &58    &SA(s)m    &32.6 \\
NGC\,4395 &12:25:48.9 &33:32:48 &2.6\footnotemark[4]  &13.2{$\times$}11.0 &38   &82    &SA(s)m    &17.6 \\
NGC\,4449 &12:28:11.9 &44:05:40 &4.2\footnotemark[5]  &6.2{$\times$}4.4   &56   &72    &IBm       &28.5 \\
NGC\,3077 &10:03:19.1 &68:44:02 &3.6\footnotemark[6]  &5.4{$\times$}4.5   &     &42    &I0 pec    &24.4 \\
NGC\,4214 &12:15:39.2 &36:19:37 &4.7\footnotemark[7]  &8.5{$\times$}6.6   &37   &78    &IAB(s)m   &31.9 \\
\hline
\end{tabular}
\footnotetext{1.\,\citet{Tully88} apart from NGC 4449 \citep{Summers03}}
\footnotetext{2.\,Physical radius scales for 5-pixel radius SNRs (corresponding to 1.4$\arcsec$) at the distance of each galaxy.}
\footnotetext{3.\,\citet{Freedman88}}
\footnotetext{4.\,\citet{Saha94}}
\footnotetext{5.\,\citet{Annibali08}}
\footnotetext{6.\,\citet{Freedman94}}
\footnotetext{7.\,\citet{Tully88}}
\end{minipage}
\end{table*}

\begin{table*}
 \centering
 \begin{minipage}{220mm}
  \caption{Interference Filter Characteristics}
  \begin{tabular}{cccc}
  \hline
  Filter  & $\lambda_{c}$  & $\Delta\lambda$  & T$_{peak}$\\
   &(\AA) &(\AA) &(\%) \\
\hline
\Ha+\NII 6548 \& 6584 \AA     &6570 &75  &80      \\
\SII 6716 \& 6731 \AA         &6720 &27  &80      \\
\OIII 5007 \AA                &5010 &20  &63      \\
Continuum red                 &6096 &134 &-       \\
Continuum blue                &5470 &230 &-       \\
\hline
\end{tabular}
\end{minipage}
\end{table*}

\begin{table*}
\caption{\,\,Properties of photometric SNRs in NGC 2403}
\centering
\begin{minipage}{400mm}
\begin{tabular}{@{}lcccccccccc@{}}
\hline
 SourceID &RA & Dec & Rad &An & Dan &F(\Ha) &F(\SII) &(\SII/\Ha)$_{phot}$ &Spectra & Classification \\
 & (h:m:s) &(d:m:s) &(pix) &(pix) &(pix) &(erg sec$^{-1}$ cm$^{-2}$) &(erg sec$^{-1}$ cm$^{-2}$) &&&\\
 & (J2000) &(J2000) & & & &($\times$ 10$^{-15}$) &($\times$ 10$^{-15}$) & & & \\
\hline
LBZ\,1      &07:36:30.4 &65:35:43.4  &8  &10  &10 &107.0 $\pm$0.4  &40.8 $\pm$0.3   &0.38$\pm$0.01 &yes (M) & SNR \footnote{Photometrically detected optical SNR by \citet{MF97} denoted as SNR-3. This source is also a possible superbubble (see \S5.3.4)}\\
LBZ\,2      &07:36:35.2 &65:37:17.5  &5  &10  &10 &19.4  $\pm$0.5  &12.9 $\pm$0.4   &0.66$\pm$0.03 &yes (M) &SNR\\
LBZ\,3      &07:36:41.1 &65:36:18.0  &6  &10  &10 &12.8  $\pm$0.4  &11.7 $\pm$0.3   &0.91$\pm$0.05 &yes (M) &SNR\\
LBZ\,4      &07:36:48.3 &65:34:40.3  &8  &40  &5  &39.3  $\pm$0.5  &14.1 $\pm$0.3   &0.36$\pm$0.01 &yes (M) & SNR$^{g}$\\ 
LBZ\,5      &07:36:50.6 &65:35:35.8  &6  &10  &10 &101.0 $\pm$0.6  &36.2 $\pm$0.4   &0.36$\pm$0.01 &yes (M) & SNR\\
\hline
\end{tabular}
\end{minipage}
\footnotetext[7]{Possible superbubble (see \S 5.3.4)}
\end{table*}

\begin{table*}
 \contcaption{\,\,}
\centering
 \begin{minipage}{400mm}
  \begin{tabular}{@{}lcccccccccc@{}}
\hline
 SourceID &RA & Dec & Rad &An & Dan &F(\Ha) &F(\SII) &(\SII/\Ha)$_{phot}$ &Spectra & Classification \\
 & (h:m:s) &(d:m:s) &(pix) &(pix) &(pix) &(erg sec$^{-1}$ cm$^{-2}$) &(erg sec$^{-1}$ cm$^{-2}$) &&&\\
 &(J2000) &(J2000) & & & &($\times$ 10$^{-15}$) &($\times$ 10$^{-15}$) & & & \\
\hline
LBZ\,28     &07:36:26.0 &65:37:58.8  &8  &10  &10 &18.4 $\pm$0.5   &18.2 $\pm$0.4   &0.99 $\pm$0.03 &no &candidate SNR$^{b}$ \\
LBZ\,29     &07:36:27.0 &65:37:04.8  &6  &10  &10 &23.0 $\pm$0.4   &10.9 $\pm$0.3   &0.47 $\pm$0.02 &no &candidate SNR \\
LBZ\,30     &07:36:27.2 &65:38:53.4  &6  &10  &10 &22.0 $\pm$0.4   &15.5 $\pm$0.3   &0.70 $\pm$0.02 &no &candidate SNR \\
LBZ\,31     &07:36:27.4 &65:38:48.2  &6  &10  &10 &17.4 $\pm$0.4   &10.9 $\pm$0.3   &0.63 $\pm$0.02 &no &candidate SNR \\
LBZ\,32     &07:36:29.2 &65:37:00.1  &5  &5   &2  &3.68 $\pm$0.3   &4.3  $\pm$0.3   &1.17 $\pm$0.12 &no &candidate SNR \\
\hline
\end{tabular}
\end{minipage}
\footnotetext[2]{Possible superbubble (see \S 5.3.4)}
\end{table*}

\begin{table*}
\centering
 \contcaption{\,\,}
 \begin{minipage}{400mm}
\begin{tabular}{@{}lcccccccccc@{}}
\hline
 SourceID &RA & Dec & Rad &An & Dan &F(\Ha) &F(\SII) &(\SII/\Ha)$_{phot}$ &Spectra & Classification \\
 & (h:m:s) &(d:m:s) &(pix) &(pix) &(pix) &(erg sec$^{-1}$ cm$^{-2}$) &(erg sec$^{-1}$ cm$^{-2}$) &&&\\
 & (J2000) &(J2000) & & & &($\times$ 10$^{-15}$) &($\times$ 10$^{-15}$) & & & \\
\hline
LBZ\,59     &07:36:42.5 &65:39:03.7  &6  &10  &10 &18.9 $\pm$0.2   &7.9  $\pm$0.2   &0.42 $\pm$0.01 &no &candidate SNR \\
LBZ\,60     &07:36:42.9 &65:34:51.9  &6  &10  &10 &8.9  $\pm$0.3   &5.9  $\pm$0.2   &0.67 $\pm$0.03 &no &candidate SNR\footnote{Optically detected, photometric SNR (SNR-5) by \citet{MF97}} \\
LBZ\,61     &07:36:44.1 &65:39:10.9  &6  &10  &10 &16.9 $\pm$0.3   &8.3  $\pm$0.2   &0.49 $\pm$0.02 &no &candidate SNR \\
LBZ\,62     &07:36:44.2 &65:36:45.7  &6  &10  &10 &18.8 $\pm$0.4   &18.2 $\pm$0.3   &0.97 $\pm$0.04 &no &candidate SNR \\
LBZ\,63     &07:36:44.2 &65:37:20.5  &6  &10  &10 &18.9 $\pm$0.6   &10.3 $\pm$0.5   &0.54 $\pm$0.03 &no &candidate SNR \\
\hline
\end{tabular}
\end{minipage}
\footnotetext[6]{Possible superbubble (see \S 5.3.4)}
\end{table*}

\begin{table*}
\centering
\begin{minipage}{400mm}
\contcaption{\,\,}
\begin{tabular}{@{}lcccccccccc@{}}
\hline
 SourceID &RA & Dec & Rad &An & Dan &F(\Ha) &F(\SII) &(\SII/\Ha)$_{phot}$ &Spectra & Classification \\
 & (h:m:s) &(d:m:s) &(pix) &(pix) &(pix) &(erg sec$^{-1}$ cm$^{-2}$) &(erg sec$^{-1}$ cm$^{-2}$) &&&\\
 & (J2000) &(J2000) & & & &($\times$ 10$^{-15}$) &($\times$ 10$^{-15}$) & & & \\
\hline
LBZ\,89     &07:36:53.4 &65:35:59.8  &6  &10  &10 &58.4 $\pm$0.6   &34.2 $\pm$0.5   &0.59 $\pm$0.02 &no &candidate SNR\footnote{Optically detected, photometric SNR (SNR-11) by \citet{MF97}} \\ 
LBZ\,90     &07:36:53.8 &65:33:41.7  &8  &10  &10 &68.8 $\pm$0.2   &30.3 $\pm$0.2   &0.44 $\pm$0.01 &no &candidate SNR\footnote{Spectroscopically observed, optical SNR (SNR-12) by \citet{MF97}. Also possible superbubble (see \S 5.3.4)} \\
LBZ\,91     &07:36:54.3 &65:34:04.0  &5  &10  &10 &3.2  $\pm$0.2   &2.5  $\pm$0.1   &0.80 $\pm$0.06 &no &candidate SNR\\
LBZ\,92     &07:36:54.4 &65:35:11.0  &5  &10  &10 &4.0  $\pm$0.4   &2.5  $\pm$0.2   &0.63 $\pm$0.07 &no &candidate SNR\\
LBZ\,93     &07:36:55.1 &65:35:38.1  &10 &15  &5  &78.1 $\pm$0.9   &43.0 $\pm$0.5   &0.55 $\pm$0.01 &no &candidate SNR\footnote{Optically detected, photometric SNR (SNR-14) by \citet{MF97}. Also possible superbubble (see \S 5.3.4)} \\
\hline
\end{tabular}
\end{minipage}
\footnotetext[13]{Possible superbubble (see \S 5.3.4)}
\end{table*}

\begin{table*}
\centering
\begin{minipage}{400mm}
\contcaption{\,\,}
\begin{tabular}{@{}lcccccccccc@{}}
\hline
 SourceID &RA & Dec & Rad &An & Dan &F(\Ha) &F(\SII) &(\SII/\Ha)$_{phot}$ &Spectra & Classification \\
 & (h:m:s) &(d:m:s) &(pix) &(pix) &(pix) &(erg sec$^{-1}$ cm$^{-2}$) &(erg sec$^{-1}$ cm$^{-2}$) &&&\\
 &(J2000) &(J2000) & & & &($\times$ 10$^{-15}$) &($\times$ 10$^{-15}$) & & & \\
\hline
LBZ\,116    &07:36:30.7 &65:35:48.6  &8 &10 &10 &74.2 $\pm$0.3  &24.4 $\pm$0.2  &0.33 $\pm$0.01  &no &probable candidate SNR$^{j}$ \\
LBZ\,117    &07:36:37.0 &65:36:35.4  &6 &10 &10 &110.0$\pm$0.4  &35.2 $\pm$0.3  &0.32 $\pm$0.01  &no &probable candidate SNR \\
LBZ\,118    &07:36:37.0 &65:36:39.1  &8 &10 &10 &311.0$\pm$0.6  &114.0$\pm$0.5  &0.37 $\pm$0.01  &no &probable candidate SNR\footnote{Spectroscopically observed, optical SNR (SNR-4) by \citet{MF97}. Also possible superbubble (see \S 5.3.4)}  \\
LBZ\,119    &07:36:38.1 &65:36:26.2  &6 &10 &10 &151.0$\pm$0.5  &44.8 $\pm$0.4  &0.30 $\pm$0.01  &no &probable candidate SNR \\
LBZ\,120    &07:36:40.2 &65:39:22.0  &6 &10 &10 &23.0 $\pm$0.2  &8.60 $\pm$0.2  &0.37 $\pm$0.01  &no &probable candidate SNR \\
\hline
\end{tabular}
\end{minipage}
\footnotetext[10]{Possible superbubble (see \S 5.3.4)}
\end{table*}

\begin{table*}
\centering
\begin{minipage}{400mm}
\caption{\,\,Properties of photometric SNRs in NGC\,3077}
\begin{tabular}{@{}lcccccccccc@{}}
\hline
 SourceID &RA & Dec & Rad &An & Dan &F(\Ha) &F(\SII) &(\SII/\Ha)$_{phot}$ &Spectra & Classification \\
 & (h:m:s) &(d:m:s) &(pix) &(pix) &(pix) &(erg sec$^{-1}$ cm$^{-2}$) &(erg sec$^{-1}$ cm$^{-2}$) &&&\\
 & (J2000) &(J2000) & & & &($\times$ 10$^{-15}$) &($\times$ 10$^{-15}$) & & & \\
\hline
LBZ\,1     &10:03:14.2  &68:43:52.1  &4   &10   &10   &10.3$\pm$0.7  &4.6$\pm$0.3  &0.45$\pm$0.08  &yes (M) &SNR \\
LBZ\,2     &10:03:16.5  &68:44:41.8  &4   &10   &10   &7.0$\pm$0.4   &3.6$\pm$0.2  &0.51$\pm$0.08  &yes (M) &SNR \\
LBZ\,3     &10:03:17.1  &68:44:37.5  &5   &10   &10   &24.5$\pm$0.6  &6.8$\pm$0.3  &0.28$\pm$0.02  &yes (M) &SNR  \\
LBZ\,4     &10:03:18.4  &68:43:19.3  &5   &5    &10   &13.7$\pm$0.6  &3.8$\pm$0.3  &0.28$\pm$0.04  &yes (M) &SNR  \\
LBZ\,5     &10:03:18.8  &68:43:37.9  &4   &5    &5    &3.6$\pm$0.7   &2.7$\pm$0.3  &0.75$\pm$0.34  &yes (M) &SNR \\
\hline
\end{tabular}
\end{minipage}
\end{table*}

\begin{table*}
\centering
\begin{minipage}{400mm}
\caption{\,\,Properties of photometric SNRs in NGC\,4214}
\begin{tabular}{@{}lcccccccccc@{}}
\hline
SourceID  &RA & Dec & Rad &An & Dan &F(\Ha) &F(\SII) &(\SII/\Ha)$_{phot}$ &Spectra & Classification \\
 & (h:m:s) &(d:m:s) &(pix) &(pix) &(pix) &(erg sec$^{-1}$ cm$^{-2}$) &(erg sec$^{-1}$ cm$^{-2}$) &&&\\
 & (J2000) &(J2000) & & & &($\times$ 10$^{-15}$) &($\times$ 10$^{-15}$) & & &  \\
\hline
LBZ\,1   &12:15:33.6 &36:19:26.9  &5  &ext&ext  &18.8$\pm$0.2  &4.7$\pm$0.3  &0.25$\pm$0.03  &yes (S) &SNR \\
LBZ\,2   &12:15:33.8 &36:19:12.0  &3  &20 &10   &2.4$\pm$0.1   &1.7$\pm$0.2  &0.71$\pm$0.15  &yes (M) &SNR \\
LBZ\,3   &12:15:33.8 &36:19:30.9  &5  &ext&ext  &22.2$\pm$0.2  &5.3$\pm$0.3  &0.24$\pm$0.02  &yes (S) &SNR \\
LBZ\,4   &12:15:35.7 &36:17:41.6  &5  &10 &10   &16.8$\pm$0.2  &9.3$\pm$0.2  &0.55$\pm$0.03  &yes (M) &SNR \\
LBZ\,5   &12:15:37.6 &36:20:12.3  &3  &10 &10   &4.1$\pm$0.2   &3.0$\pm$0.3  &0.73$\pm$0.16  &yes (S) &SNR \\
\hline
\end{tabular}
\end{minipage}
\footnotetext[4]{Possible superbubble (see \S 5.3.4)}
\end{table*}

\begin{table*}
\centering
\begin{minipage}{400mm}
\contcaption{\,\,}
\begin{tabular}{@{}lcccccccccc@{}}
\hline
 SourceID  &RA & Dec & Rad &An & Dan &F(\Ha) &F(\SII) &(\SII/\Ha)$_{phot}$ &Spectra & Classification \\
 & (h:m:s) &(d:m:s) &(pix) &(pix) &(pix) &(erg sec$^{-1}$ cm$^{-2}$) &(erg sec$^{-1}$ cm$^{-2}$) &&&\\
 & (J2000) &(J2000) & & & &($\times$ 10$^{-15}$) &($\times$ 10$^{-15}$) & & &  \\
\hline
LBZ\,36  &12:15:34.0 &36:21:53.5  &5  &10 &10   &5.3$\pm$0.1  &2.3$\pm$0.2  &0.43$\pm$0.07  &no  &candidate SNR\\
LBZ\,37  &12:15:34.5 &36:18:23.9  &5  &10 &10   &6.2$\pm$0.2  &4.1$\pm$0.2  &0.67$\pm$0.08  &no  &candidate SNR\\
LBZ\,38  &12:15:35.2 &36:21:18.4  &5  &10 &10   &5.8$\pm$0.2  &2.7$\pm$0.2  &0.46$\pm$0.07  &no  &candidate SNR\\
LBZ\,39  &12:15:35.2 &36:22:16.2  &7  &15 &5    &8.7$\pm$0.2  &8.5$\pm$0.3  &0.98$\pm$0.08  &no  &candidate SNR$^{d}$\\
LBZ\,40  &12:15:35.4 &36:21:12.2  &5  &10 &10   &4.6$\pm$0.2  &2.0$\pm$0.2  &0.43$\pm$0.09  &no  &candidate SNR\\
\hline
\end{tabular}
\end{minipage}
\footnotetext[4]{Possible superbubble (see \S 5.3.4)}
\end{table*}

\begin{table*}
\centering
\begin{minipage}{400mm}
\contcaption{\,\,}
\begin{tabular}{@{}lcccccccccc@{}}
\hline
 SourceID  &RA & Dec & Rad &An & Dan &F(\Ha) &F(\SII) &(\SII/\Ha)$_{phot}$ &Spectra & Classification \\
 & (h:m:s) &(d:m:s) &(pix) &(pix) &(pix) &(erg sec$^{-1}$ cm$^{-2}$) &(erg sec$^{-1}$ cm$^{-2}$) &&&\\
 & (J2000) &(J2000) & & & &($\times$ 10$^{-15}$) &($\times$ 10$^{-15}$) & & &  \\
\hline
LBZ\,68  &12:15:47.1 &36:17:14.8  &5  &10 &10   &4.8$\pm$0.1   &2.7$\pm$0.2  &0.56$\pm$0.06  &no  &candidate SNR\\
LBZ\,69-1&12:15:47.6 &36:17:37.2  &5  &10 &10   &8.2$\pm$0.1   &5.4$\pm$0.2  &0.65$\pm$0.05  &no  &candidate SNR\\
LBZ\,70-2&12:15:47.7 &36:17:35.5  &5  &10 &10   &10.1$\pm$0.1  &7.5$\pm$0.2  &0.74$\pm$0.04  &no  &candidate SNR\\
LBZ\,71-3&12:15:47.8 &36:17:37.1  &5  &10 &10   &7.9$\pm$0.1   &5.6$\pm$0.2  &0.71$\pm$0.05  &no  &candidate SNR\\
LBZ\,72-4&12:15:47.9 &36:17:36.7  &5  &10 &10   &8.6$\pm$0.1   &5.9$\pm$0.2  &0.69$\pm$0.05  &no  &candidate SNR\\
\hline
\end{tabular}
\end{minipage}
\footnotetext[6]{Possible superbubble (see \S 5.3.4)}
\end{table*}

\begin{table*}
\centering
\begin{minipage}{400mm}
\caption{\,\,Properties of photometric SNRs in NGC\,4395}
\begin{tabular}{@{}lcccccccccc@{}}
\hline
 SourceID  &RA & Dec & Rad &An & Dan &F(\Ha) &F(\SII) &(\SII/\Ha)$_{phot}$ &Spectra & Classification \\
 & (h:m:s) &(d:m:s) &(pix) &(pix) &(pix) &(erg sec$^{-1}$ cm$^{-2}$) &(erg sec$^{-1}$ cm$^{-2}$) &&\\
 & (J2000) &(J2000) & & & &($\times$ 10$^{-15}$) &($\times$ 10$^{-15}$) & & &  \\
\hline
LZB\,1   &12:25:44.9  &33:30:36.3  &6   &10   &10   &14.0$\pm$0.2 &3.5$\pm$0.1  &0.25$\pm$0.02  &yes (M)&SNR       \\
LBZ\,2   &12:25:46.1  &33:30:47.2  &4   &10   &10   &4.3$\pm$0.1  &1.2$\pm$0.1  &0.27$\pm$0.04  &yes (M)&SNR       \\
LBZ\,3   &12:25:47.2  &33:32:45.3  &3   &10   &10   &5.5$\pm$0.1  &1.4$\pm$0.1  &0.26$\pm$0.04  &yes (M)&SNR       \\
LBZ\,4   &12:25:53.4  &33:31:03.9  &5   &20   &30   &14.0$\pm$0.2 &3.3$\pm$0.1  &0.24$\pm$0.02  &yes (M)&SNR       \\
LBZ\,5   &12:25:54.1  &33:30:57.6  &6   &15   &5    &17.0$\pm$0.3 &4.3$\pm$0.1  &0.25$\pm$0.02  &yes (S)&SNR       \\
\hline
\end{tabular}
\end{minipage}
\end{table*}

\clearpage

\begin{table*}
\centering
\begin{minipage}{400mm}
\contcaption{\,\,}
\begin{tabular}{@{}lcccccccccc@{}}
\hline
 SourceID  &RA & Dec & Rad &An & Dan &F(\Ha) &F(\SII) &(\SII/\Ha)$_{phot}$ &Spectra & Classification \\
 & (h:m:s) &(d:m:s) &(pix) &(pix) &(pix) &(erg sec$^{-1}$ cm$^{-2}$) &(erg sec$^{-1}$ cm$^{-2}$) &&&\\
 & (J2000) &(J2000) & & & &($\times$ 10$^{-15}$) &($\times$ 10$^{-15}$) & & &  \\
\hline
LBZ\,33 &12:25:51.9  &33:30:32.8  &4   &10   &10   &4.3$\pm$0.1  &1.1$\pm$0.1  &0.26$\pm$0.04  &no &probable candidate SNR \\
LBZ\,34 &12:25:53.6  &33:35:09.9  &4   &10   &10   &3.4$\pm$0.1  &0.9$\pm$0.1  &0.26$\pm$0.04  &no &probable candidate SNR \\
LBZ\,35 &12:25:53.7  &33:30:57.4  &4   &10   &10   &5.8$\pm$0.2  &1.8$\pm$0.1  &0.32$\pm$0.05  &no &probable candidate SNR \\
LBZ\,36 &12:25:53.9  &33:30:49.5  &4   &10   &10   &3.6$\pm$0.2  &1.1$\pm$0.1  &0.29$\pm$0.09  &no &probable candidate SNR \\
LBZ\,37 &12:25:55.7  &33:31:30.5  &4   &10   &10   &8.7$\pm$0.2  &2.6$\pm$0.1  &0.30$\pm$0.03  &no &probable candidate SNR \\
\hline
\end{tabular}
\end{minipage}
\end{table*}

\begin{table*}
\centering
\begin{minipage}{400mm}
\caption{\,\,Properties of photometric SNRs in NGC\,4449}
\begin{tabular}{@{}lcccccccccc@{}}
\hline
SourceID &RA & Dec & Rad &An & Dan &F(\Ha) &F(\SII) &(\SII/\Ha)$_{phot}$ &Spectra & Classification \\
 & (h:m:s) &(d:m:s) &(pix) &(pix) &(pix) &(erg sec$^{-1}$ cm$^{-2}$) &(erg sec$^{-1}$ cm$^{-2}$) &&&\\
 & (J2000) &(J2000) & & & &($\times$ 10$^{-15}$) &($\times$ 10$^{-15}$) & & &  \\
\hline
LBZ\,1    &12:28:07.3     &44:04:44.9    &4    &4   &5    &3.8$\pm$0.2   &3.6$\pm$0.3   &0.93$\pm$0.21  &yes (M)&SNR   \\
LBZ\,2    &12:28:07.4     &44:05:51.6    &5    &10  &10   &8.0$\pm$0.3   &4.4$\pm$0.3   &0.54$\pm$0.08  &yes (M)&SNR   \\
LBZ\,3    &12:28:09.5     &44:05:49.2    &5    &10  &10   &19.0$\pm$0.4  &7.0$\pm$0.5   &0.39$\pm$0.05  &yes (M)&SNR   \\ 
LBZ\,4    &12:28:11.8     &44:05:13.9    &5    &10  &10   &17.0$\pm$0.5  &5.4$\pm$0.6   &0.30$\pm$0.06  &yes (M)&SNR   \\
LBZ\,5    &12:28:11.9     &44:04:49.4    &5    &5   &5    &7.5$\pm$0.3   &4.4$\pm$0.3   &0.59$\pm$0.10  &yes (M)&SNR   \\
\hline
\end{tabular}
\end{minipage}
\end{table*}

\begin{table*}
\centering
\begin{minipage}{400mm}
\contcaption{\,\,}
\begin{tabular}{@{}lcccccccccc@{}}
\hline
 SourceID &RA & Dec & Rad &An & Dan &F(\Ha) &F(\SII) &(\SII/\Ha)$_{phot}$ &Spectra & Classification \\
 & (h:m:s) &(d:m:s) &(pix) &(pix) &(pix) &(erg sec$^{-1}$ cm$^{-2}$) &(erg sec$^{-1}$ cm$^{-2}$) &&&\\
 & (J2000) &(J2000) & & & &($\times$ 10$^{-15}$) &($\times$ 10$^{-15}$) & & &  \\
\hline
LBZ\,41   &12:28:12.2     &44:06:47.4    &5    &10  &10   &8.2$\pm$0.4   &4.4$\pm$0.4   &0.54$\pm$0.10  &no  &candidate SNR \\
LBZ\,42   &12:28:12.2     &44:06:50.9    &5    &10  &10   &7.5$\pm$0.4   &2.7$\pm$0.3   &0.36$\pm$0.09  &no  &candidate SNR \\
LBZ\,43-1 &12:28:13.8     &44:06:35.2    &5    &10  &10   &6.5$\pm$0.4   &6.2$\pm$0.4   &0.85$\pm$0.31  &no  &candidate SNR \\
LBZ\,44-2 &12:28:14.0     &44:06:33.7    &5    &10  &10   &10.0$\pm$0.4  &3.4$\pm$0.4   &0.44$\pm$0.07  &no  &candidate SNR \\
LBZ\,45-3 &12:28:14.0     &44:06:35.4    &5    &5   &2    &11.0$\pm$0.4  &3.1$\pm$0.4   &0.51$\pm$0.13  &no  &candidate SNR \\
\hline
\end{tabular}
\end{minipage}
\end{table*}

\begin{table*}
\centering
\begin{minipage}{400mm}
\caption{\,\,Properties of photometric SNRs in NGC\,5204}
\begin{tabular}{@{}lcccccccccc@{}}
\hline
 SourceID &RA & Dec & Rad &An & Dan &F(\Ha) &F(\SII) &(\SII/\Ha)$_{phot}$ &Spectra & Classification  \\
 & (h:m:s) &(d:m:s) &(pix) &(pix) &(pix) &(erg sec$^{-1}$ cm$^{-2}$) &(erg sec$^{-1}$ cm$^{-2}$) &&&\\
 & (J2000) &(J2000) & & & &($\times$ 10$^{-15}$) &($\times$ 10$^{-15}$) & & &  \\
\hline
LBZ\,1      &13:29:32.9  &58:24:53.3  &5   &6    &5    &5.1$\pm$0.3   &1.3$\pm$0.2  &0.26$\pm$0.08  &yes (S)&SNR \\
LBZ\,2      &13:29:33.2  &58:24:48.4  &5   &10   &10   &4.6$\pm$0.4   &1.4$\pm$0.2  &0.29$\pm$0.10  &yes (S)&SNR \\
LBZ\,3      &13:29:33.7  &58:25:09.4  &7   &8    &2    &20.5$\pm$0.6  &2.9$\pm$0.3  &0.15$\pm$0.03  &yes (S)&SNR$^{d}$\\
LBZ\,4      &13:29:34.5  &58:24:23.8  &5   &10   &10   &10.5$\pm$0.3  &2.9$\pm$0.2  &0.28$\pm$0.04  &yes (M)&SNR\footnote{Optically, spectroscopically-verified SNR by \citet{MF97}, denoted as SNR 2} \\
LBZ\,5      &13:29:38.8  &58:25:34.8  &5   &10   &10   &10.9$\pm$0.4  &2.7$\pm$0.2  &0.25$\pm$0.04  &yes (M)&SNR \\
\hline
\end{tabular}
\end{minipage}
\footnotetext[4]{Possible superbubble (see \S 5.3.4)}
\end{table*}

\begin{table*}
\centering
\begin{minipage}{220mm}
\caption{\,\,Photometric properties of spectroscopically-verified non-SNRs in our sample of galaxies}
\begin{tabular}{@{}cccccccccc@{}}
\hline
 Galaxy&SourceID &RA & Dec & Rad &An & Dan &F(\Ha) &F(\SII) &(\SII/\Ha)$_{phot}$ \\
& & (h:m:s) &(d:m:s) &(pix) &(pix) &(pix) &(erg sec$^{-1}$ cm$^{-2}$) &(erg sec$^{-1}$ cm$^{-2}$) & \\
& & (J2000) &(J2000) & & & &($\times$ 10$^{-15}$) &($\times$ 10$^{-15}$) & \\
\hline 
NGC\,2403 &LBZ\,1059 &07:36:13.6 &65:37:57.5  &5 &10 &10  &10.7 $\pm$0.2  &9.6 $\pm$0.2  &0.90$\pm$0.03  \\
-         &LBZ\,731  &07:36:27.1 &65:36:30.2  &5 &10 &10  &41.5 $\pm$0.2  &17.3$\pm$0.2  &0.42$\pm$0.01  \\
-         &LBZ\,561  &07:36:28.4 &65:37:02.6  &5 &10 &10  &3.3  $\pm$0.4  &5.6 $\pm$0.3  &1.72$\pm$0.22  \\
-         &LBZ\,484  &07:36:32.1 &65:35:42.7  &5 &10 &10  &17.0 $\pm$0.4  &4.1 $\pm$0.2  &0.24$\pm$0.01  \\
-         &LBZ\,982  &07:36:35.1 &65:37:35.7  &5 &10 &10  &61.3 $\pm$0.7  &25.8$\pm$0.6  &0.42$\pm$0.01  \\
\hline
\end{tabular}
\end{minipage}
\end{table*}

\begin{table*}
 \centering
\begin{minipage}{140mm}
  \caption{Slit centres for Skinakas spectroscopy}
  \begin{tabular}{cccc}
  \hline
\multicolumn{1}{c}{} & \multicolumn{2}{c}{Slit center}\\
\cline{2-3}\\
Galaxy & RA & DEC &Total Exposure Time\footnotemark[1] \\
 & (h:m:s) &(d:m:s) &(sec)\\
&(J2000)& (J2000) & \\ 

\hline
NGC 3077 -- Slit 1  &10:03:23.3 &+68:44:27.1 &3000 (1)\\
NGC 4214 -- Slit 1  &12:15:33.7 &+36:19:07.2 &5400 (3)\\
NGC 4214 -- Slit 2  &12:15:37.6 &+36:18:30.2 &3600 (2)\\
NGC 4395 -- Slit 1  &12:25:54.1 &+33:31:39.6 &5400 (3)\\
NGC 4395 -- Slit 2  &12:25:41.1 &+33:30:56.0 &3600 (2)\\
NGC 4449 -- Slit 1  &12:28:13.3 &+44:05:20.8 &5400 (3)\\
NGC 5204 -- Slit 1  &13:29:33.3 &+58:25:06.1 &3600 (2)\\
NGC 5204 -- Slit 2  &13:29:40.0 &+58:24:40.9 &3600 (2)\\
\hline
\end{tabular}
\footnotetext{1.\,Numbers in parenthesis denote the number of exposures.}
\end{minipage}
\end{table*}

\begin{table*}
\centering
\begin{minipage}{400mm}
\caption{\,Emission Line Fluxes of spectroscopically observed SNRs in our sample of galaxies}
 \begin{tabular}{@{}cccccccccccccccc@{}}
 \hline
Line &\Hb &\OIII &\OIII &\OI &\HeI &\HeII &\OI &\OI &\NII &\Ha &\NII &\HeI &\SII &\SII &\ArIII\\
 (\AA)&(4861) &(4959) &(5007) &(5577) &(5876) &(6234) &(6300) &(6364) &(6548) &(6563) &(6584) &(6678) &(6716) &(6731) &(7136) \\
\hline
\multicolumn{16}{c}{NGC\,2403 -  LBZ\,1} \\
\hline
F   &31 &- &14 &- &- &- &- &- &7 &100 &26 &- &28 &22 &- \\
I   &35 &- &16 &- &- &- &- &- &7 &100 &26 &- &28 &21 &- \\
S/N &8  &- &6  &- &- &- &- &- &7 &30  &14 &- &15 &12 &- \\
\hline
\multicolumn{16}{c}{NGC\,2403 -  LBZ\,2}\\
\hline
F   &14 &- &8 &-  &- &- &3  &- &7 &100 &27 &- &24 &18 &- \\
\hline
\end{tabular}
\end{minipage}
\end{table*}

\clearpage

\begin{table*}
\centering
\begin{minipage}{400mm}
\contcaption{\,\,}
 \begin{tabular}{@{}cccccccccccccccc@{}}
 \hline
Line &\Hb &\OIII &\OIII &\OI &\HeI &\HeII &\OI &\OI &\NII &\Ha &\NII &\HeI &\SII &\SII &\ArIII\\
 (\AA)&(4861) &(4959) &(5007) &(5577) &(5876) &(6234) &(6300) &(6364) &(6548) &(6563) &(6584) &(6678) &(6716) &(6731) &(7136) \\
\hline
\multicolumn{16}{c}{NGC\,2403 -  LBZ\,9}\\
\hline
F   &34 &11 &32 &16 &5 &- &10 &4 &9 &100 &21 &2 &24 &17 &3 \\
I   &35 &11 &32 &16 &5 &- &10 &4 &9 &100 &21 &2 &24 &17 &3 \\
S/N &10 &5  &13 &12 &5 &- &12 &8 &9 &37  &16 &3 &17 &14 &5 \\
\hline
\multicolumn{16}{c}{NGC\,2403 -  LBZ\,10}\\
\hline
F   &16 &- &32 &- &- &- &- &- &9 &100 &24 &- &35 &26 &- \\
\hline
\end{tabular}
\end{minipage}
\end{table*}

\clearpage

\begin{table*}
\centering
 \begin{minipage}{400mm}
\contcaption{\,\,}
 \begin{tabular}{@{}cccccccccccccccc@{}}
 \hline
Line &\Hb &\OIII &\OIII &\OI &\HeI &\HeII &\OI &\OI &\NII &\Ha &\NII &\HeI &\SII &\SII &\ArIII\\
 (\AA)&(4861) &(4959) &(5007) &(5577) &(5876) &(6234) &(6300) &(6364) &(6548) &(6563) &(6584) &(6678) &(6716) &(6731) &(7136) \\
\hline
\multicolumn{16}{c}{NGC\,3077 -  LBZ\,5}\\
\hline
F   &16 &- &17 &22 &- &- &4 &- &10&100 &35&- &26 &20 &- \\
I   &35 &- &34 &32 &- &- &5 &- &11&100 &35&- &24 &19 &- \\ 
S/N &6  &- &5  &10 &- &- &6 &- &6 &33  &16&- &15 &10 &- \\ 
\hline
\multicolumn{16}{c}{NGC\,3077 -  LBZ\,6}\\
\hline
F   &- &- &- &- &- &- &- &- &17 &100 &27 &- &27 &27 &- \\
\hline
\end{tabular}
\end{minipage}
\end{table*}

\begin{table*}
\centering
\begin{minipage}{400mm}
\contcaption{\,\,}
 \begin{tabular}{@{}cccccccccccccccc@{}}
 \hline
Line &\Hb &\OIII &\OIII &\OI &\HeI &\HeII &\OI &\OI &\NII &\Ha &\NII &\HeI &\SII &\SII &\ArIII\\
 (\AA)&(4861) &(4959) &(5007) &(5577) &(5876) &(6234) &(6300) &(6364) &(6548) &(6563) &(6584) &(6678) &(6716) &(6731) &(7136) \\
\hline
\multicolumn{16}{c}{NGC\,4214 -  LBZ\,7}\\
\hline
F   &15 &- &27 &- &- &- &- &- &- &100 &13 &- &44 &34 &- \\
I   &35 &- &56 &- &- &- &- &- &- &100 &13 &- &41 &32 &- \\
S/N &8  &- &14 &- &- &- &- &- &- &41  &12 &- &22 &16 &- \\
\hline
\multicolumn{16}{c}{NGC\,4214 -  LBZ\,8}\\
\hline
F   &- &- &- &- &- &- &18 &- &- &100 &- &- &34 &29 &- \\
\hline
\end{tabular}
\end{minipage}
\end{table*}

\clearpage

\begin{table*}
\centering
\begin{minipage}{400mm}
\contcaption{\,\,}
 \begin{tabular}{@{}cccccccccccccccc@{}}
 \hline
Line &\Hb &\OIII &\OIII &\OI &\HeI &\HeII &\OI &\OI &\NII &\Ha &\NII &\HeI &\SII &\SII &\ArIII\\
 (\AA)&(4861) &(4959) &(5007) &(5577) &(5876) &(6234) &(6300) &(6364) &(6548) &(6563) &(6584) &(6678) &(6716) &(6731) &(7136) \\
\hline
\multicolumn{16}{c}{NGC\,4214 -  LBZ\,15}\\
\hline
F   &28 &12 &47 &- &- &- &4 &- &3 &100 &16 &- &26 &18 &- \\
I   &35 &15 &57 &- &- &- &4 &- &3 &100 &16 &- &25 &18 &- \\
S/N &11 &10 &19 &- &- &- &8 &- &6 &41  &14 &- &19 &15 &- \\
\hline
\multicolumn{16}{c}{NGC\,4214 -  LBZ\,16}\\
\hline
F   &24 &26 &82  &- &3  &- &14 &4  &4  &100 &13 &- &34 &24 &3 \\
\hline
\end{tabular}
\end{minipage}
\end{table*}

\clearpage

\begin{table*}
\centering
\begin{minipage}{400mm}
\contcaption{\,\,}
 \begin{tabular}{@{}cccccccccccccccc@{}}
 \hline
Line &\Hb &\OIII &\OIII &\OI &\HeI &\HeII &\OI &\OI &\NII &\Ha &\NII &\HeI &\SII &\SII &\ArIII\\
 (\AA)&(4861) &(4959) &(5007) &(5577) &(5876) &(6234) &(6300) &(6364) &(6548) &(6563) &(6584) &(6678) &(6716) &(6731) &(7136) \\
\hline
\multicolumn{16}{c}{NGC\,4395 -  LBZ\,5}\\
\hline
F   &22 &12 &53 &- &- &- &- &- &- &100 &15 &- &35 &28 &- \\
I   &35 &18 &78 &- &- &- &- &- &- &100 &14 &- &34 &26 &- \\
S/N &9  &8  &15 &- &- &- &- &- &- &38  &13 &- &20 &16 &- \\
\hline
\multicolumn{16}{c}{NGC\,4395 -  LBZ\,6}\\
\hline
F   &17 &10 &49 &- &- &- &- &- &- &100 &16 &- &35 &27 &- \\
\hline
\end{tabular}
\end{minipage}
\end{table*}

\begin{table*}
\centering
 \begin{minipage}{400mm}
\contcaption{\,\,}
 \begin{tabular}{@{}cccccccccccccccc@{}}
 \hline
Line &\Hb &\OIII &\OIII &\OI &\HeI &\HeII &\OI &\OI &\NII &\Ha &\NII &\HeI &\SII &\SII &\ArIII\\
 (\AA)&(4861) &(4959) &(5007) &(5577) &(5876) &(6234) &(6300) &(6364) &(6548) &(6563) &(6584) &(6678) &(6716) &(6731) &(7136) \\
\hline
\multicolumn{16}{c}{NGC\,4449 -  LBZ\,6b}\\
\hline
F   &28 &14 &48 &21 &- &- &4 &- &3 &100 &15 &- &32 &23 &- \\
I   &35 &17 &59 &23 &- &- &4 &- &3 &100 &15 &- &32 &22 &- \\
S/N &10 &6  &14 &16 &- &- &7 &- &7 &38  &12 &- &19 &16 &- \\
\hline
\multicolumn{16}{c}{NGC\,4449 -  LBZ\,7}\\
\hline
F   &11 &18 &49  &- &3  &- &3 &- &3  &100 &14 &- &26 &21 &- \\
\hline
\end{tabular}
\end{minipage}
\end{table*}

\clearpage

\begin{table*}
\centering
\begin{minipage}{400mm}
\contcaption{\,\,}
 \begin{tabular}{@{}cccccccccccccccc@{}}
 \hline
Line &\Hb &\OIII &\OIII &\OI &\HeI &\HeII &\OI &\OI &\NII &\Ha &\NII &\HeI &\SII &\SII &\ArIII\\
 (\AA)&(4861) &(4959) &(5007) &(5577) &(5876) &(6234) &(6300) &(6364) &(6548) &(6563) &(6584) &(6678) &(6716) &(6731) &(7136) \\
\hline
\multicolumn{16}{c}{NGC\,4449 -  LBZ\,14}\\
\hline
F   &23 &-  &- &- &- &- &- &- &4 &100 &12 &- &38 &29 &- \\
I   &35 &-  &- &- &- &- &- &- &4 &100 &12 &- &37 &28 &- \\
S/N &4  &-  &- &- &- &- &- &- &3 &14  &5  &- &9  &6  &- \\
\hline
\multicolumn{16}{c}{NGC\,4449 -  LBZ\,15}\\
\hline
F   &13 &- &16 &- &- &- &- &- &2 &100 &15 &- &35 &25 &- \\
\hline
\end{tabular}
\end{minipage}
\end{table*}

\begin{table*}
\centering
 \begin{minipage}{400mm}
\contcaption{\,\,}
 \begin{tabular}{@{}cccccccccccccccc@{}}
 \hline
\footnotetext{F: Observed emission line fluxes normalized to F(\Ha)=100, not corrected for interstellar extinction}
\footnotetext{I: Extinction corrected emission line fluxes, normalized to I(\Ha)=100}
\footnotetext{S/N: Signal-to-noise ratios of the quoted fluxes}
Line &\Hb &\OIII &\OIII &\OI &\HeI &\HeII &\OI &\OI &\NII &\Ha &\NII &\HeI &\SII &\SII &\ArIII\\
 (\AA)&(4861) &(4959) &(5007) &(5577) &(5876) &(6234) &(6300) &(6364) &(6548) &(6563) &(6584) &(6678) &(6716) &(6731) &(7136) \\
\hline
\multicolumn{16}{c}{NGC\,5204 -  LBZ\,4}\\
\hline
F   &14 &- &- &- &- &- &12 &- &5 &100 &7 &- &38 &30 &- \\
I   &35 &- &- &- &- &- &13 &- &5 &100 &7 &- &36 &28 &- \\
S/N &7  &- &- &- &- &- &11 &- &8 &32  &8 &- &19 &15 &- \\
\hline
\multicolumn{16}{c}{NGC\,5204 -  LBZ\,5}\\
\hline
F   &25 &- &16 &- &- &- &- &- &3 &100 &12 &- &23 &19 &- \\
I   &35 &- &21 &- &- &- &- &- &3 &100 &12 &- &22 &18 &- \\
S/N &6  &- &5  &- &- &- &- &- &3 &24  &7  &- &9  & 8 &- \\
\hline
\multicolumn{16}{c}{NGC\,5204 -  LBZ\,6}\\
\hline
F   &25 &3  &7  &- &- &- &3 &- &4 &100 &11 &- &22 &17 &- \\
I   &35 &4  &9  &- &- &- &3 &- &4 &100 &11 &- &22 &16 &- \\
S/N &12 &3  &8  &- &- &- &5 &- &7 &50  &14 &- &22 &18 &- \\
\hline
\multicolumn{16}{c}{NGC\,5204 -   LBZ\,7}\\
\hline
F   &20 &- &- &- &- &- &- &- &17 &100 &2  &- &24  &20  &- \\
I   &35 &- &- &- &- &- &- &- &17 &100 &2  &- &23  &19  &- \\
S/N &16 &- &- &- &- &- &- &- &20 &43  &13 &- &21  &19  &- \\
\hline
\end{tabular}
\end{minipage}
\end{table*}

\clearpage

\begin{table*}
\centering
\begin{minipage}{400mm}
\caption{\,\,Emission Line Parameters of spectroscopically observed SNRs in our sample of galaxies}
\begin{tabular}{@{}llcccccccc@{}}
\hline
Galaxy &SourceID &F(\Ha) & c(\Hb) & E$_{(B-V)}$ &\Ha / \Hb &\SII(6716+6731)  & \NII(6548+6584)   &\SII(6716)     &\OIII (4959+5007) \\
&&&&&&/ \Ha&/ \Ha&/\SII(6731)&/ \Hb \\
\hline
NGC\,2403     &LBZ\,1    &0.17   &0.15$\pm$0.17  &0.12$\pm$0.13  &3.25$\pm$0.44   &0.49$\pm$0.03 &0.32$\pm$0.02  &1.32$\pm$0.14 &0.46$\pm$0.10\thanks{*}       \\
-             &LBZ\,2    &0.29   &1.19$\pm$0.17  &0.92$\pm$0.13  &7.42$\pm$0.99   &0.39$\pm$0.02 &0.33$\pm$0.01  &1.34$\pm$0.10 &0.50$\pm$0.10\thanks{*}       \\
-             &LBZ\,3    &0.73   &1.19$\pm$0.10  &0.92$\pm$0.08  &7.43$\pm$0.59   &0.41$\pm$0.01 &0.37$\pm$0.01  &1.39$\pm$0.06 &0.80$\pm$0.11     \\
-             &LBZ\,4    &0.48   &0.19$\pm$0.10  &0.15$\pm$0.07  &3.35$\pm$0.26   &0.56$\pm$0.02 &0.37$\pm$0.01  &1.39$\pm$0.07 &-                 \\
-             &LBZ\,5    &0.92   &0.92$\pm$0.09  &0.71$\pm$0.07  &5.97$\pm$0.42   &0.43$\pm$0.01 &0.47$\pm$0.01  &1.40$\pm$0.05 &0.22$\pm$0.03\thanks{*}       \\
-             &LBZ\,6    &0.53   &0.95$\pm$0.11  &0.73$\pm$0.09  &6.14$\pm$0.56   &0.61$\pm$0.02 &0.47$\pm$0.02  &1.40$\pm$0.07 &0.51$\pm$0.13       \\
\hline
\end{tabular}
\end{minipage}
\end{table*}

\clearpage

\begin{table*}
\centering
\begin{minipage}{400mm}
\contcaption{\,\,}
\begin{tabular}{@{}llcccccccc@{}}
\hline
\footnotetext{Note -- F(\Ha) is the absolute flux in units of 10$^{-14}$ erg sec$^{-1}$ cm$^{-2}$. The extinction c(\Hb) is derived from the relation c = (0.346 $\times$ log((\Ha/\Hb)$_{spec}$ / 2.87))$^{-1}$,\\ using the absorbed \Ha, \Hb\ fluxes. For the calculation of the \Ha / \Hb\ ratios the absorbed measured fluxes were used. The rest of the ratios, and when the \Hb\ emission \\line was available, were measured from the extinction-corrected emission line fluxes. All errors were calculated through standard error propagation. The asterisk denotes \\that one of the two lines of \NII\ or \OIII\ was detected and thus used.}
Galaxy &SourceID &F(\Ha) & c(\Hb) & E$_{(B-V)}$ &\Ha / \Hb &\SII(6716+6731)  & \NII(6548+6584)   &\SII(6716)     &\OIII (4959+5007) \\
&&&&&&/ \Ha&/ \Ha&/\SII(6731)&/ \Hb \\
\hline
NGC\,4395     &LBZ\,1   &0.22 &0.20$\pm$0.10  &0.15$\pm$0.08  &3.37$\pm$0.28   &0.42$\pm$0.02 &0.11$\pm$0.01  &1.41$\pm$0.12 &0.42$\pm$0.06\thanks{*}     \\
-             &LBZ\,2   &0.02 &0.80$\pm$0.58  &0.62$\pm$0.45  &5.42$\pm$2.50   &0.59$\pm$0.11 &-              &0.96$\pm$0.29 &1.45$\pm$0.82\thanks{*}    \\
-             &LBZ\,3   &0.07 &1.48$\pm$0.37  &1.14$\pm$0.29  &9.35$\pm$2.78   &0.40$\pm$0.04 &0.09$\pm$0.02\thanks{*}  &1.26$\pm$0.22 &0.82$\pm$1.01     \\
-             &LBZ\,4   &0.13 &0.42$\pm$0.15  &0.32$\pm$0.11  &4.00$\pm$0.48   &0.41$\pm$0.02 &0.19$\pm$0.02  &1.43$\pm$0.15 &1.21$\pm$0.24     \\
-             &LBZ\,5   &0.33 &0.57$\pm$0.14  &0.44$\pm$0.11  &4.50$\pm$0.50   &0.61$\pm$0.03 &0.14$\pm$0.01\thanks{*}  &1.29$\pm$0.10 &2.24$\pm$0.36    \\
-             &LBZ\,6   &0.21 &0.92$\pm$0.15  &0.71$\pm$0.11  &5.97$\pm$0.69   &0.59$\pm$0.04 &0.16$\pm$0.02\thanks{*}  &1.31$\pm$0.12 &2.68$\pm$0.45     \\
\hline
\end{tabular}
\end{minipage}
\end{table*}

\clearpage

\begin{table*}
\centering
\begin{minipage}{400mm}
\caption{\,\,Emission Line Parameters of spectroscopically-verified non-SNRs in our sample of galaxies}
\begin{tabular}{@{}llcccccccc@{}}
\hline
Galaxy &SourceID &F(\Ha) & c(\Hb) & E$_{(B-V)}$ &\Ha / \Hb &\SII(6716+6731)  & \NII(6548+6584)   &\SII(6716)     &\OIII (4959+5007) \\
&&&&&&/ \Ha&/ \Ha&/\SII(6731)&/ \Hb \\
\hline
NGC\,2403 &LBZ\,1059 &0.15 &-              &-             &-             &0.35$\pm$0.02  &-              &1.17$\pm$0.14  &-                       \\
-         &LBZ\,731  &0.11 &0.39$\pm$0.20  &0.30$\pm$0.16 &3.93$\pm$0.63 &0.19$\pm$0.02  &0.28$\pm$0.03  &1.24$\pm$0.24  &0.89$\pm$0.18\thanks{*}  \\
-         &LBZ\,561  &0.22 &0.30$\pm$0.10  &0.23$\pm$0.08 &3.64$\pm$0.30 &0.23$\pm$0.02  &0.25$\pm$0.02  &1.32$\pm$0.17  &1.16$\pm$0.15  \\
-         &LBZ\,484  &0.11 &0.74$\pm$0.21  &0.57$\pm$0.16 &5.17$\pm$0.87 &0.26$\pm$0.03  &0.32$\pm$0.03  &0.94$\pm$0.17  &5.99$\pm$1.43  \\
-         &LBZ\,982  &0.64 &0.43$\pm$0.09  &0.33$\pm$0.07 &4.03$\pm$0.27 &0.21$\pm$0.01  &0.28$\pm$0.01  &1.36$\pm$0.12  &0.49$\pm$0.07  \\
-         &LBZ\,1291 &0.34 &0.82$\pm$0.16  &0.63$\pm$0.13 &5.49$\pm$0.71 &0.30$\pm$0.02  &0.32$\pm$0.02  &1.20$\pm$0.13  &0.56$\pm$0.10\thanks{*}  \\
\hline
\end{tabular}
\end{minipage}
\end{table*}

\begin{table*}
\centering
\begin{minipage}{400mm}
\contcaption{\,\,}
\begin{tabular}{@{}llcccccccc@{}}
\hline
\footnotetext{Note -- F(\Ha) is the absolute flux in units of 10$^{-14}$ erg sec$^{-1}$ cm$^{-2}$. The extinction c(\Hb) is derived from the relation c = (0.346 $\times$ log((\Ha/\Hb)$_{spec}$ / 2.87))$^{-1}$, \\using the absorbed \Ha, \Hb\ fluxes. For the calculation of the \Ha / \Hb\ ratios the absorbed measured fluxes were used. The rest of the ratios, and when the \Hb\ emission \\line was available, were measured from the extinction-corrected emission line fluxes. All errors were calculated through standard error propagation. The asterisk denotes \\that one of the two lines of \NII\ or \OIII\ was detected and thus used.}
Galaxy &SourceID &F(\Ha) & c(\Hb) & E$_{(B-V)}$ &\Ha / \Hb &\SII(6716+6731)  & \NII(6548+6584)   &\SII(6716)     &\OIII (4959+5007) \\
&&&&&&/ \Ha&/ \Ha&/\SII(6731)&/ \Hb \\
\hline
NGC\,4449     &LBZ\,581  &0.82 &0.39$\pm$0.06  &0.30$\pm$0.04 &3.90$\pm$0.18 &0.22$\pm$0.01  &0.18$\pm$0.01 &1.40$\pm$0.08 &2.77$\pm$0.18  \\         
-             &LBZ\,593  &0.15 &0.32$\pm$0.14  &0.25$\pm$0.11 &3.72$\pm$0.42 &0.24$\pm$0.02  &0.25$\pm$0.02 &0.77$\pm$0.12 &1.64$\pm$0.26  \\
-             &LBZ\,567  &0.20 &0.62$\pm$0.17  &0.48$\pm$0.13 &4.71$\pm$0.63 &0.20$\pm$0.02  &0.11$\pm$0.01 &1.40$\pm$0.23 &0.41$\pm$0.08\thanks{*}  \\
-             &LBZ\,527  &0.71 &0.35$\pm$0.07  &0.27$\pm$0.05 &3.81$\pm$0.20 &0.31$\pm$0.01  &0.19$\pm$0.01 &1.41$\pm$0.07 &0.83$\pm$0.07  \\
-             &LBZ\,503  &0.27 &0.64$\pm$0.12  &0.49$\pm$0.09 &4.78$\pm$0.45 &0.36$\pm$0.02  &0.23$\pm$0.01 &1.29$\pm$0.10 &1.82$\pm$0.25  \\
-             &LBZ\,500  &0.20 &0.84$\pm$0.15  &0.65$\pm$0.12 &5.62$\pm$0.68 &0.24$\pm$0.02  &0.15$\pm$0.02 &1.36$\pm$0.16 &-              \\
\hline
\end{tabular}
\end{minipage}
\end{table*}

\begin{table*}
\centering
 \begin{minipage}{180mm}
 \caption{Census of SNRs in our sample of galaxies and Success Rates}
 \begin{tabular}{cccccc}
 \hline
\footnotetext{Note--Column 1: (\SII/\Ha)phot ratios splitted into three categories: $>$0.4 (candidate SNRs), 0.3 - 0.4 (probable candidate SNRs) and $<$0.3, Column 2: number of the photomeric sources that were detected in each category (within their error-bars), Column 3: photometric SNRs presented in Tables 3-8. These numbers result from the detected SNRs (Column 2) if we subtract the spectroscopically observed, non-SNRs (Column 4 minus Column 5). In the case of $<$0.3 for this column, we quoted only the sources that were spectroscopically verified as SNRs, Column 4: number of spectroscopically observed sources, Column 5: number of sources that were spectroscopically verified as SNRs, Column 5: percentage of the spectroscopically confirmed SNRs (success rate).\\The numbers in the {\it sum} section result from the addition of the individual numbers in each column from the {\it All galaxies} section.}
Photometric Ratio  & Detected sources &Photometric SNRs & Obtained Spectra  & Spectroscopic Classification in SNRs & Success in SNRs\\
\hline
&&&NGC\,2403&&\\
\hline
 $>$0.4    &111 &102  &14 &5  &36$\%$   \\
 0.3 - 0.4 &48  &47   &8  &7  &88$\%$   \\
 $<$0.3    &...\footnotemark[1] &0    &2  &0  &0 $\%$   \\
\hline
&&&NGC\,3077&&\\
\hline
 $>$0.4    &16  &16  &4  &4  &100$\%$   \\
 0.3 - 0.4 &8   &8   &2  &2  &100$\%$   \\
\hline
\end{tabular}
\footnotetext{1.\, We do not present any detected sources with \SII/\Ha $<$ 0.3 since they are not considered as SNRs.}
\footnotetext{2.\, These three dots refer to the sum of  detected sources with \SII/\Ha $<$ 0.3.}
\end{minipage}
\end{table*}

\begin{table*}
 \centering
\begin{minipage}{140mm}
  \caption{\,\, Nitrogen and Oxygen Abundances in the galaxies of our sample}
  \begin{tabular}{ccc}
  \hline
Galaxy & 12 + log (N/H) & 12 + log (O/H) \\
\hline
NGC\,2403         &7.48\footnotemark[1]    &8.52\footnotemark[1] \\
NGC\,3077         &7.48\footnotemark[2]    &8.64\footnotemark[2] \\
NGC\,4214         &-                       &8.22\footnotemark[3] \\
NGC\,4395         &7.18\footnotemark[1]    &8.27\footnotemark[1] \\
NGC\,4449         &7.10\footnotemark[4]    &8.30\footnotemark[4] \\
NGC\,5204         &-                       &-                    \\
\hline
NGC\,6946         &8.15\footnotemark[1]    &8.70\footnotemark[1] \\ 
NGC\,5585         &-                       &-                    \\
M\,81 (NGC\,3031) &7.96\footnotemark[1]    &8.69\footnotemark[1] \\
M\,101 (NGC\,5457)&8.23\footnotemark[1]    &8.80\footnotemark[1] \\
\hline
LMC               &6.45\footnotemark[5]    &8.35\footnotemark[6] \\
SMC               &-                       &8.03\footnotemark[6] \\
Milky Way         &-                       &8.52\footnotemark[1] \\
\hline
\end{tabular}
\footnotetext{1.\, \citet{Pilyugin04}}
\footnotetext{2.\, \citet{Storchi-Bergmann94}}
\footnotetext{3.\, \citet{Martin97}}
\footnotetext{4.\, \citet{Sabbadin84}}
\footnotetext{5.\, \citet{Lequeux79}}
\footnotetext{6.\, \citet{Richer95}}
\end{minipage}
\end{table*}

\begin{table*}
 \centering
\begin{minipage}{400mm}
\caption{\, Associations of known, multiwavelength SNRs in NGC\,2403}
  \begin{tabular}{cccccccccc}
  \hline
\hline
Source ID & Classification & RA &Dec &Optical  &Offset       &X-ray & Offset      &Radio &Offset \\
&  & (h:m:s) &(d:m:s)   &associate\footnotemark[1]    &($\arcsec$)  &associate\footnotemark[2] &($\arcsec$)  &associate &($\arcsec$)\\
\hline
\footnotetext{NOTE -- The classification 'Frame-4' corresponds to sources detected in Frame-4 (2$\times$2 mosaic of NGC\,2403) but not used since the observing night was not \\photometric ($\S$ 2.4).}
out of field of view  &-                       &07:36:21.2 &65:40:56.4 &SNR-1    &-     &-  &-  &-  &-  \\
LBZ\,22               &candidate SNR           &07:36:24.1 &65:36:07.2 &SNR-2    &2.29  &-  &-  &-  &-  \\
LBZ\,1                &SNR                     &07:36:30.4 &65:35:43.4 &SNR-3    &3.44  &-  &-  &-  &-  \\
LBZ\,118              &probable candidate SNR  &07:36:37.0 &65:36:39.1 &SNR-4    &1.24  &-  &-  &-  &-  \\
LBZ\,60               &candidate SNR           &07:36:42.9 &65:34:51.9 &SNR-5    &1.01  &-  &-  &-  &-  \\
LBZ\,67               &candidate SNR           &07:36:45.8 &65:36:36.0 &SNR-6    &0.69  &-  &-  &-  &-  \\
\hline 
\end{tabular}
\end{minipage}
\end{table*}

\begin{table*}
 \centering
\begin{minipage}{400mm}
  \contcaption{\,\,}
  \begin{tabular}{cccccccccc}
  \hline
  \hline
Source ID & Classification & RA &Dec &Optical  &Offset       &X-ray  & Offset      &Radio  &Offset \\
&  & (h:m:s) &(d:m:s)   &associate\footnotemark[1]    &($\arcsec$)  &associate\footnotemark[2] &($\arcsec$)  &associate &($\arcsec$)\\
\hline
\footnotetext{Notes -- Column 1: Source ID. For the meaning of questionmark see \S 5.1, Column 2: Source classification of this study. Sources with (\SII/\Ha)$_{phot} < $ 0.3 are \\denoted as SNR/\HII, Cols 3,4: RA and Dec in J2000 of the detected sources by this study. For sources that were not detected we present the coordinates of the \\associated multi-wavelength SNR, Column 5: Optically associated SNR by other studies, Column 6: Offset in arcseconds between the source's center coordinates \\of the present and other optical studies, Column 7: X-ray associated SNR, Column 8: Offset in arcseconds between the source's center coordinates of the present \\and the X-ray study, Column 9: Radio associated SNR, Column 10: Offset in arcseconds of the source's center coordinates between this study and the radio study.}
LBZ\,8               &SNR           &07:37:03.2 &65:37:13.7 &-        &- &SNR (LZB\,81)           &2.82  &-  &-   \\
LBZ\,74              &candidate SNR &07:36:47.9 &65:36:23.9 &-        &- &probable SNR (LZB\,120) &2.09  &-  &-   \\
LBZ\,56              &candidate SNR &07:36:41.9 &65:36:51.7 &-        &- &SNR (LZB\,107)          &3.05  &SNR (TH2)\footnotemark[6]  &   \\
LBZ\,635.2           &SNR/\HII\     &07:36:52.3 &65:36:40.3 &-        &- &probable SNR (LZB\,5)   &1.72  &-  &-   \\
out of field of view &-             &07:37:08.0 &65:39:20.6 &-        &- &probable SNR (LZB\,78)  & -    &-  &-   \\
not detected         &nothing       &07:37:17.9 &65:36:24.2 &-        &- &probable SNR (LZB\,76)  &-     &-  &-   \\
\hline 
\end{tabular}
\footnotetext{1.\, Optically detected SNRs by \citet{MFBL97}}
\footnotetext{2.\, X-ray selected SNRs in Paper I}
\footnotetext{3.\, Radio SNR by \citet{Eck02}}
\footnotetext{4.\, Although the offset is large, the size of the source ($\sim 3.6\arcsec$) makes the association eligible.}
\footnotetext{5.\, Although the offset is large, the size of the source ($\sim 4\arcsec\times 2.5\arcsec$) makes the association eligible.}
\footnotetext{6.\, Radio SNR by \citet{Turner94}}
\end{minipage}
\end{table*}

\begin{table*}
 \centering
\begin{minipage}{200mm}
  \caption{\,Associations of known, multiwavelength SNRs in NGC\,3077}
  \begin{tabular}{c@{}c@{}cc@{}cc@{}cccc}
  \hline
  \hline
Source ID & Classification & RA &Dec &Optical  &Offset       &X-ray & Offset      &Radio &Offset \\
&  & (h:m:s) &(d:m:s)   &associate    &($\arcsec$)  &associate &($\arcsec$)  &associate &($\arcsec$)\\
\hline
\footnotetext{Notes -- Column 1: Source ID. For the meaning of questionmark see \S 5.1, Column 2: Source classification of this study. Sources with (\SII/\Ha)$_{phot} < $ 0.3 are denoted \\as SNR/\HII, Cols 3,4: RA and Dec in J2000 of the detected sources by this study. For sources that were not detected we present the coordinates of the associated multi-\\wavelength SNR, Column 5: Optically associated SNR by other studies, Column 6: Offset in arcseconds between the source's center coordinates of the present and other \\optical studies, Column 7: X-ray associated SNR, Column 8: Offset in arcseconds between the source's center coordinates of the present and the X-ray study, Column 9:\\ Radio associated SNR, Column 10: Offset in arcseconds of the source's center coordinates between this study and the radio study.}
LBZ\,236 ?    &SNR/\HII\           &10:03:18.2    &68:44:02.4  &- &- &SNR (LZB\,6\footnotemark[1], S6\footnotemark[2])&1.86  &-   &-\\
not detected  &nothing             &10:03:21.8    &68:45:03.3  &- &- &SNR (LZB\,12)\footnotemark[1]             &-     &-             &-\\
not detected  &nothing             &10:03:12.1    &68:43:19.1  &- &- &SNR (LZB\,13)\footnotemark[1]             &-     &-             &-\\
LBZ\,24       &probable candidate SNR &10:03:20.8    &68:41:40.1  &- &- &SNR (LZB\,15)\footnotemark[1]             &1.22  &-             &-\\
LBZ\,299\footnotemark[4]/LBZ\,300\footnotemark[4]  &SNR/\HII  &10:03:19/10:03:19    &68:43:54/68:43:59  &- &- &SNR (LZB\,18\footnotemark[1], S1\footnotemark[2])&2.85/2.35  &SNR (S1)\footnotemark[3]      &1.97/2.62    \\                                                 
LBZ\,303 ?    &SNR/\HII  &10:03:18.1    &68:43:57.0  &- &- &SNR (S5)\footnotemark[2]  &1.33  &-             &-\\
\hline 
\end{tabular}
\footnotetext{1.\, X-ray detected SNR from Paper I}
\footnotetext{2.\, X-ray detected SNR by\citet{Ott03}}
\footnotetext{3.\,Radio selected SNR by \citet{Rosa05}}
\footnotetext{4.\, For this association see \S5.1}
\end{minipage}
\end{table*}

\begin{table*}
 \centering
\begin{minipage}{400mm}
  \caption{\,Associations of known, multiwavelength SNRs in NGC\,4214}
  \begin{tabular}{cccccccccc}
  \hline
  \hline
Source ID & Classification & RA &Dec &Optical  &Offset       &X-ray  & Offset      &Radio  &Offset \\
&  & (h:m:s) &(d:m:s)   &associate\footnotemark[1] &($\arcsec$)  &associate\footnotemark[2] &($\arcsec$)  &associate\footnotemark[3] &($\arcsec$)\\
\hline
\footnotetext{Notes -- Column 1: Source ID.  For the meaning of questionmark see \S 5.1, Column 2: Source classification. Sources with (\SII/\Ha)$_{phot} < $ 0.3 are denoted as SNR/\HII,\\ Cols 3,4: RA and Dec in J2000 of the detected sources by this study. In the case of not detected sources, the quoted coordinates belong to the associated multi-wavelength \\SNRs, Column 5: Optically associated SNR by other studies, Column 6: Offset in arcseconds between the source's center coordinates of the present and other optical \\studies, Column 7: X-ray associated SNR, Column 8: Offset in arcseconds between the source's center coordinates of the present and the X-ray study, Column 9: Radio \\associated SNR, Column 10: Offset in arcseconds of the source's center coordinates between this study and the radio study.}
LBZ\,35  ?         &candidate SNR  &12:15:33.4 &36:19:01.0 &-                     &-     &SNR (LZB\,7)             &2.16  &-                            &-     \\
not detected       &nothing        &12:15:49.7 &36:18:46.7 &-                     &-     &candidate SNR (LZB\,10)  &-     &-                            &-     \\
LBZ\,47 ?          &candidate SNR  &12:15:38.0 &36:22:22.4 &-                     &-     &candidate SNR (LZB\,11)  &1.48  &-                            &-     \\
not detected       &diffused       &12:15:40.2 &36:19:25.2  &-                    &-     &candidate SNR (LZB\,16)  &-     &-                            &-     \\
LBZ\,73            &candidate SNR  &12:15:48.8 &36:17:02.3 &-                     &-     &candidate SNR (LZB\,23)  &0.95  &-                            &-     \\ 
LBZ\,1073          &SNR/\HII       &12:15:34.7 &36:20:17.2 &\HII\ region           &-     &-                        &-     &SNR-2                        &0.20  \\
LBZ\,80            &probable candidate SNR  &12:15:38.2 &36:19:45.2 &-                     &-     &XRB (LZB\,26)            &1.44  &SNR/\HII-3                   &0.29  \\
LBZ\,82            &probable candidate SNR  &12:15:38.9 &36:18:58.9 &SNR-1                 &0.34  &-                        &-     &SNR-4                        &0.59  \\
not detected       &diffused       &12:15:39.7 &36:19:34.3 &-                     &-     &-                        &-     &SNR/\HII-8                   &-    \\
LBZ\,57            &candidate SNR  &12:15:40.0 &36:18:39.4 &SNR-2                 &0.47  &SNR (LZB\,30)            &0.32  &SNR-9                        &0.85  \\
LBZ\,56            &candidate SNR  &12:15:39.4 &36:20:54.1 &-                     &-     &probable SNR (LZB\,31)   &0.00  &-                            &-     \\
LBZ\,1098          &SNR/\HII       &12:15:40.0 &36:19:35.8 &SNR-3                 &0.28  &probable SNR (LZB\,34)   &0.78  &SNR-10                       &0.52  \\
LBZ\,936           &SNR/\HII       &12:15:37.2 &36:22:19.6 &-                     &-     &probable SNR (LZB\,35)   &0.97  &-                            &-     \\ 
LBZ\,83            &probable candidate SNR  &12:15:40.2 &36:19:30.2 &SNR-4                 &0.30  &-                        &-     &SNR-11                       &0.72  \\
LBZ\,1099          &SNR/\HII       &12:15:40.5 &36:19:31.5 &-                     &-     &-                        &-     &SNR-12                       &0.00  \\
not detected       &diffused       &12:15:41.6 &36:19:09.7 &-                     &-     &-                        &-     &SNR/\HII-18                  &-  \\
LBZ\,87            &probable candidate SNR  &12:15:41.9 &36:19:15.5 &SNR-5                 &0.37  &probable SNR (LZB\,28)   &1.06  &SNR-19, $\rho$\footnotemark[4]&0.26,1.43\footnotemark[5] \\
LBZ\,16            &SNR            &12:15:42.5 &36:19:47.7 &SNR-6\footnotemark[6] &0.08  &-                        &-     &\HII\ region                  &-     \\
LBZ\,18            &SNR            &12:15:45.7 &36:19:41.8 &SNR-7\footnotemark[6] &0.30  &probable SNR (LZB\,38)   &0.43  &\HII\ region                  &-     \\
LBZ\,832           &SNR/\HII       &12:15:41.0 &36:19:03.8 &-                     &-     &-                        &-     &$\alpha$\footnotemark[4]     &1.91  \\
LBZ\,857           &SNR/\HII\footnotemark[7] &12:15:40.7 &36:19:11.9 &-                     &-     &-  &-   &$\beta$\footnotemark[4]           &1.99  \\
\hline 
\end{tabular}
\footnotetext{1.\,\citet{Dopita10}}
\footnotetext{2.\,Paper I}
\footnotetext{3.\,\citet{Chomiuk09}}
\footnotetext{4.\,\citet{Vukotic05}}
\footnotetext{5.\, Offset between LBZ\,1097 and radio SNR $\rho$} 
\footnotetext{6.\, Composite SNR + \HII\ source}
\footnotetext{7.\, This source also coincides with strong emission over the R continuum filter.}
\end{minipage}
\end{table*}

\begin{table*}
 \centering
\begin{minipage}{400mm}
  \caption{\,Associations of known, multiwavelength SNRs in NGC\,4395}
  \begin{tabular}{cccccccccc}
  \hline
Source ID & Classification & RA &Dec &Optical   &Offset       &X-ray  & Offset      &Radio  &Offset \\
&  & (h:m:s) &(d:m:s)   &associate    &($\arcsec$)  &associate\footnotemark[1] &($\arcsec$)  &associate\footnotemark[2] &($\arcsec$)\\
\hline
\footnotetext{Notes -- Column 1: Source ID.  For the meaning of questionmark see \S 5.1, Column 2: Source classification. Sources with (\SII/\Ha)$_{phot} < $ 0.3 are denoted as SNR/\HII,\\ Cols 3,4: RA and Dec in J2000 of the detected sources by this study. In the case of not detected sources, the quoted coordinates belong to the associated multi-wavelength \\SNRs, Column 5: Optically associated SNR by other studies, Column 6: Offset in arcseconds between the source's center coordinates of the present and other optical \\studies, Column 7: X-ray associated SNR, Column 8: Offset in arcseconds between the source's center coordinates of the present and the X-ray study, Column 9: Radio \\associated SNR, Column 10: Offset in arcseconds of the source's center coordinates between this study and the radio study.}
out of field of view &-                   &12:25:53.2  &33:38:30.4 &- &- &candidate SNR (LZB\,10) &-    &-  &-   \\
LBZ\,1503 ?          &SNR/\HII   &12:25:39.6  &33:32:04.2 &- &- &SNR (LZB\,14) &2.28 &-  &-   \\
LBZ\, 1099           &SNR/\HII   &12:25:58.1  &33:31:38.3 &- &- &-       &-    &SNR (source 3)  &1.27\\
\hline 
\end{tabular}
\footnotetext{1.\,X-ray selected SNRs from Paper I}
\footnotetext{2.\,Radio selected SNR from the work of \citet{Vukotic05}}
\end{minipage}
\end{table*}

\begin{table*}
 \centering
 \begin{minipage}{400mm}
  \caption{\,Associations of known, multiwavelength SNRs in NGC\,4449}
  \begin{tabular}{cccccccccc}
  \hline
  \hline
Source ID & Classification & RA &Dec &Optical  &Offset       &X-ray  & Offset      &Radio &Offset \\
&  & (h:m:s) &(d:m:s)   &associate    &($\arcsec$)  &associate\footnotemark[1] &($\arcsec$)  &associate\footnotemark[2] &($\arcsec$)\\
\hline
\footnotetext{Notes -- Column 1: Source ID.  For the meaning of questionmark see \S 5.1, Column 2: Source classification. Sources with (\SII/\Ha)$_{phot} < $ 0.3 are denoted as SNR/\HII,\\ Cols 3,4: RA and Dec in J2000 of the detected sources by this study. In the case of not detected sources, the quoted coordinates belong to the associated multi-wavelength \\SNRs, Column 5: Optically associated SNR by other studies, Column 6: Offset in arcseconds between the source's center coordinates of the present and other optical \\studies, Column 7: X-ray associated SNR, Column 8: Offset in arcseconds between the source's center coordinates of the present and the X-ray study, Column 9: Radio \\associated SNR, Column 10: Offset in arcseconds of the source's center coordinates between this study and the radio study.}
LBZ\,201?    &SNR/\HII               &12:28:12.1 &44:05:58.4  &-  &- &SNR (LZB\,9)  &1.29 &- &- \\
LBZ\,122     &SNR/\HII               &12:28:11.0 &44:06:47.8  &oxygen-rich SNR\footnotemark[3] &0.57 &SNR (LZB\,12)   &0.57 &SNR-12 &1.43 \\
LBZ\,241     &SNR/\HII               &12:28:11.2  &44:05:37.7 &-  &- &probable SNR (LZB\,24) &1.08 &\HII\ region   &0.79 \\
not detected &nothing                &12:28:15.6  &44:05:36.3 &-  &- &probable SNR (LZB\,26) &-  &- &- \\
LBZ\,475 ?   &SNR/\HII               &12:28:09.5  &44:05:20.4 &-  &- &-                             &-    &SNR-7  &1.92        \\
not detected &diffused               &12:28:10.9  &44:05:40.2 &-  &- &-                      &-    &SNR-11 &-       \\
LBZ363 ?     &SNR/\HII               &12:28:11.3  &44:05:38.5 &-  &- &-                             &-    &SNR-14 &2.04        \\
LBZ407 ?     &SNR/\HII               &12:28:12.8  &44:06:10.4 &-  &- &-                             &-    &SNR-17 &1.81        \\
not detected &nothing\footnotemark[4]&12:28:13.1  &44:05:37.8 &-  &- &-         &-    &SNR-19 &-         \\
LBZ\,323     &SNR/\HII               &12:28:16.2  &44:06:42.8 &-  &- &-                             &-    &SNR-24 &0.96        \\
LBZ\,57      &candidate SNR          &12:28:19.2  &44:06:55.7 &-        &- &-                       &-    &SNR-26 &0.33        \\
LBZ\,60      &probable candidate SNR &12:28:09.7  &44:05:54.8 &-  &- & XRB (LZB\,15)         &1.98  &-    &-                   \\
\hline 
\end{tabular}
\footnotetext{1.\,X-ray selected sources from Paper I}
\footnotetext{2.\,Radio selected sources from the work of \citet{Chomiuk09}}
\footnotetext{3.\,\citet{Blair83}}
\footnotetext{4.\,This source presents detectable emission on the \SII\ image. However, the source happened to be in the slit (Slit 1 in NGC\,4449, see its coordinates in Table 10) that \\was used for spectroscopy. No emission for the particular source was seen at the spectrum.}
\end{minipage}
\end{table*}

\begin{table*}
\caption{\,Associations of known, multiwavelength SNRs in NGC\,5204}
\centering
 \begin{minipage}{200mm}
\begin{tabular}{cccccccccc}
\hline
Source ID & Classification & RA &Dec &Optical  &Offset       &X-ray  & Offset      &Radio &Offset \\
& & (h:m:s) &(d:m:s)   &associate\footnotemark[1]    &($\arcsec$)  &associate &($\arcsec$)  &associate &($\arcsec$)\\
\hline
\footnotetext{Notes -- Column 1: Source ID, Column 2: Source classification, Cols 3,4: RA and Dec in J2000 of the detected sources by this study, \\Column 5: Optically associated SNR by other studies, Column 6: Offset in arcseconds between the source's center coordinates of the \\present and other optical studies, Column 7: X-ray associated SNR, Column 8: Offset in arcseconds between the source's center coor-\\rdinates of the present and the X-ray study, Column 9: Radio associated SNR, Column 10: Offset in arcseconds of the source's center \\coordinates between this study and the radio study.}
LBZ\,9    &candidate SNR     &13:29:30.3 &58:25:20.6 &SNR-1 &0.70 &- &- &- &- \\
LBZ\,4    &SNR               &13:29:34.5 &58:24:23.8 &SNR-2 &1.64 &- &- &- &- \\
LBZ\,16   &candidate SNR     &13:29:36.9 &58:24:26.9 &SNR-3 &0.31 &- &- &- &- \\
\hline 
\end{tabular}
\footnotetext{1.\,From he work of \citet{MF97}}
\end{minipage}
\end{table*}

\clearpage





\begin{figure}
\includegraphics[width=3.0in]{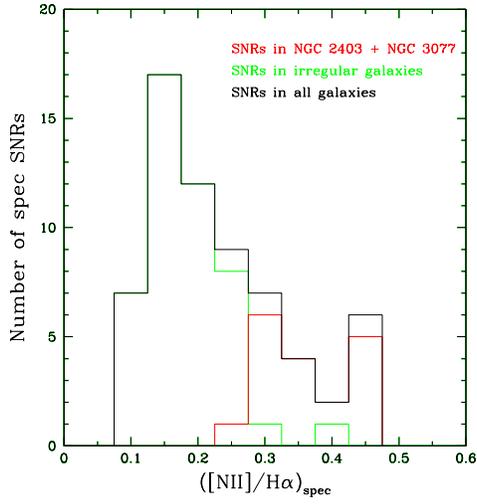}
\centering
 \caption{Number of spectroscopically-observed SNRs against their \NII/\Ha\ ratios. SNRs in the irregular galaxies, apart from NGC\,3077, extend to lower \NII/\Ha\ ratios indicating that the contamination of the \NII\ emission lines in the \Ha + \NII\ images is different in each galaxy.}
\end{figure}

\clearpage

\begin{figure}
\includegraphics[width=7in]{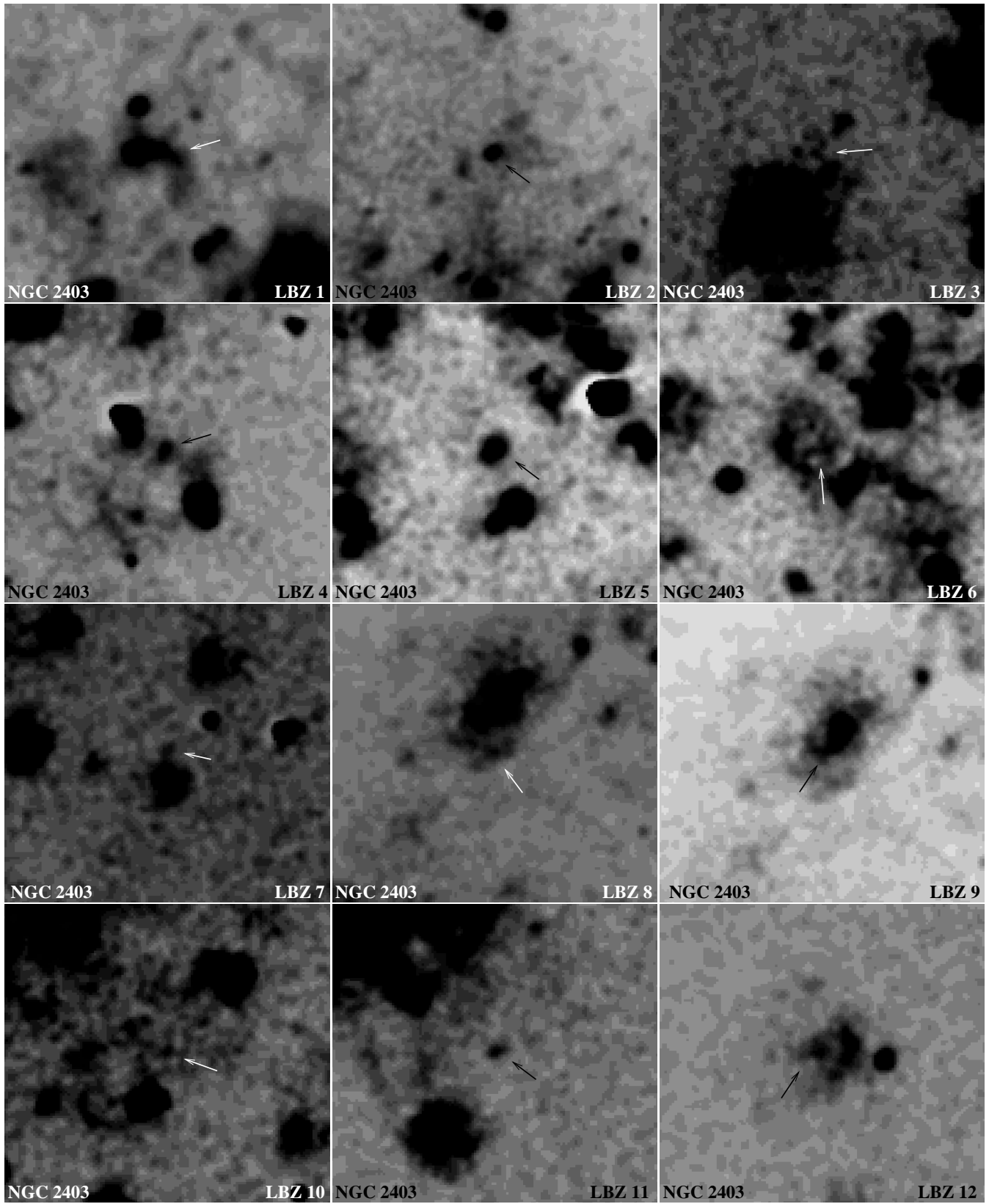}
\centering
 \caption{Zoomed-in display of the 67 spectroscopically observed SNRs in this study over the relevant \Ha\ image of each galaxy. The arrows point at the SNRs' position. North is at the top, East to the left. The images cover 30$\arcsec$$\times$30$\arcsec$ on both sides.}
\end{figure}

\clearpage
\begin{figure}
\includegraphics[width=7in]{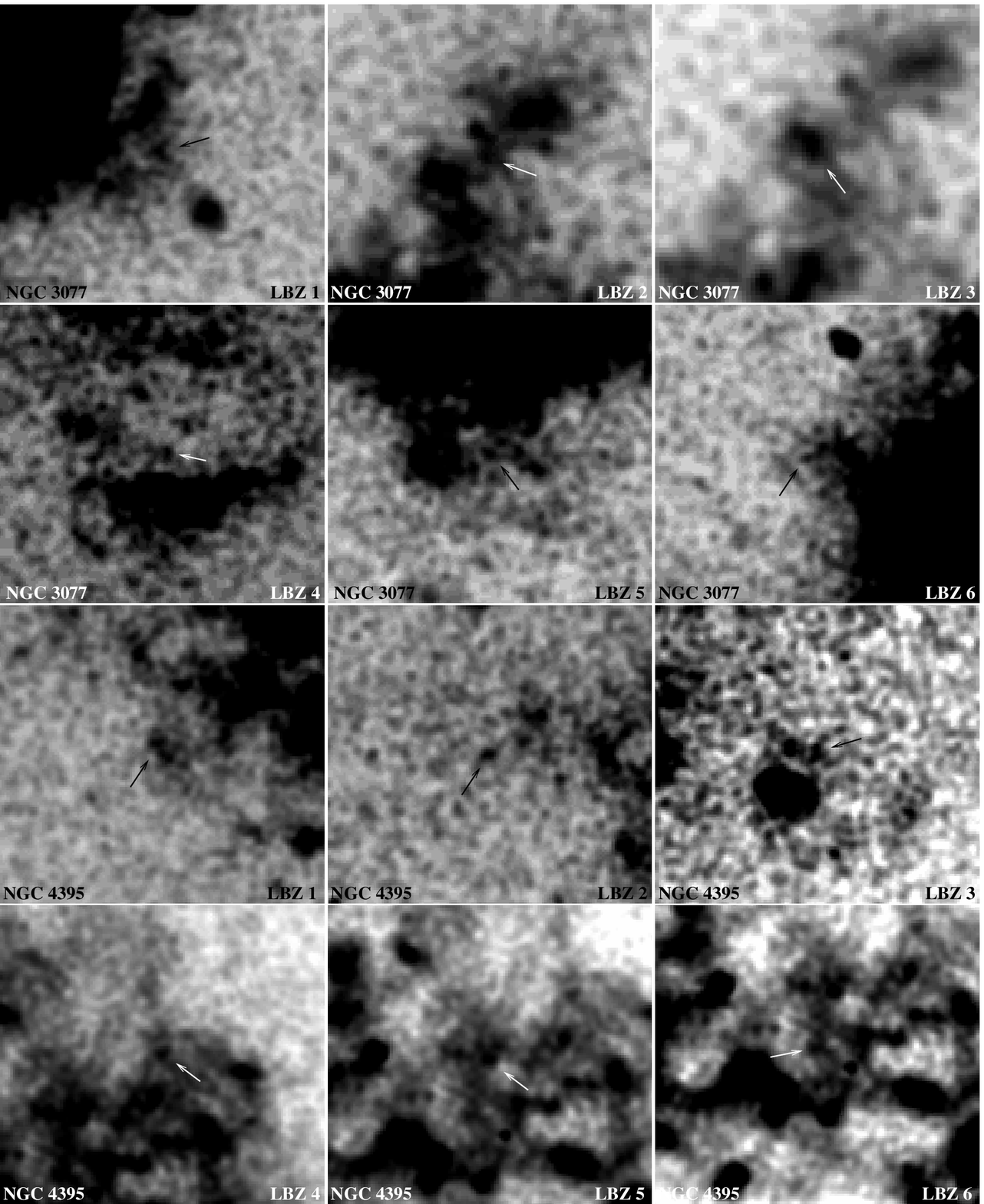}
\centering
 \contcaption{}
\end{figure}

\clearpage
\begin{figure}
\includegraphics[width=7in]{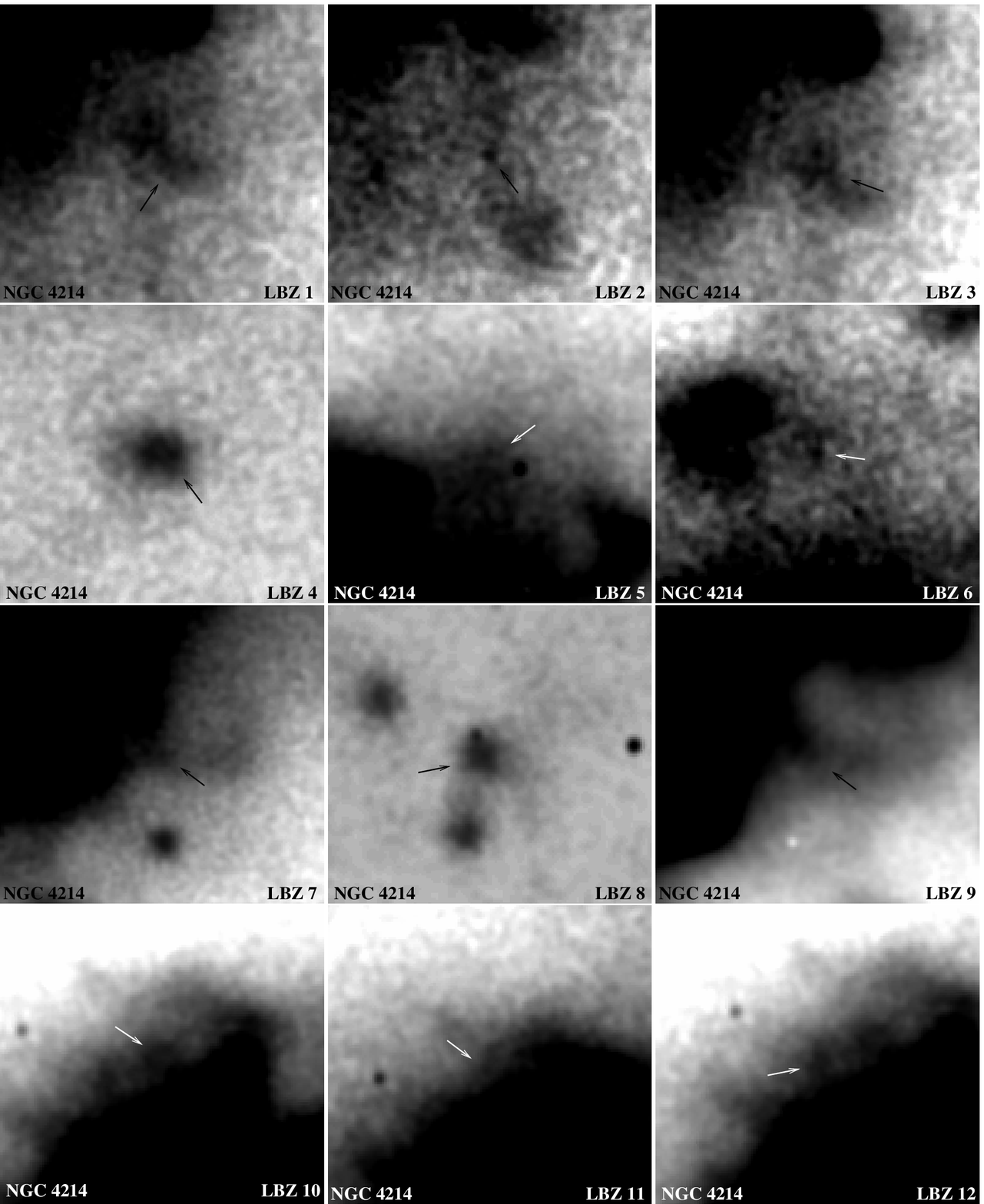}
\centering
 \contcaption{}
\end{figure}

\clearpage
\begin{figure}
\includegraphics[width=7in]{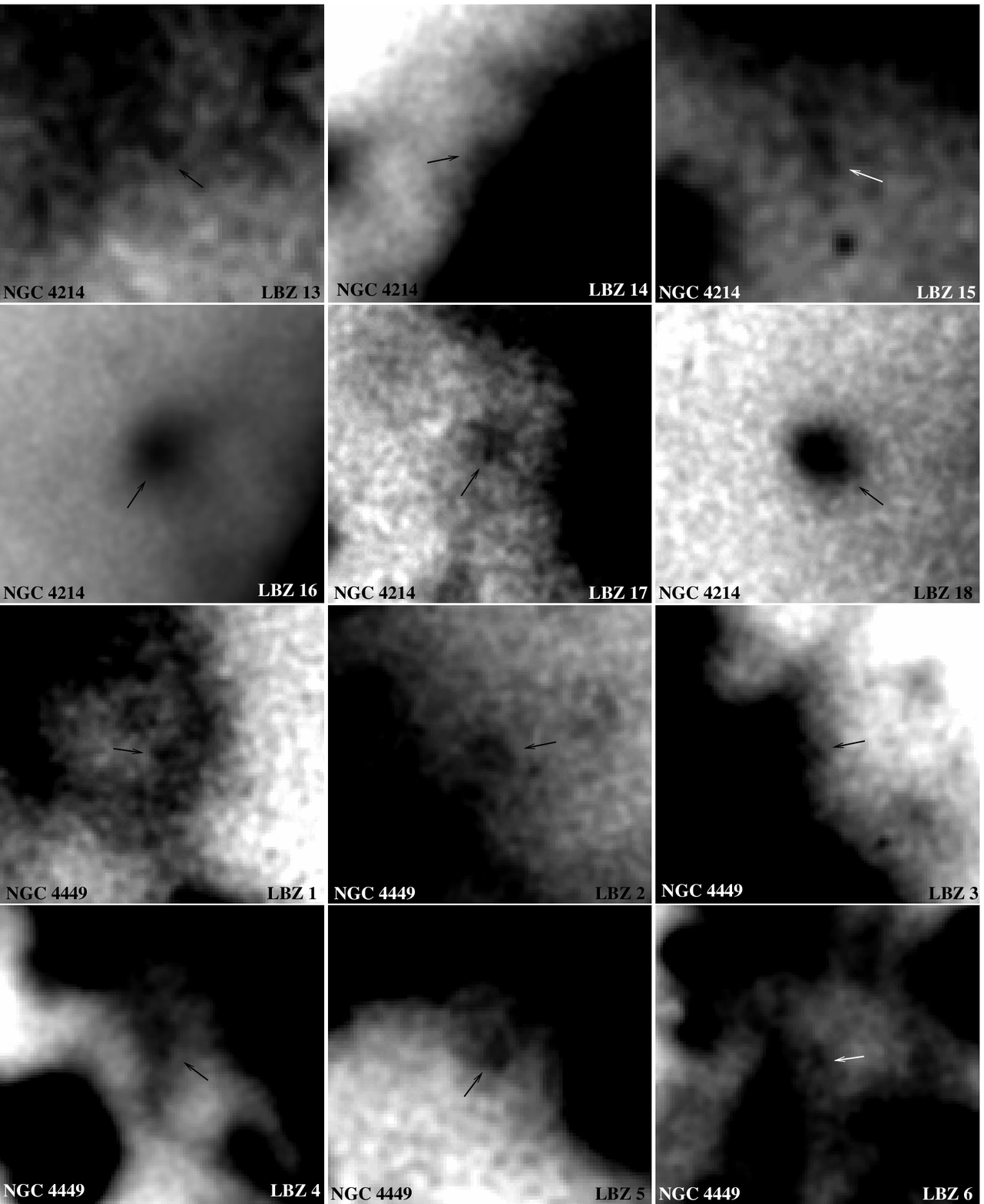}
\centering
 \contcaption{}
\end{figure}

\clearpage
\begin{figure}
\includegraphics[width=7in]{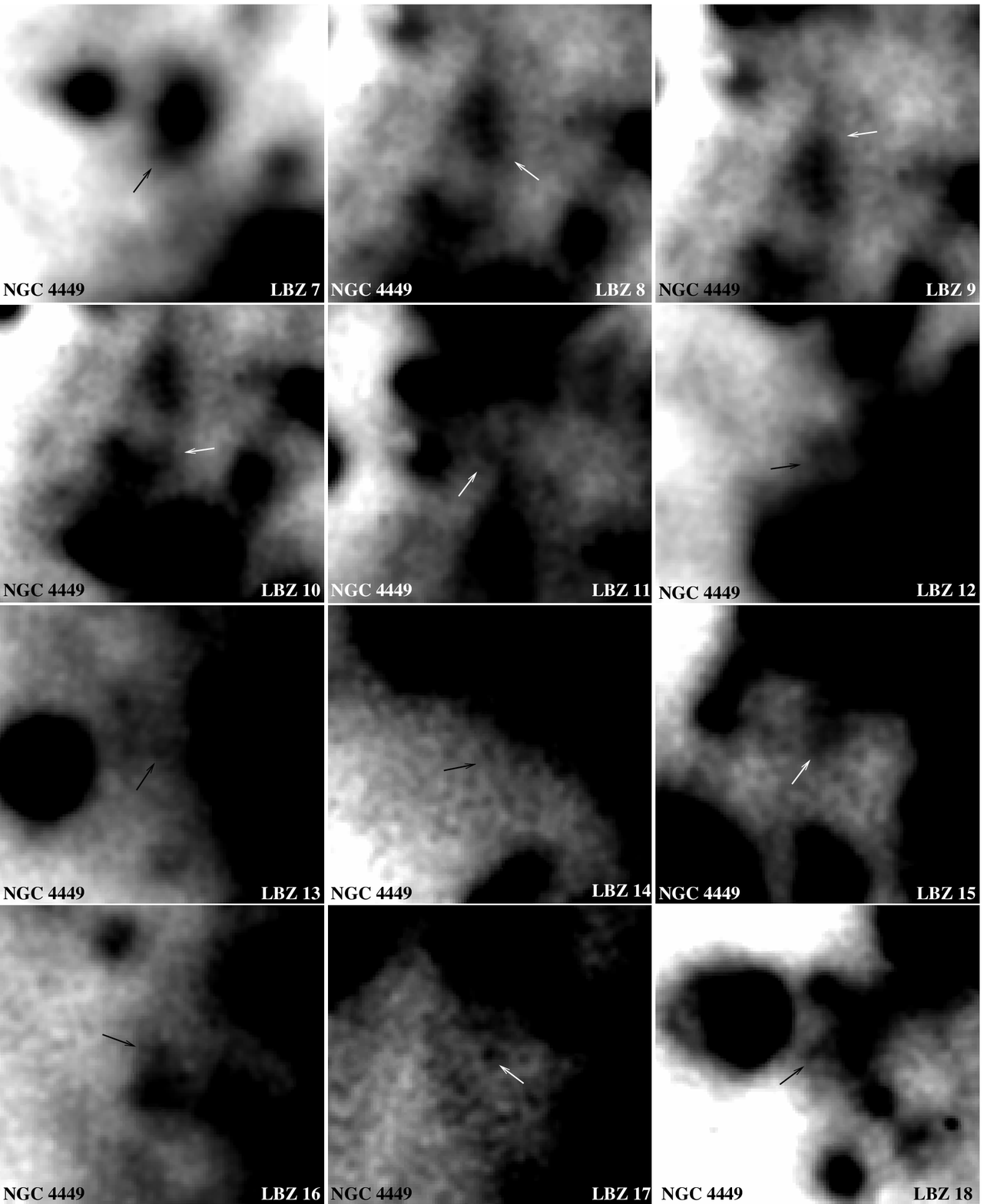}
\centering
 \contcaption{}
\end{figure}

\clearpage
\begin{figure}
\includegraphics[width=7in]{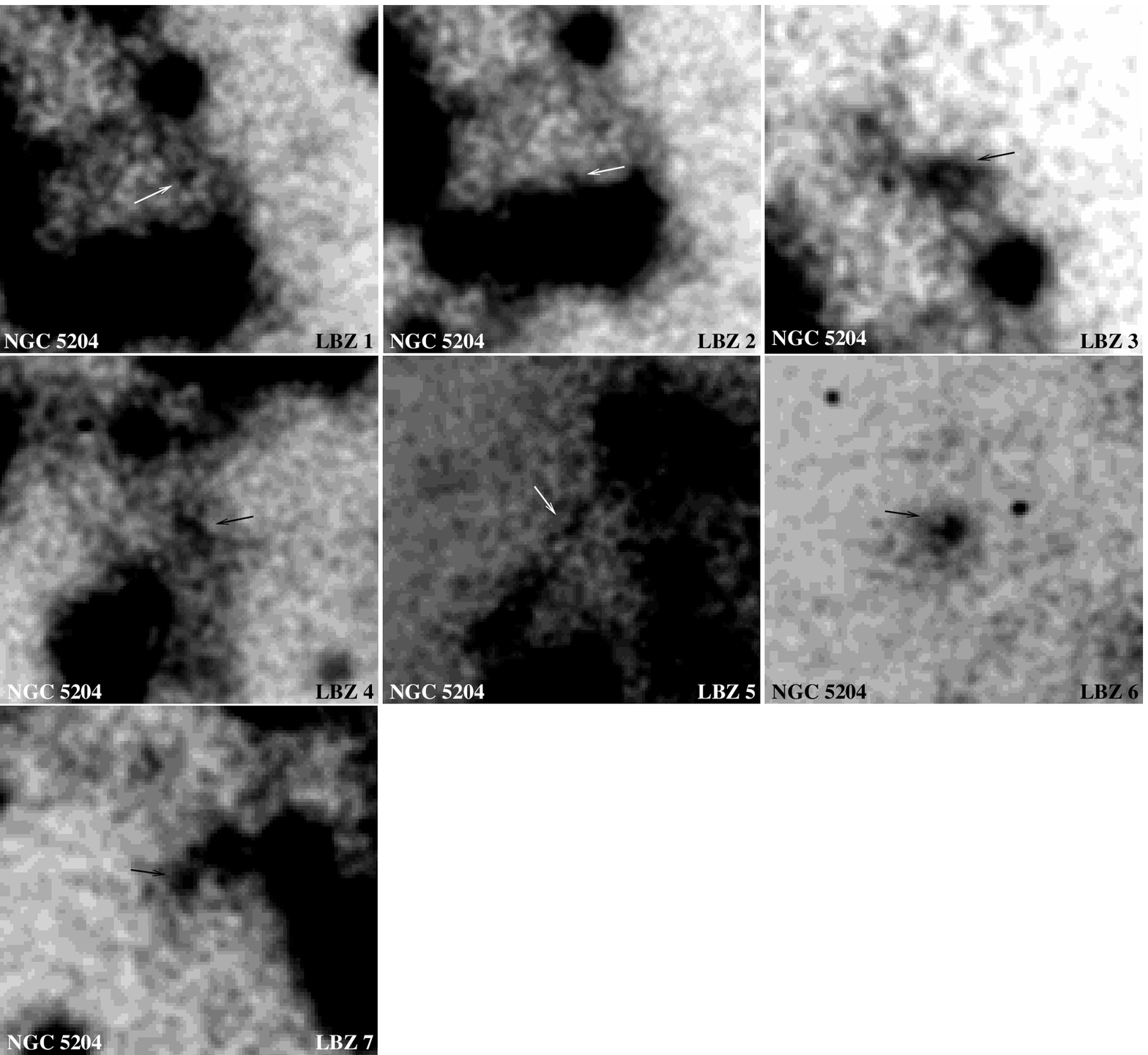}
\centering
 \contcaption{}
\end{figure}


\clearpage


\clearpage

\begin{figure}
\includegraphics[width=180mm,height=220mm]{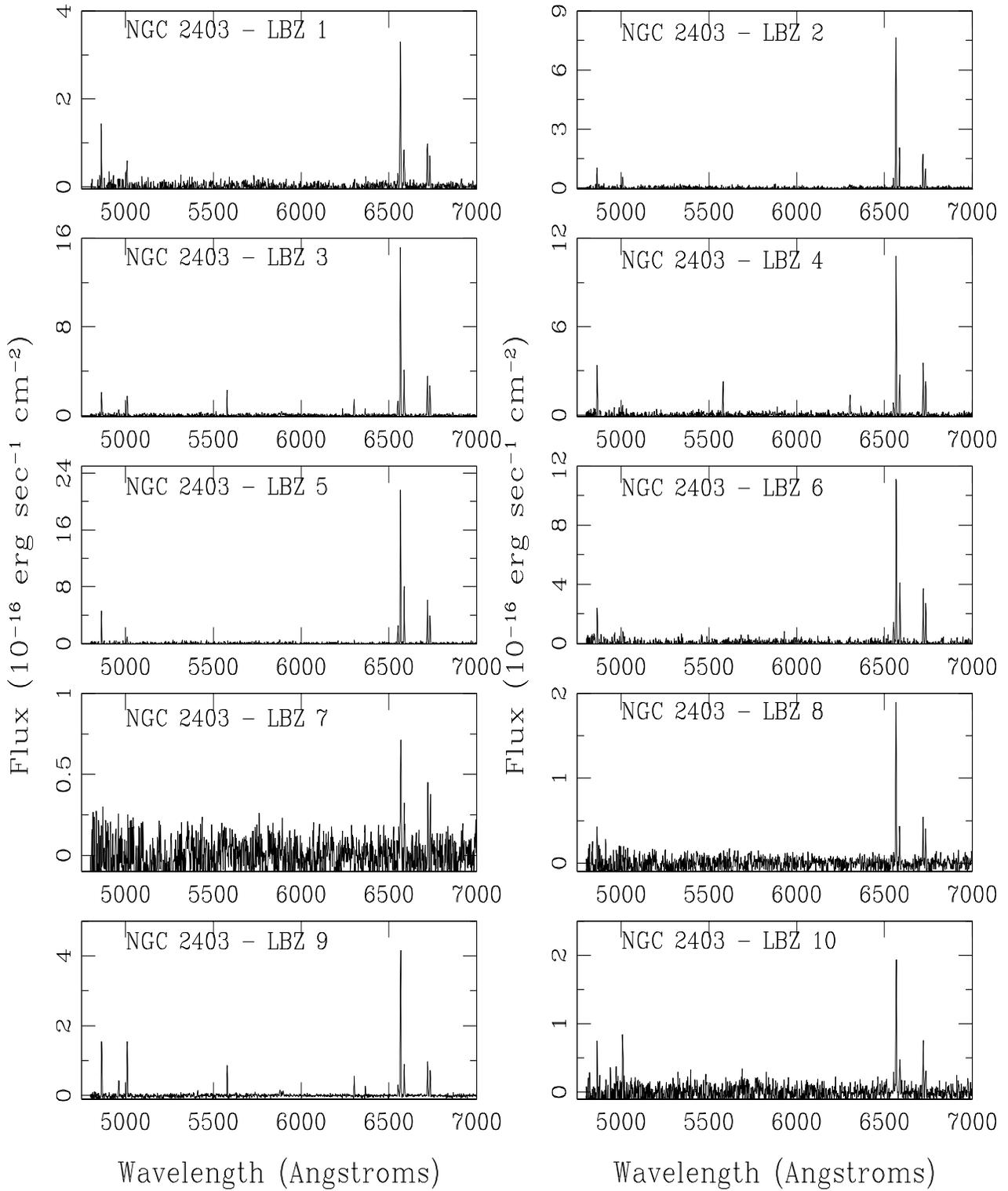}
\centering
 \caption{Spectra of the spectroscopically observed SNRs}
\end{figure}

\clearpage
\begin{figure}
\includegraphics[width=180mm,height=220mm]{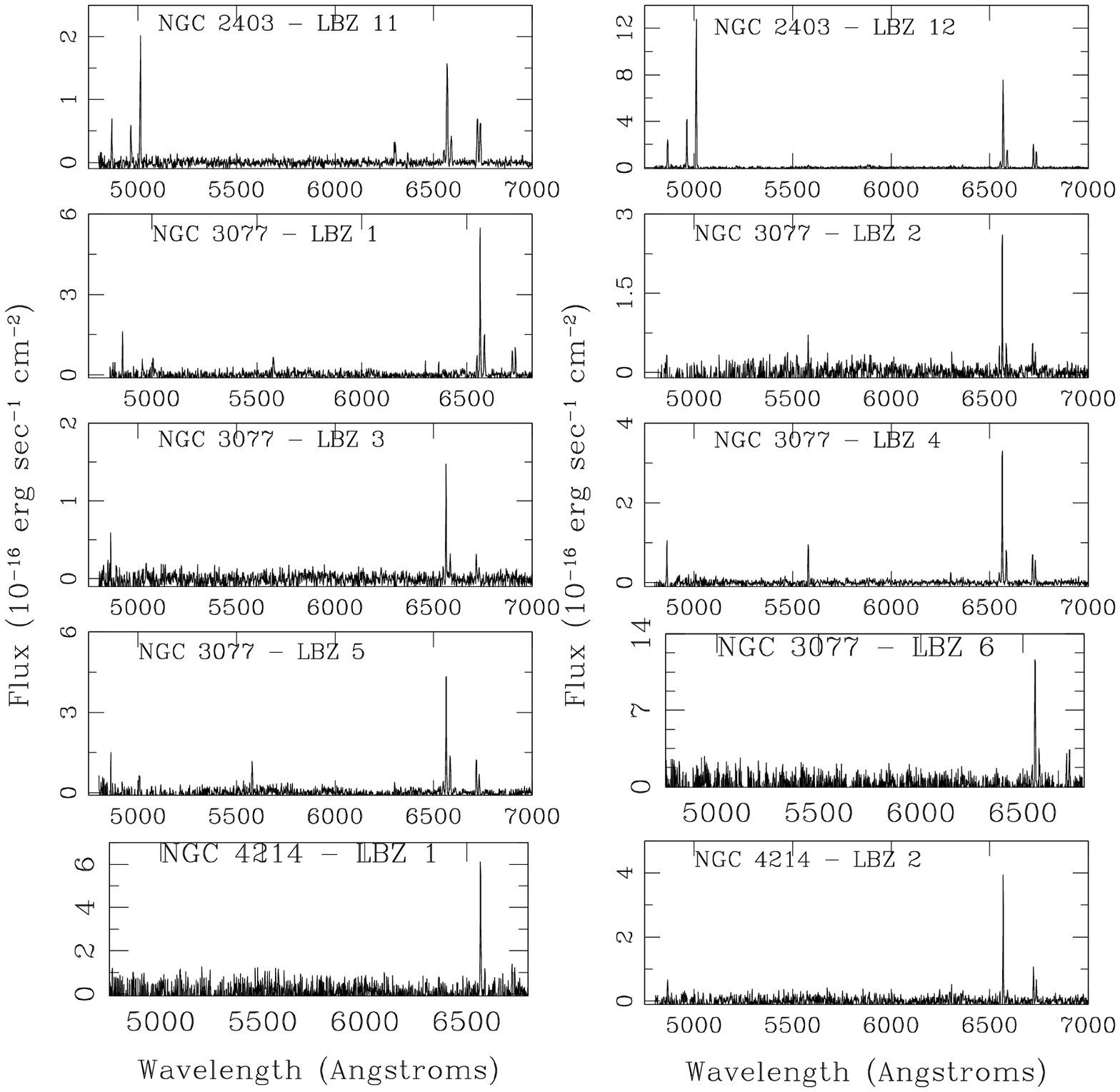}
\centering
 \contcaption{}
\end{figure}

\clearpage
\begin{figure}
\includegraphics[width=180mm,height=220mm]{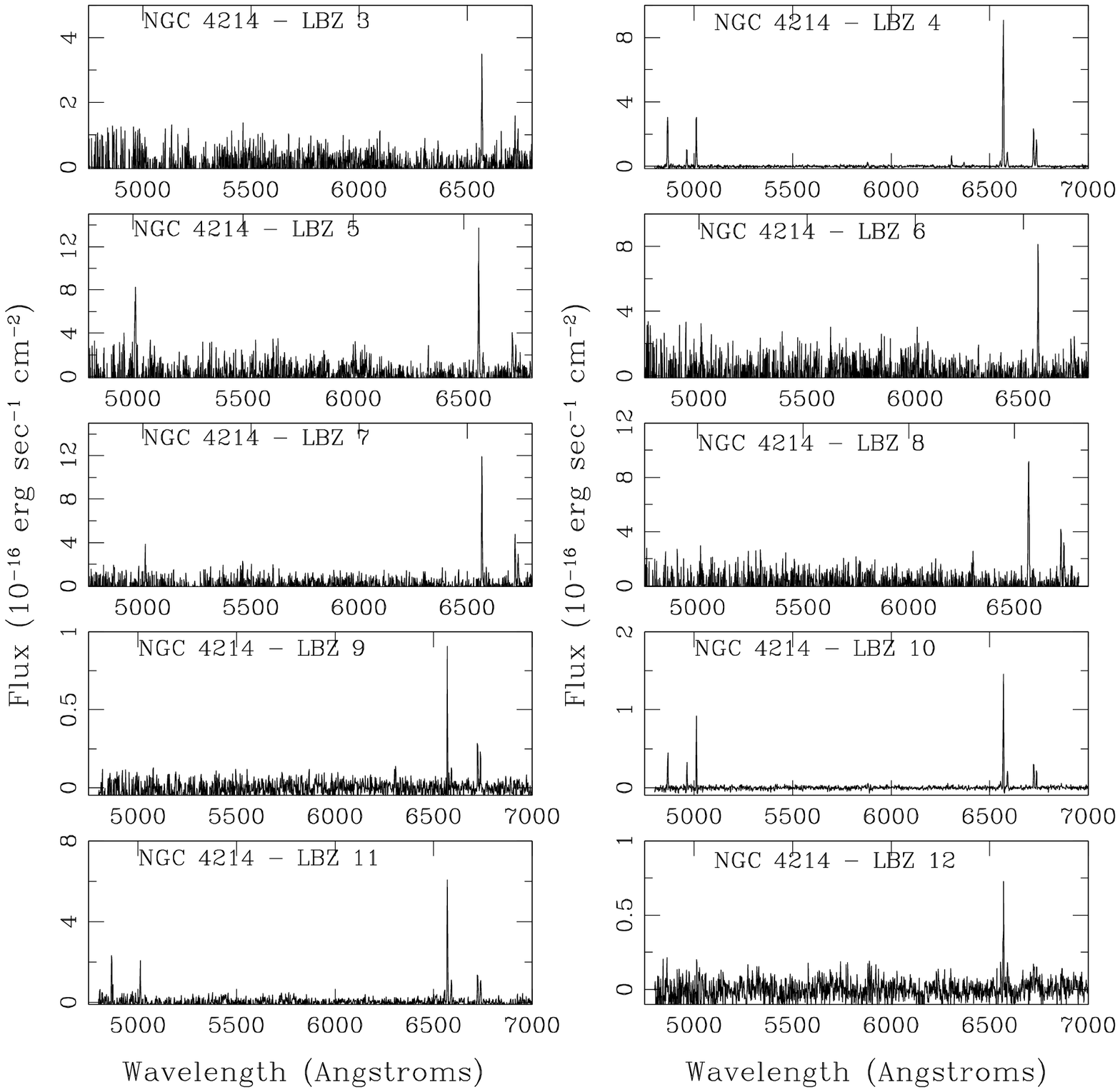}
\centering
 \contcaption{}
\end{figure}

\clearpage
\begin{figure}
\includegraphics[width=180mm,height=220mm]{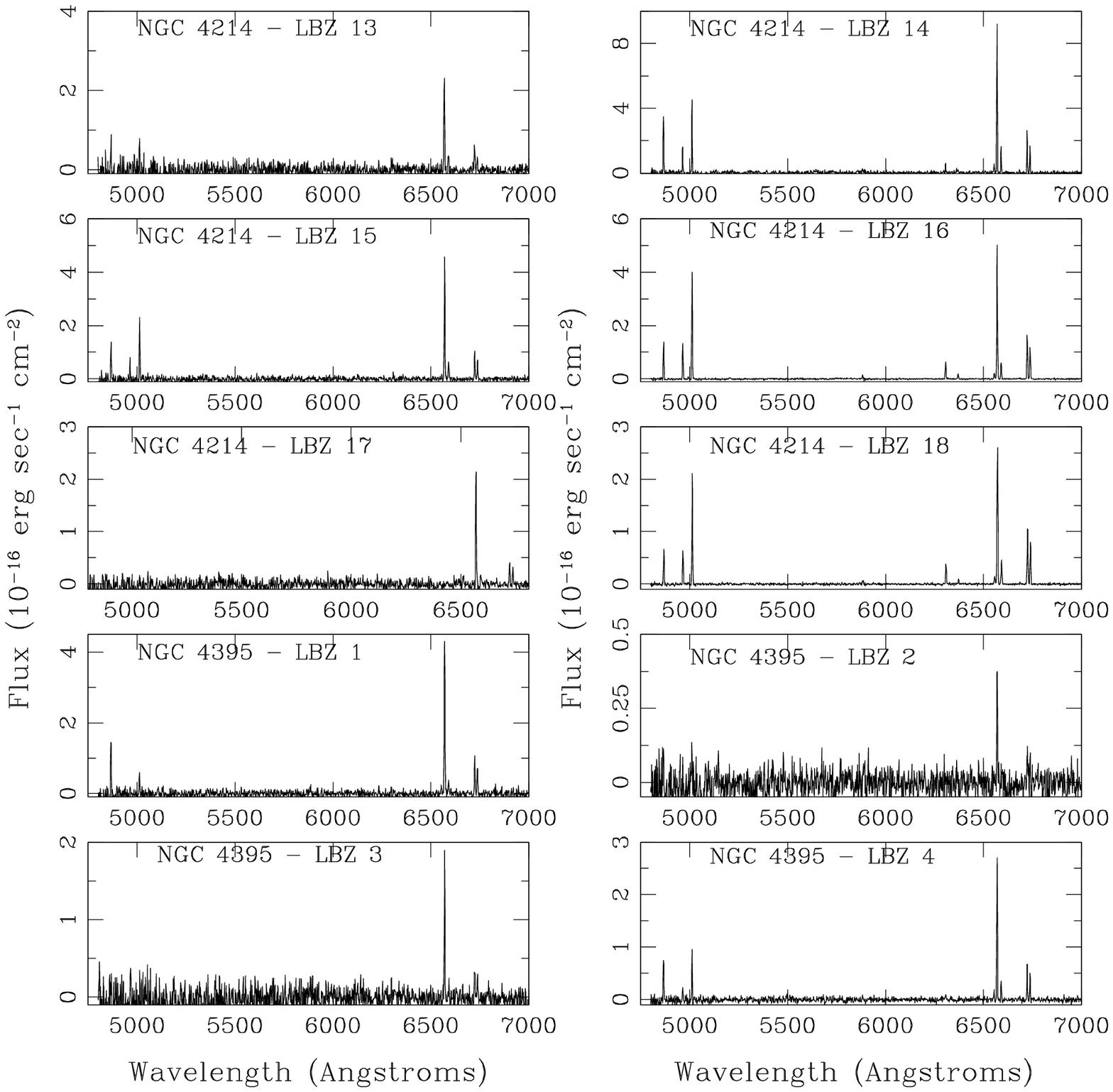}
\centering
 \contcaption{}
\end{figure}

\clearpage
\begin{figure}
\includegraphics[width=180mm,height=220mm]{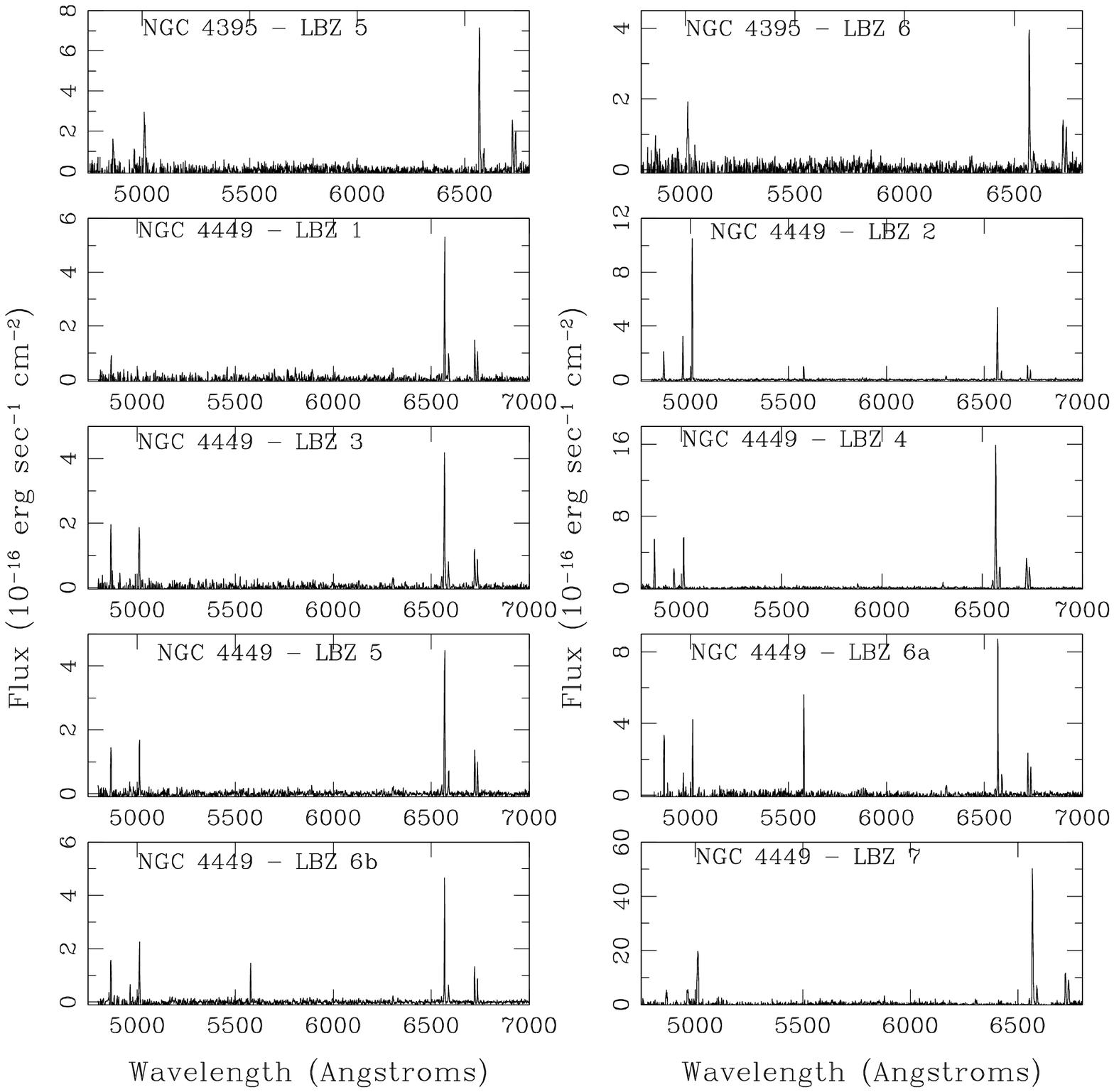}
\centering
 \contcaption{}
\end{figure}

\clearpage
\begin{figure}
\includegraphics[width=180mm,height=220mm]{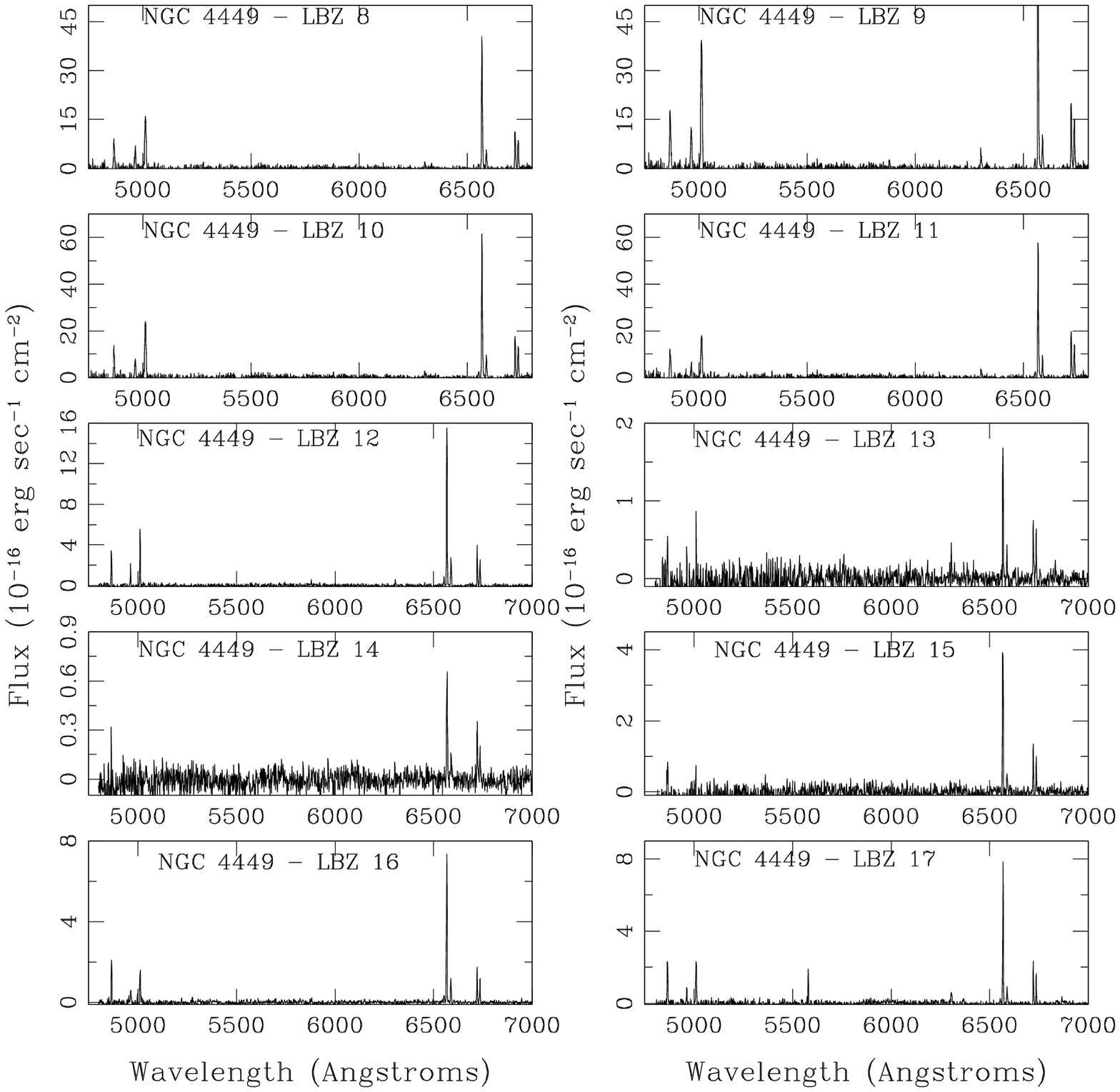}
\centering
 \contcaption{}
\end{figure}

\clearpage
\begin{figure}
\includegraphics[width=180mm,height=200mm]{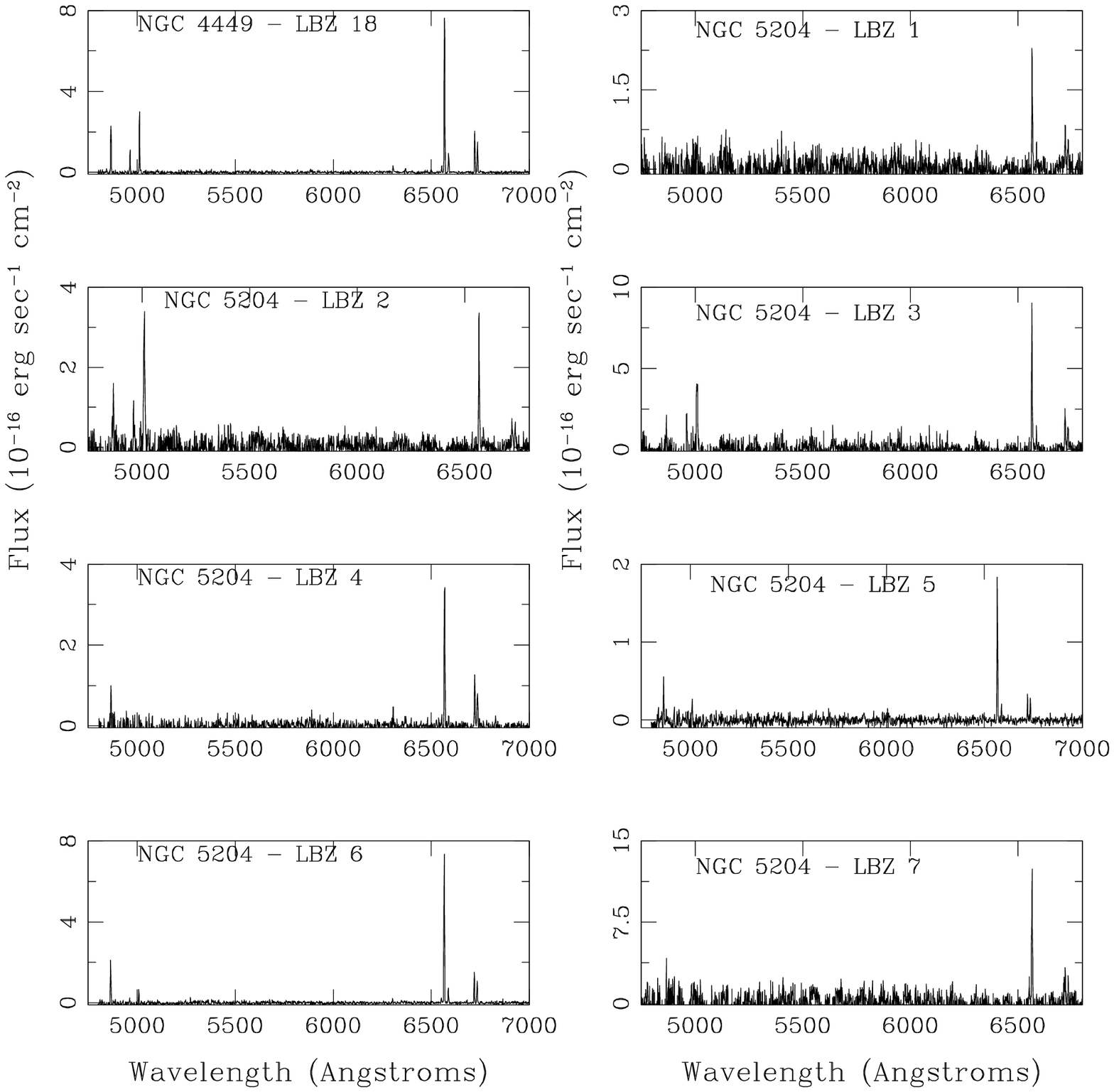}
\centering
 \contcaption{}
\end{figure}

\clearpage

\begin{figure}
    \centering
    \subfigure[]
    {
        \includegraphics[width=1.5in]{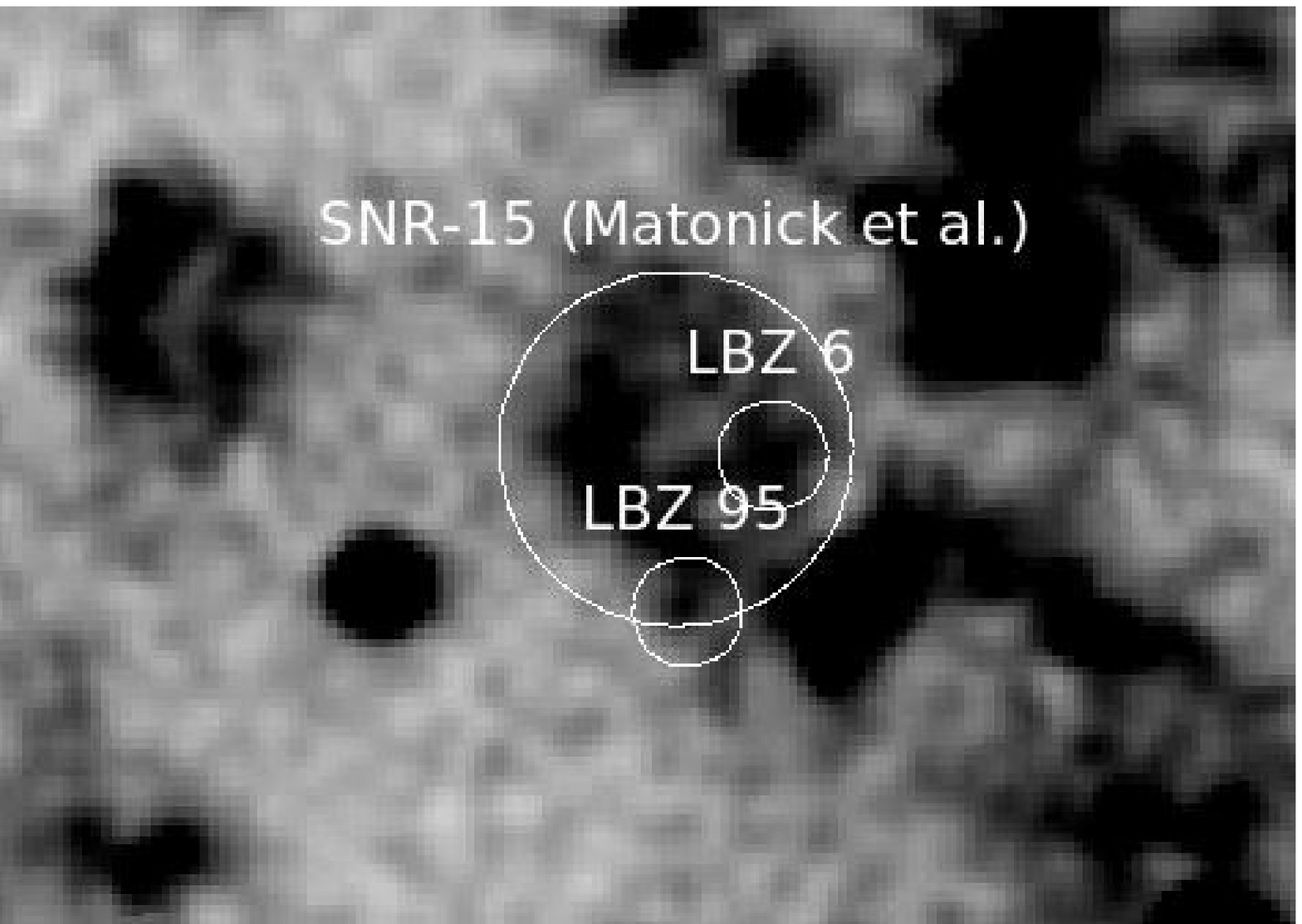}
        \label{fig:first_sub}
    }
    \vspace{-0.2in}
    \subfigure[]
    {
        \includegraphics[width=1.5in]{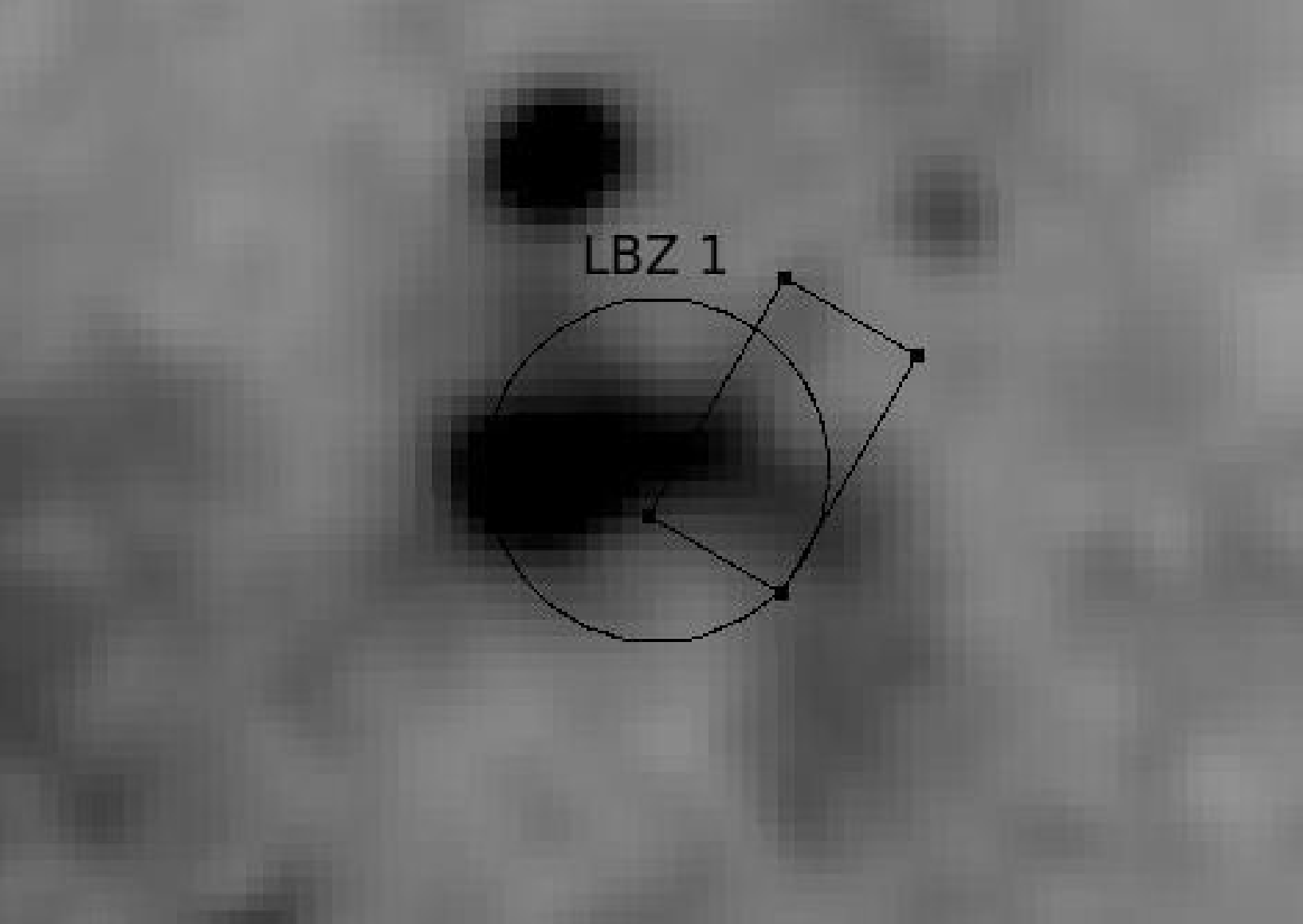}
        \label{fig:second_sub}
    }
    \subfigure[]
    {
        \includegraphics[width=1.5in]{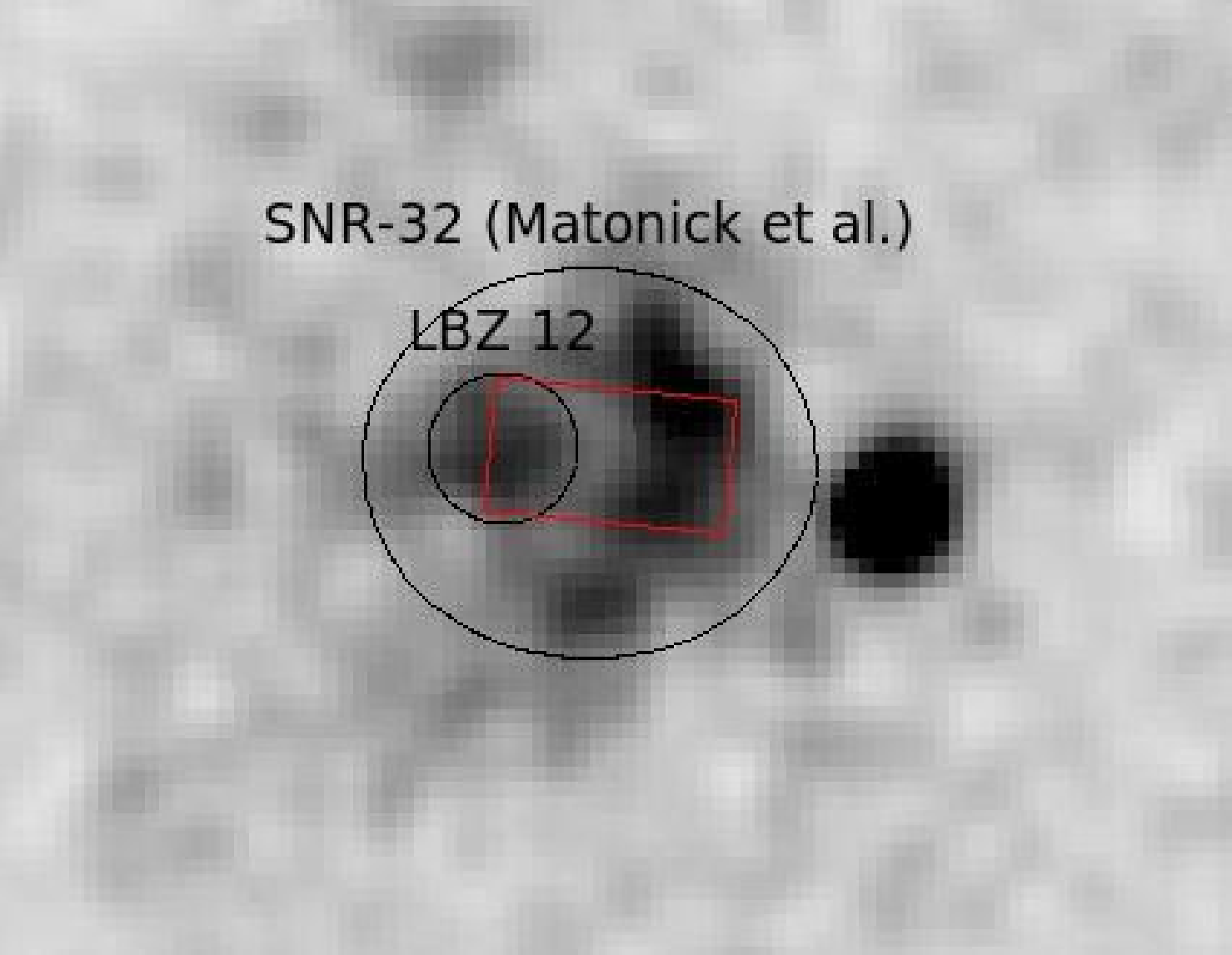}
        \label{fig:third_sub}
    }
    \caption{SNRs in NGC\,2403}
    \label{fig:sample_subfigures}
\end{figure}

\begin{figure}
\includegraphics[width=3.0in]{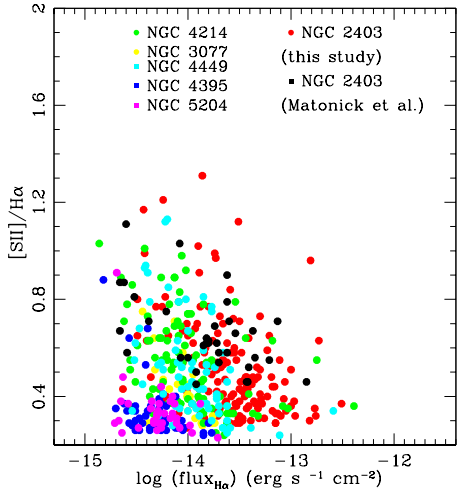}
\centering
 \caption{\SII/\Ha\ ratio of the 418 photometric SNRs (see Table 14) in our sample of galaxies against their \Ha\ flux (see Tables 3-8 and \S4.2).}
\end{figure}

\begin{figure}
\includegraphics[width=3.0in]{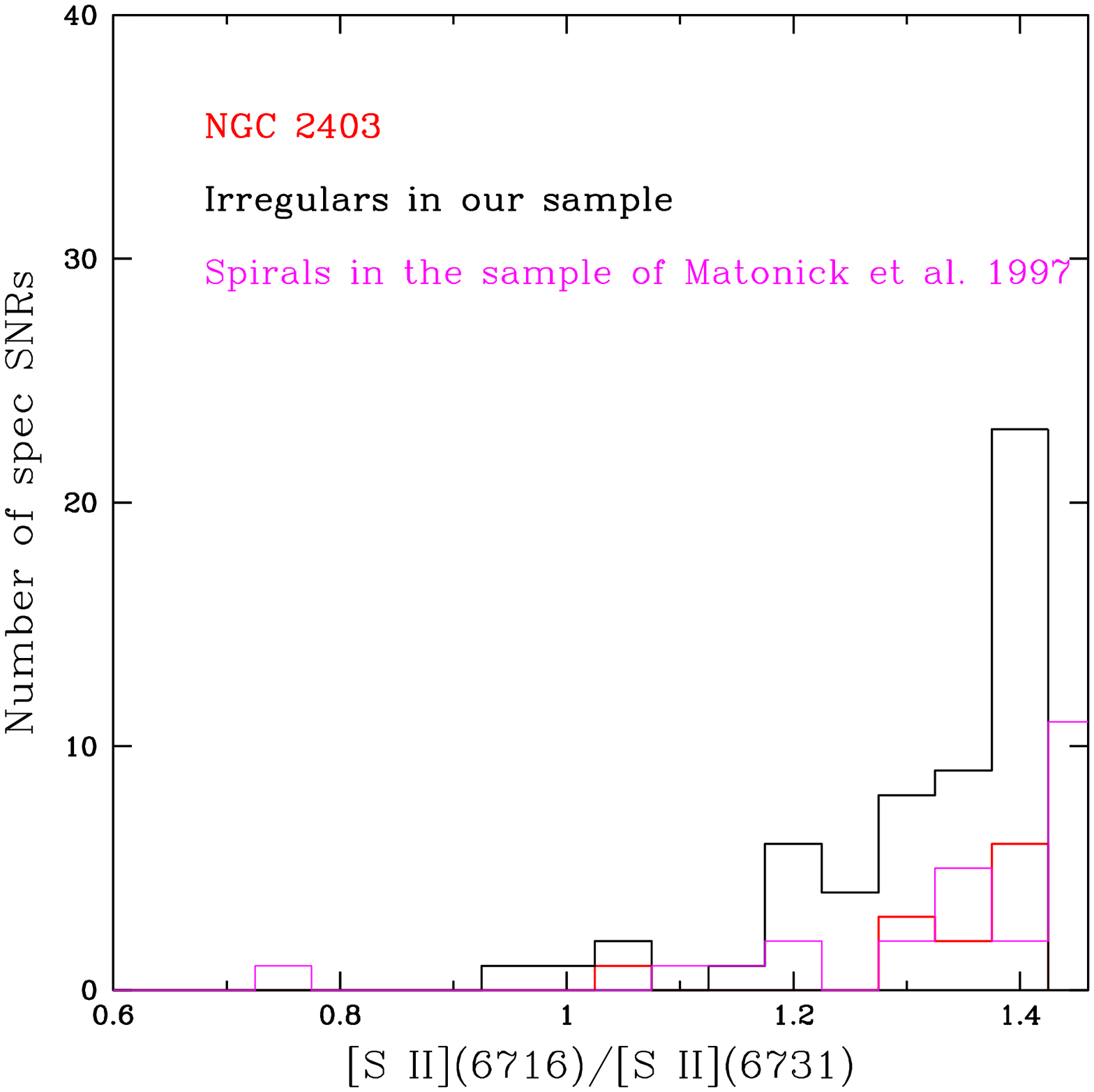}
\centering
 \caption{Number of spectroscopically-observed SNRs against their \SII(6716 \AA)/\SII(6731\AA) ratios. Red: SNRs in NGC\,2403 (the only spiral galaxy in our sample), black: SNRs in the remaining galaxies of our sample (irregulars), magenta: Spectroscopically observed SNRs by \citet{MFBL97} in four spiral galaxies (NGC\,5585, NGC\,6946, M81 and M101).  As can been seen, there is no trend in the \SII ratios between different types of galaxies. However, the majority of the spectroscopically observed SNRs present \SII(6716 \AA)/\SII(6731\AA) $>$ 1 (see \S4.2 for details).}
\end{figure}

\clearpage

\begin{figure}
\centering
\begin{tabular}{@{}c@{}c@{}}
\includegraphics[width=0.37\textwidth]{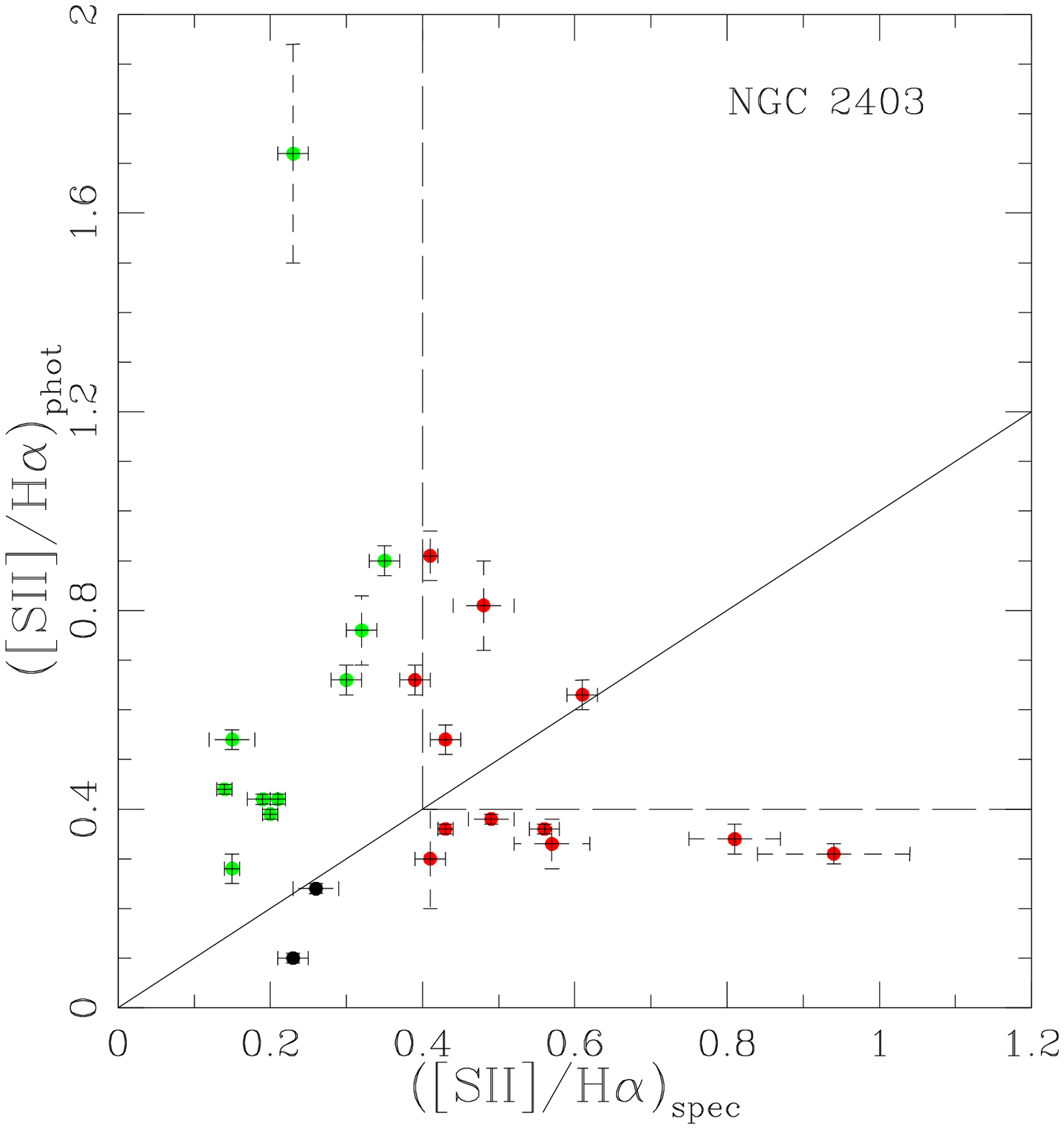} &
\includegraphics[width=0.37\textwidth]{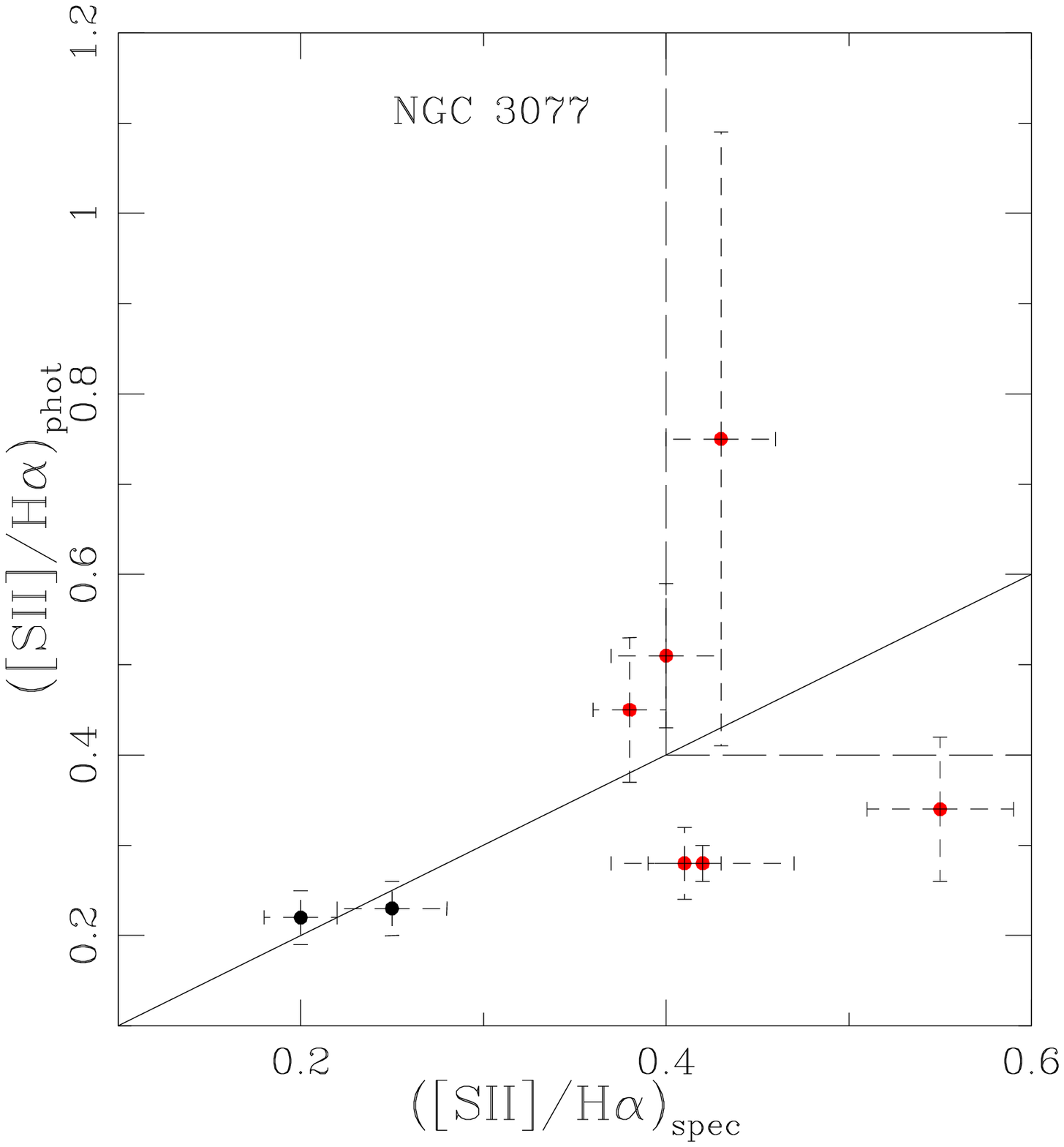} \\
\includegraphics[width=0.37\textwidth]{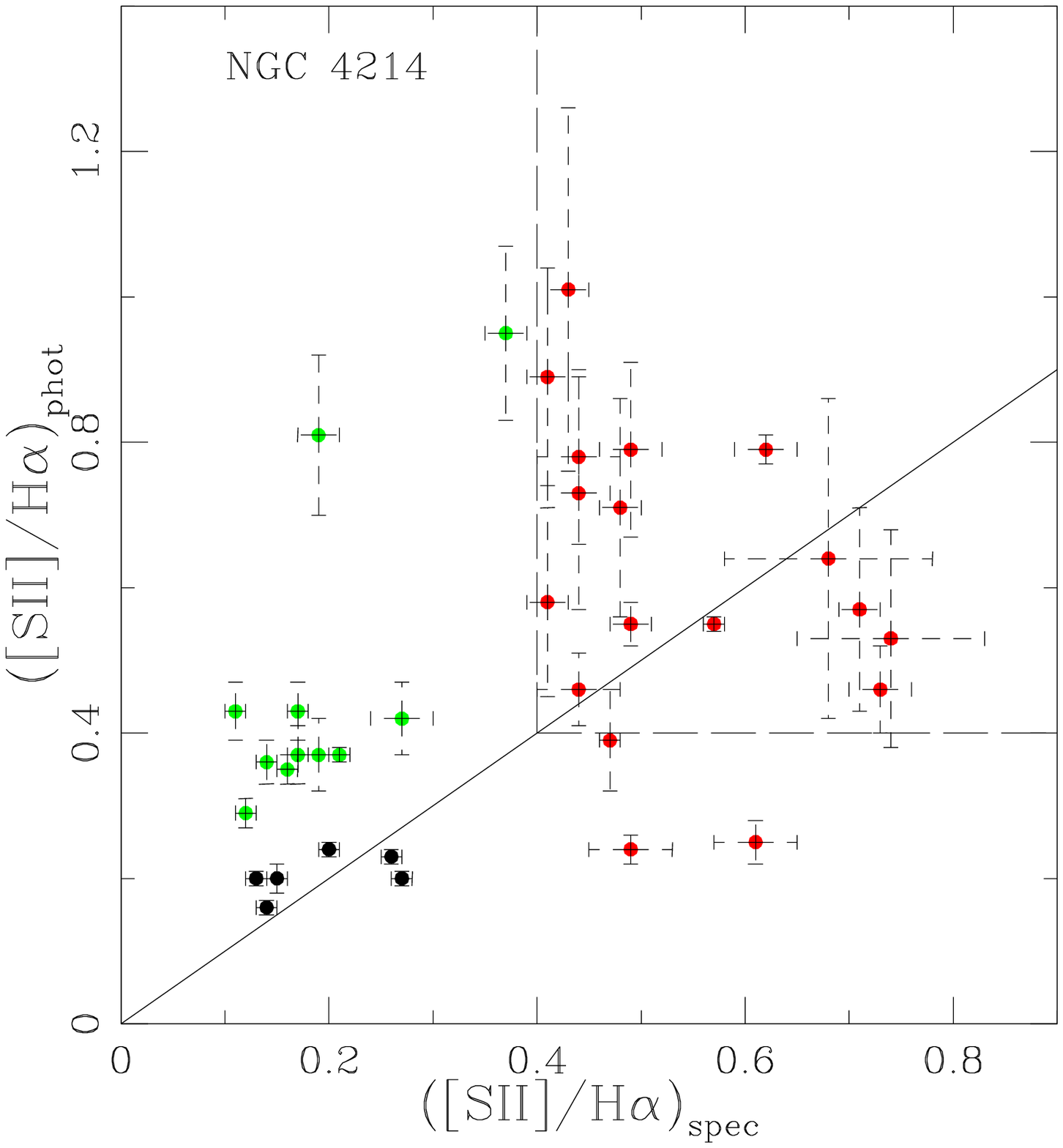} &
\includegraphics[width=0.37\textwidth]{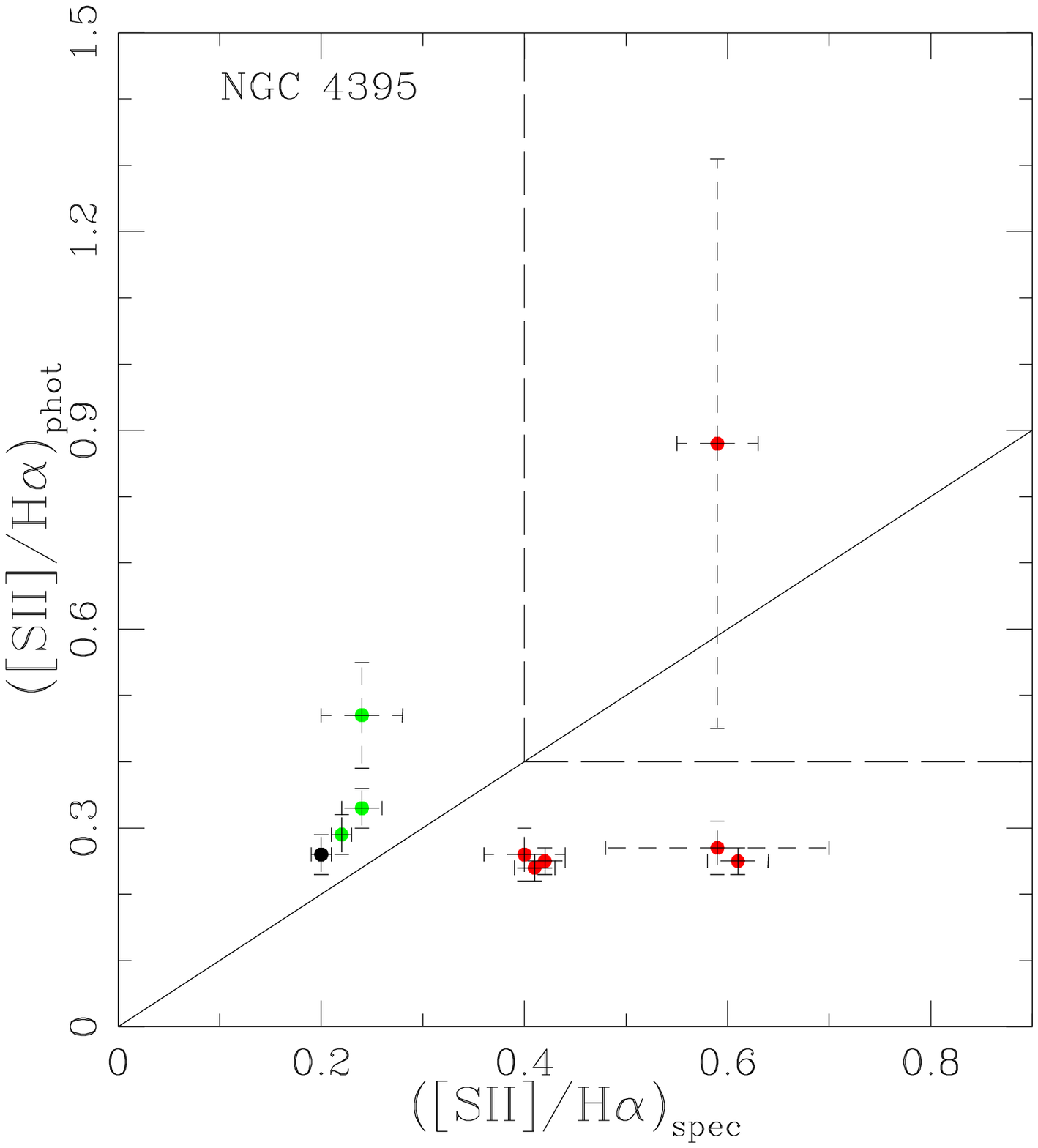} \\
\includegraphics[width=0.37\textwidth]{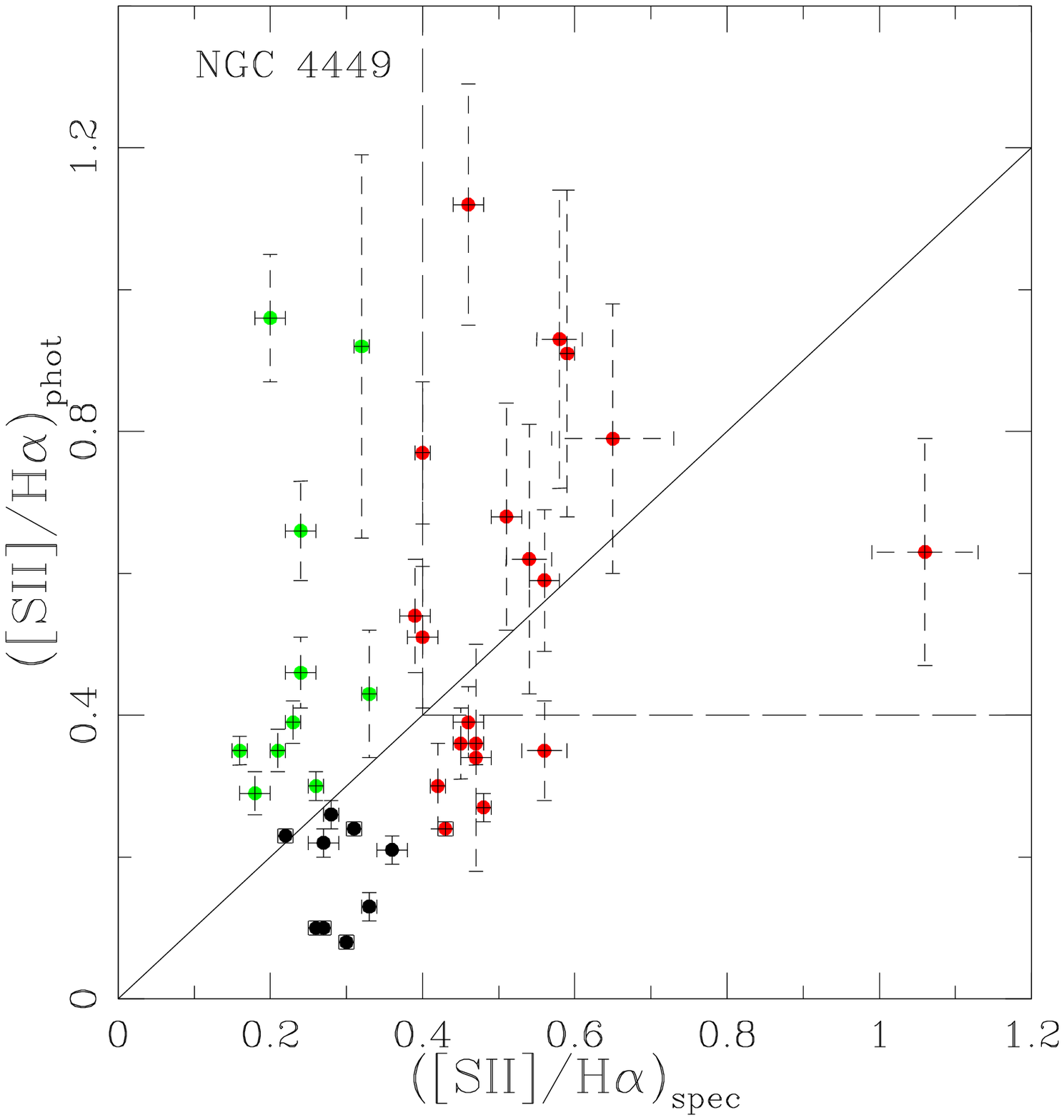} &
\includegraphics[width=0.37\textwidth]{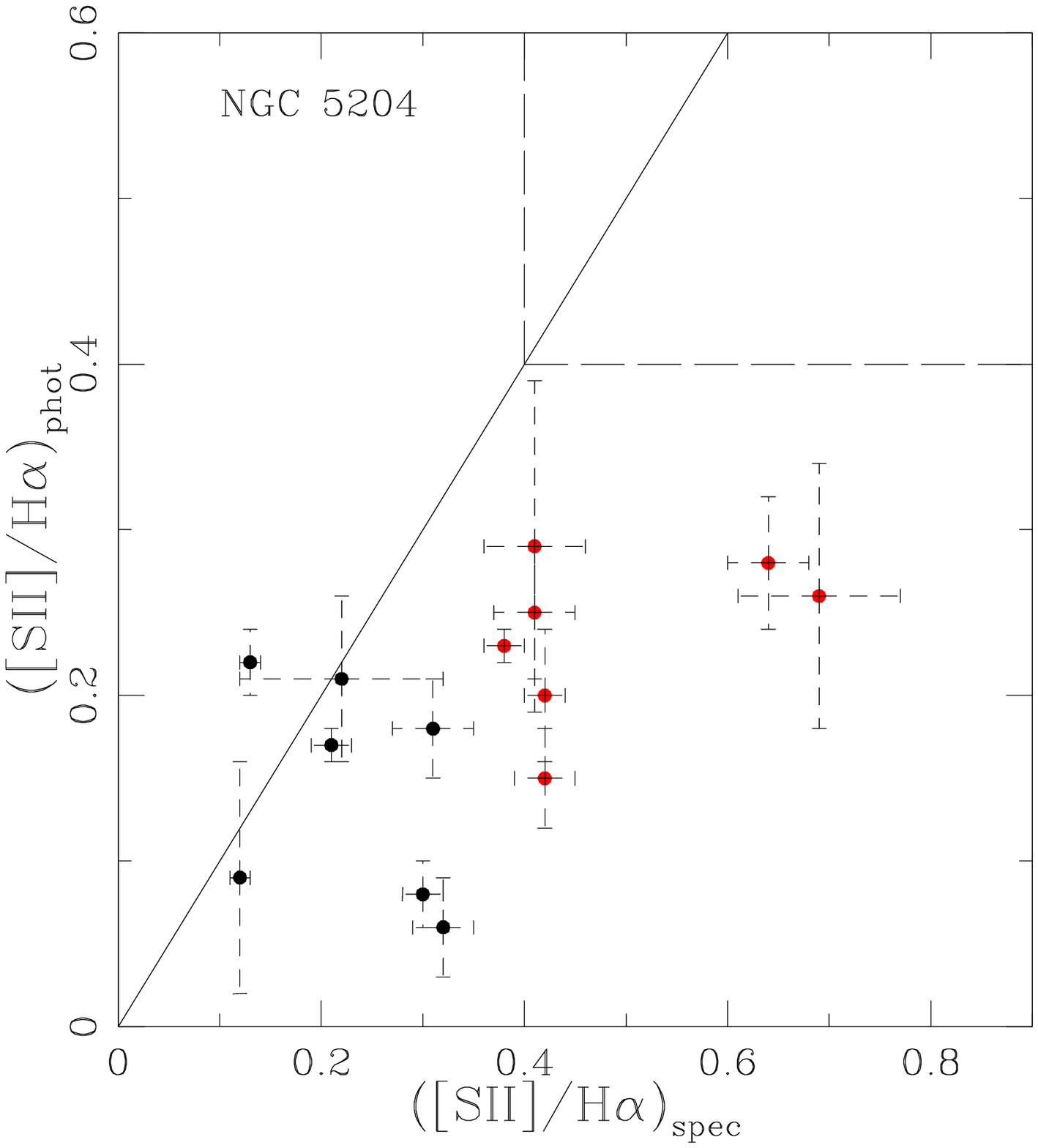} \\
\end{tabular}
\caption{The (\SII/\Ha)$_{spec}$ versus the (\SII/\Ha)$_{phot}$ ratios of all spectroscopically observed sources. The red circles denote SNRs (see $\S$4, Tables 3-8, Table 14) while the green circles indicate sources with (\SII/\Ha)$_{phot}$ $\geq$ 0.3 (within their error-bars) but were not spectroscopically verified as SNRs (see Tables 9 and 13). Black circles denote sources with (\SII/\Ha)$_{phot} \le$ 0.4 and (\SII/\Ha)$_{spec} \le$ 0.4 (see Tables 9 and 13). The solid line represents the 1:1 relation between photometric and spectroscopic \SII/\Ha\ ratios while the dashed lines denote the borderline area for SNRs (\SII/\Ha$>$0.4).}
\label{eis1}
\end{figure} 
\clearpage

\begin{figure}
\includegraphics[width=4in]{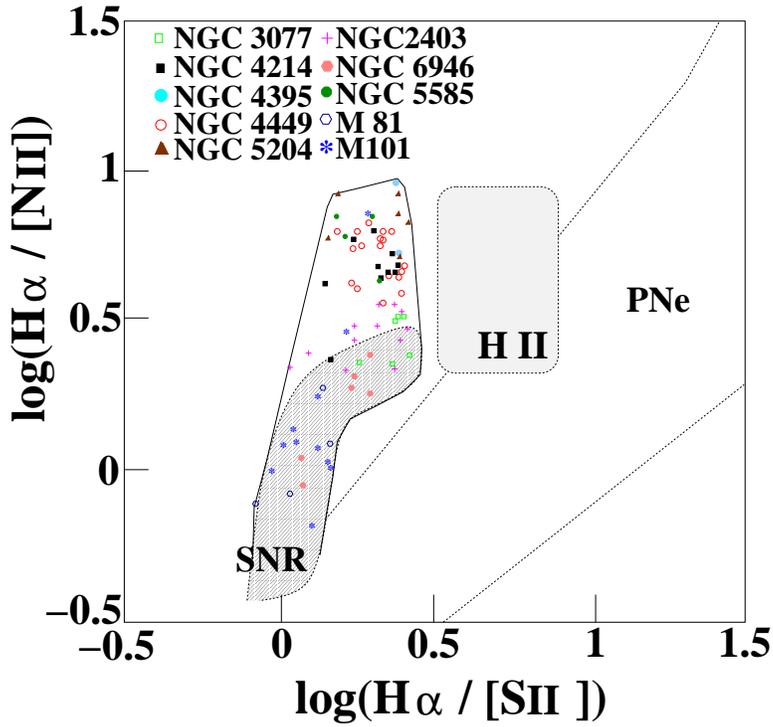}
\centering
 \caption{log(\Ha/\SII \AA\ (6716 \& 6731)) against the log(\Ha/\NII \AA\ (6548 \& 6584)) emission line ratios of the spectroscopically observed SNRs. The dashed lines have been defined using the emission line ratios of an adequate number of Galactic SNRs, \HII\ regions and planetary nebulae (PNe).}
\end{figure}

\begin{figure}
\includegraphics[width=4.5in]{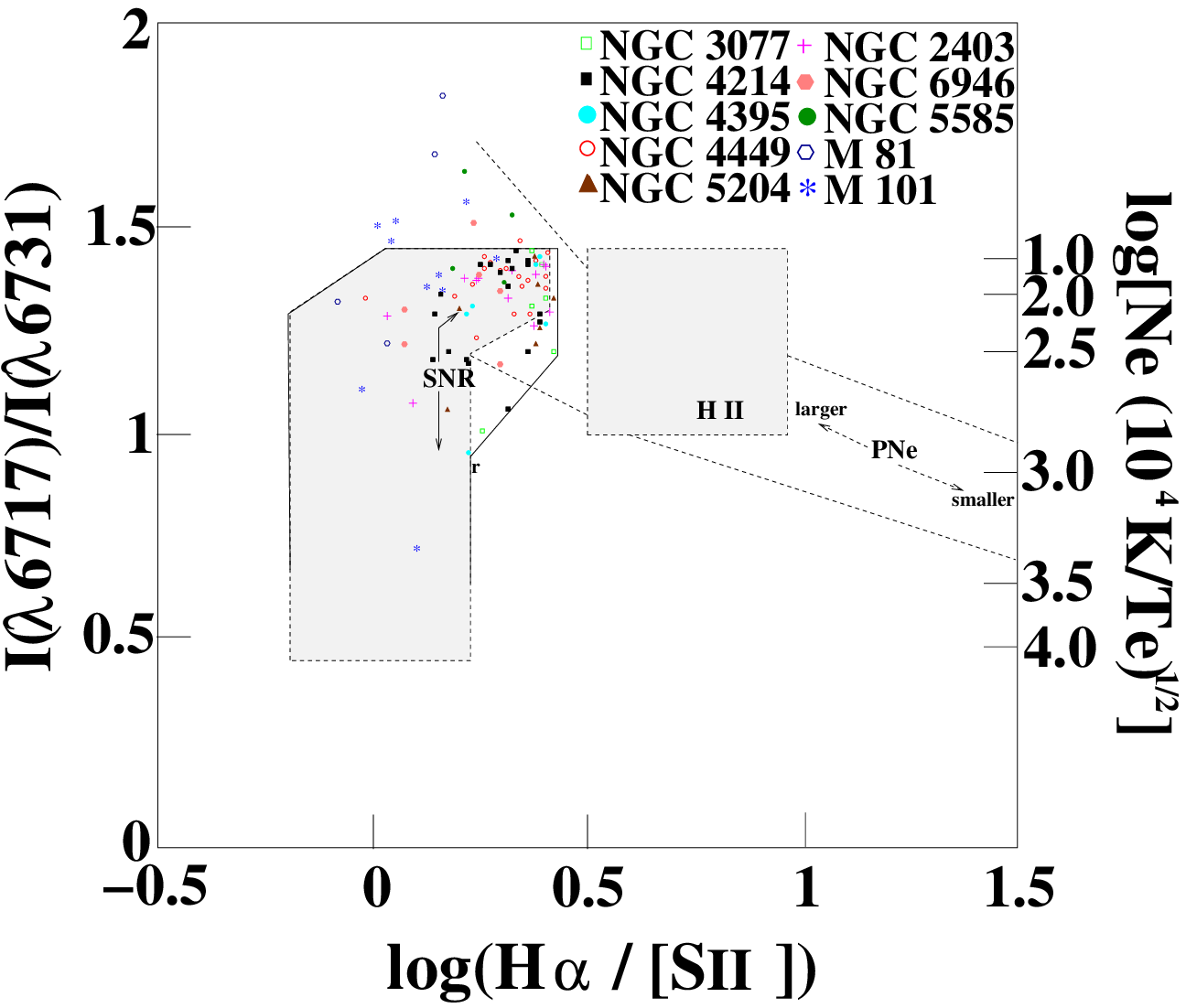}
\centering
 \caption{\SII \AA\ (6716)/\SII \AA\ (6731) line ratio versus log(\Ha/\SII \AA\ (6716 \& 6731).}
\end{figure}

\clearpage
\begin{figure}
\includegraphics[width=4.5in]{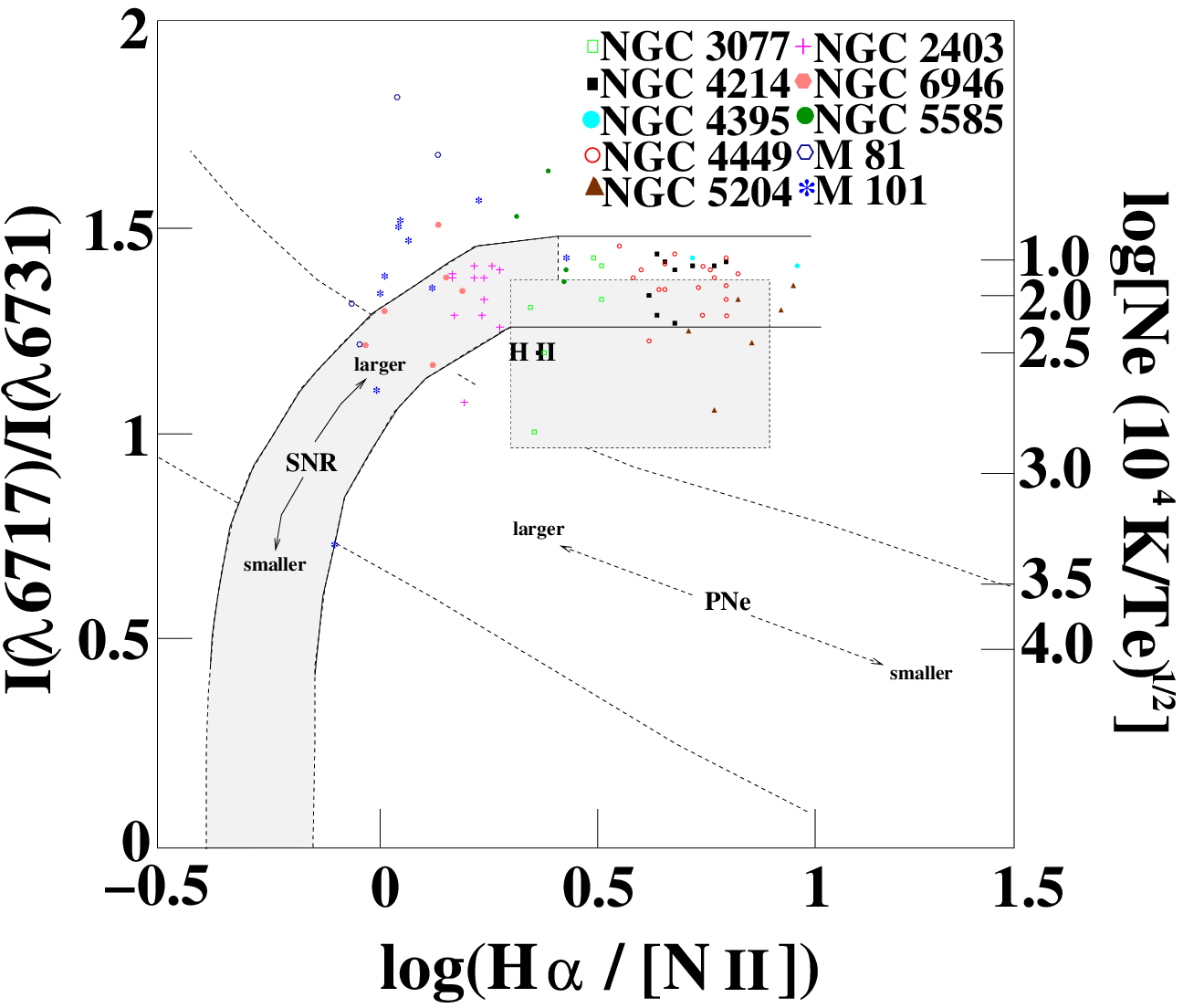}
\centering
 \caption{\SII \AA\ (6716)/\SII \AA\ (6731) line ratio versus log(\NII/\SII \AA\ (6716 \& 6731).}
\end{figure}

\begin{figure}
\includegraphics[width=4.5in]{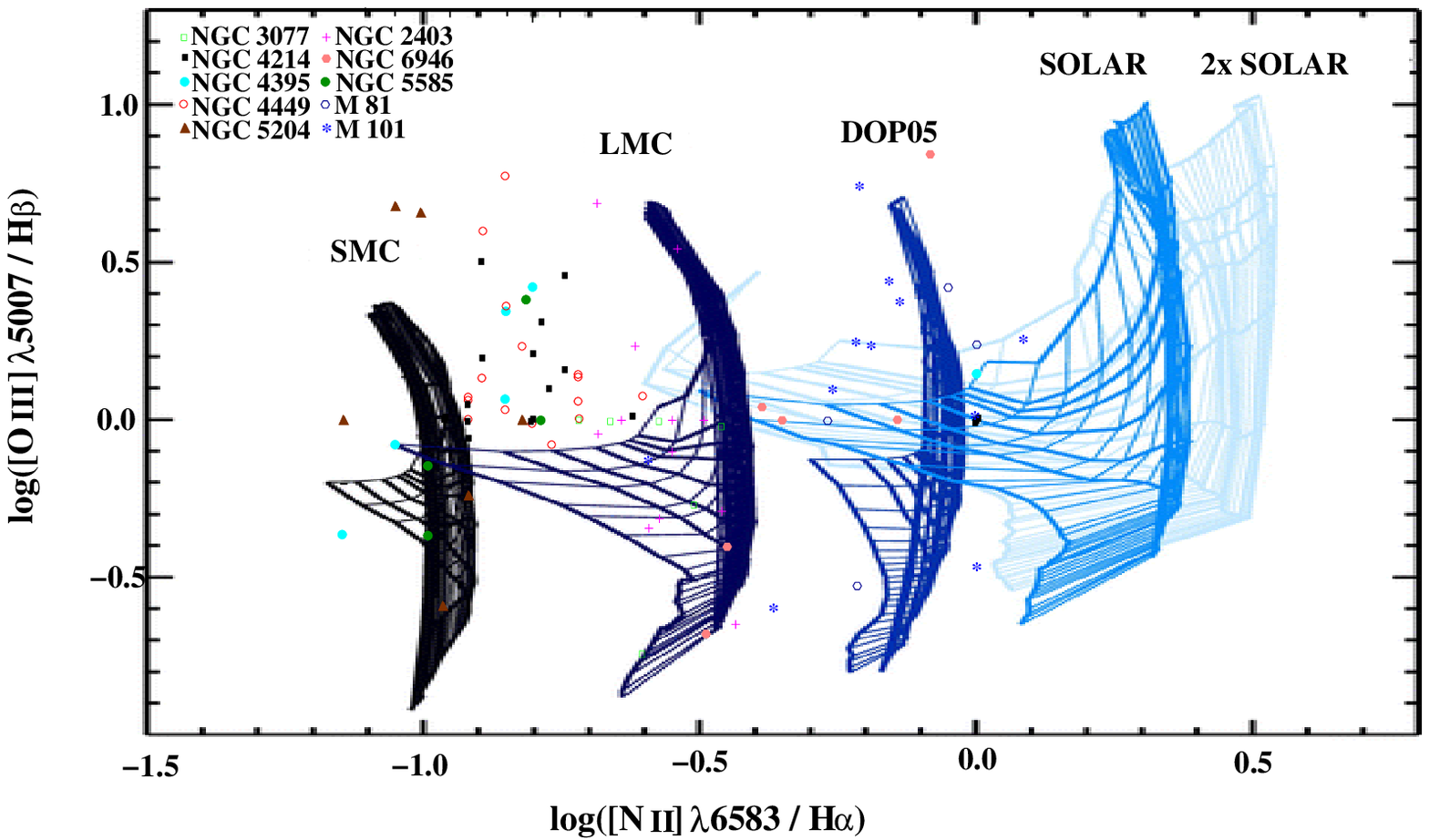}
\centering
 \caption{Diagnostic diagram of \OIII(\AA\ 5007)/\Hb\ versus \NII(\AA\ 6584)/\Ha\ for shock-only models and for five different abundance sets with n=1 cm$^{-3}$ by  \citet{Allen08}. Each grid is labeled with the abundance set that was used, moving from left to right with increasing metallicity. Each grid comprises lines of constant magnetic parameter shown with thick lines and lines of constant shock velocity shown with thin lines. The shock velocities range between 200 and 1000  km s$^{-1}$, from top to bottom with a step of 50 km s$^{-1}$.}
\end{figure}

\clearpage

\begin{figure}
\includegraphics[width=4.5in]{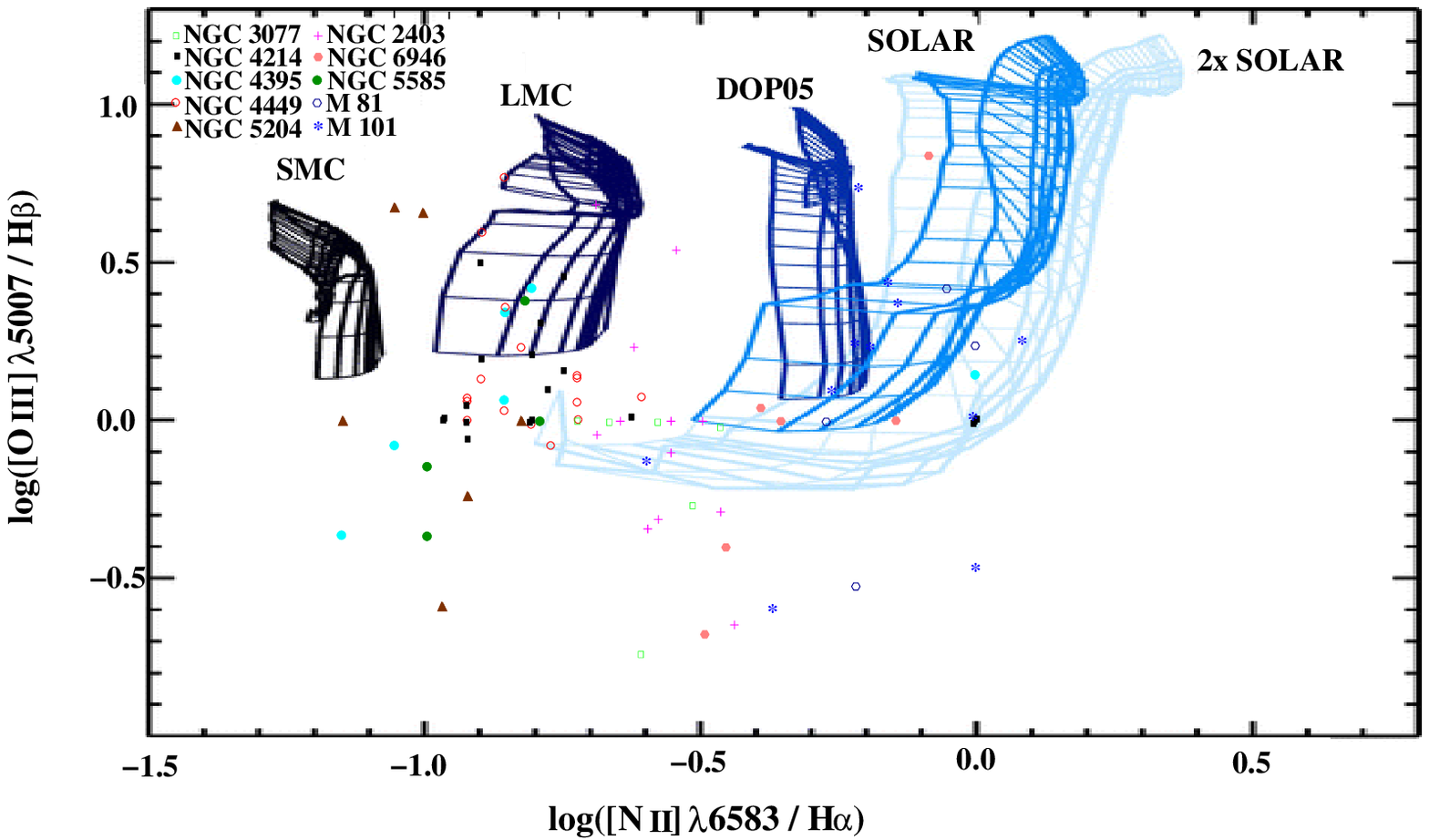}
\centering
 \caption{Diagnostic diagram of \OIII(\AA\ 5007)/\Hb\ versus \NII(\AA\ 6584)/\Ha\ for shock+precursor models and for five different abundance sets with n=1 cm$^{-3}$ by  \citet{Allen08}. Each grid is labeled with the abundance set that was used, moving from left to right with increasing metallicity. Each grid comprises lines of constant magnetic parameter shown with thick lines and lines of constant shock velocity shown with thin lines. The shock velocities range between 200 and 1000 km s$^{-1}$, from bottom to top with a step of 50 km s$^{-1}$.}
\end{figure}

\begin{figure}
    \centering
    \subfigure[]
    {
        \includegraphics[width=1.5in]{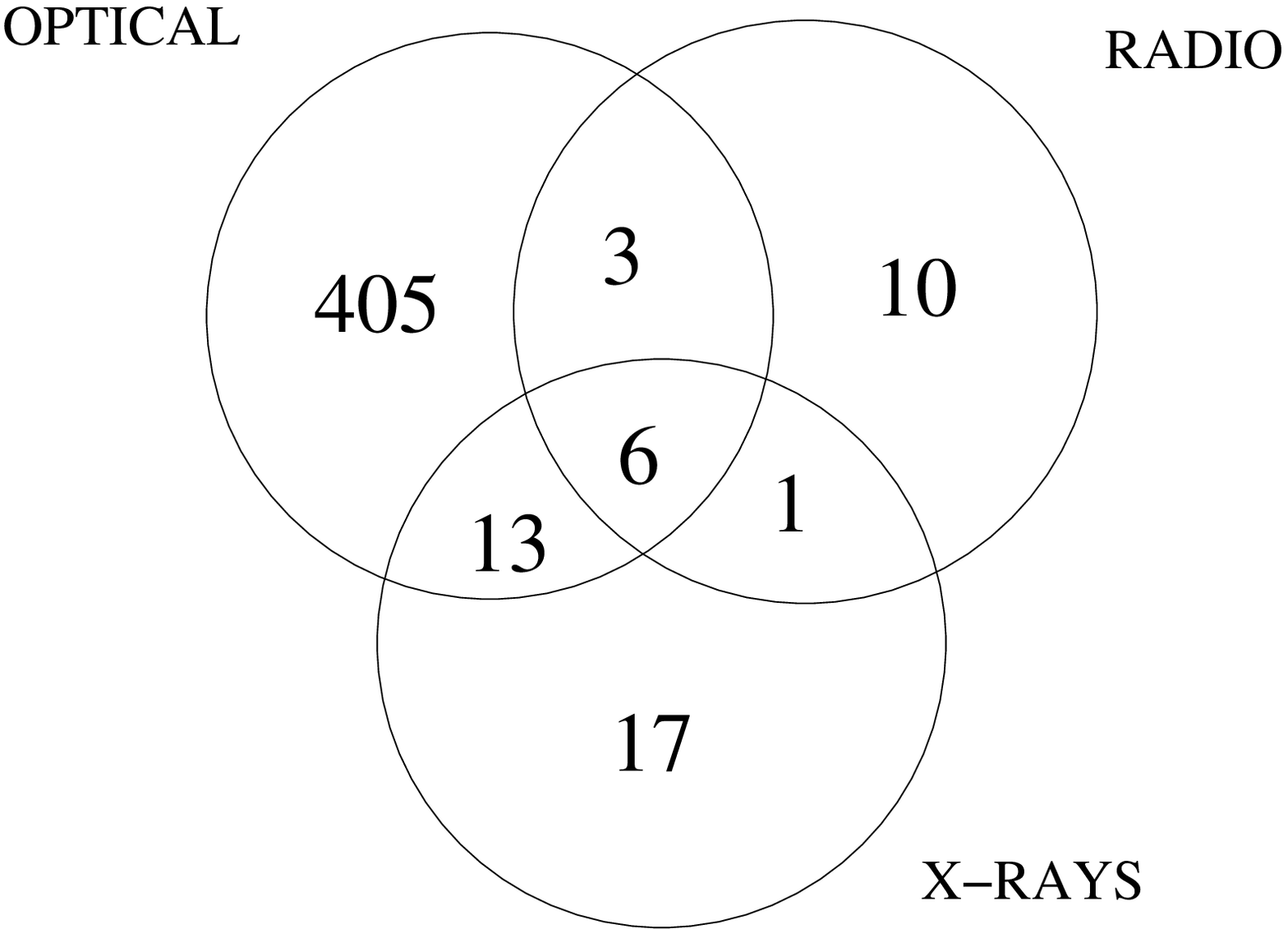}
        \label{fig:first_sub}
    }
    \vspace{-0.2in}
    \subfigure[]
    {
        \includegraphics[width=1.5in]{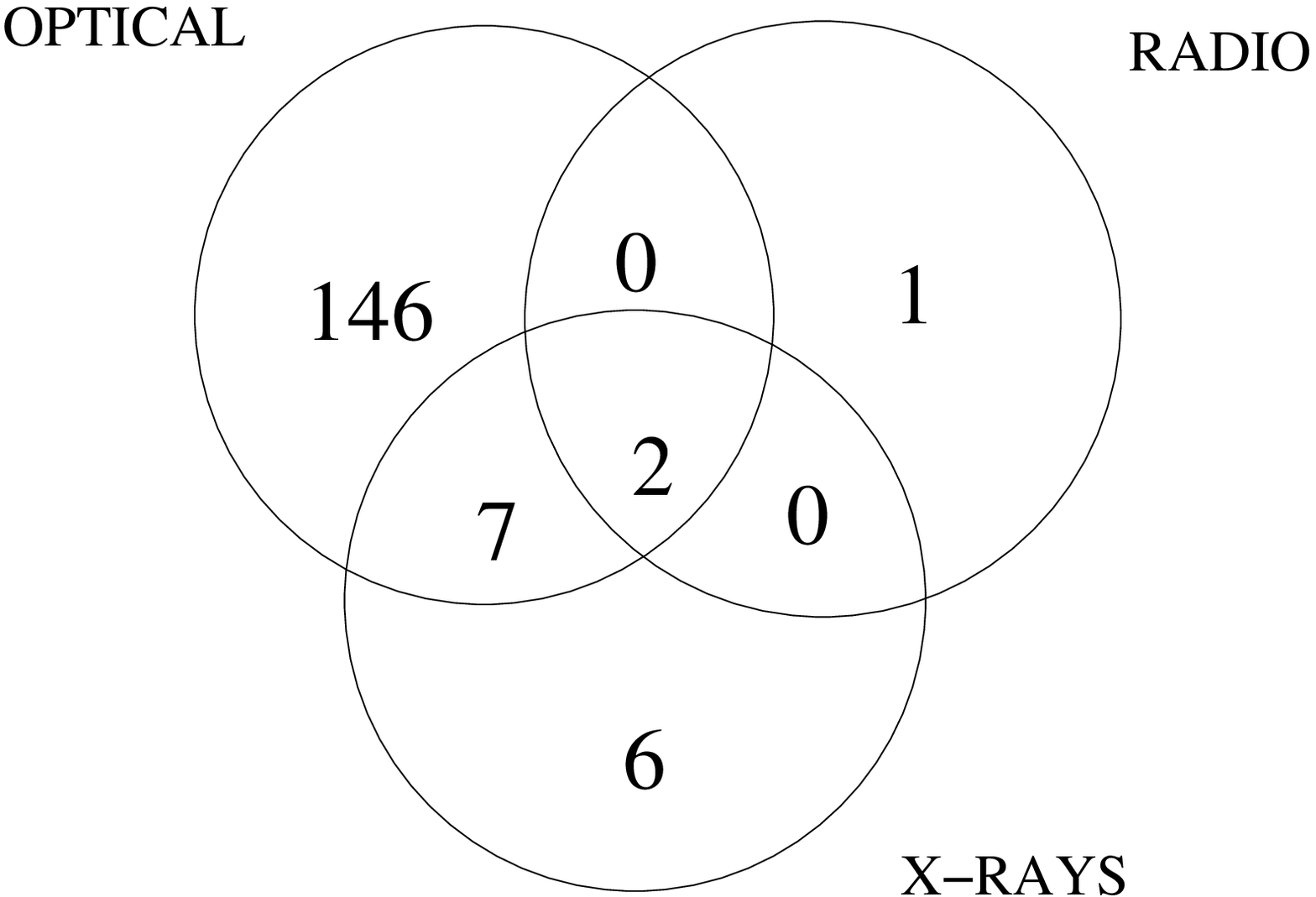}
        \label{fig:second_sub}
    }
    \caption{Venn diagrams for a) all detected SNRs in our sample and b) for SNRs in NGC\,2403.}
    \label{fig:sample_subfigures}
\end{figure}

\begin{figure}
\includegraphics[width=2.5in]{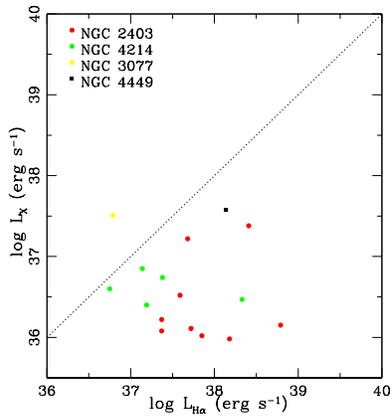}
\centering
 \caption{\Ha\ against the X-ray luminosity of the 16 optically selected, X-ray emitting SNRs. The dashed line indicates the 1:1 relation between the two luminosities.}
\end{figure}

\clearpage
\begin{figure}
\includegraphics[width=2.5in]{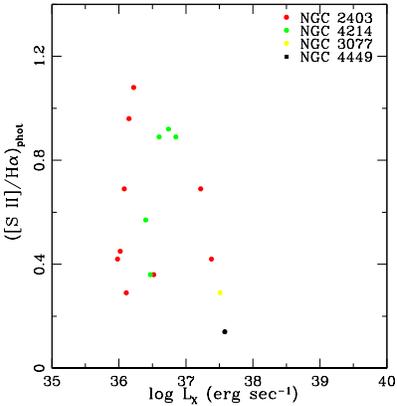}
\centering
 \caption{X-ray luminosity against the \SII/\Ha\ ratio of the 16 optically selected, X-ray emitting SNRs.}
\end{figure}

\clearpage

\begin{figure}
\includegraphics[width=5in]{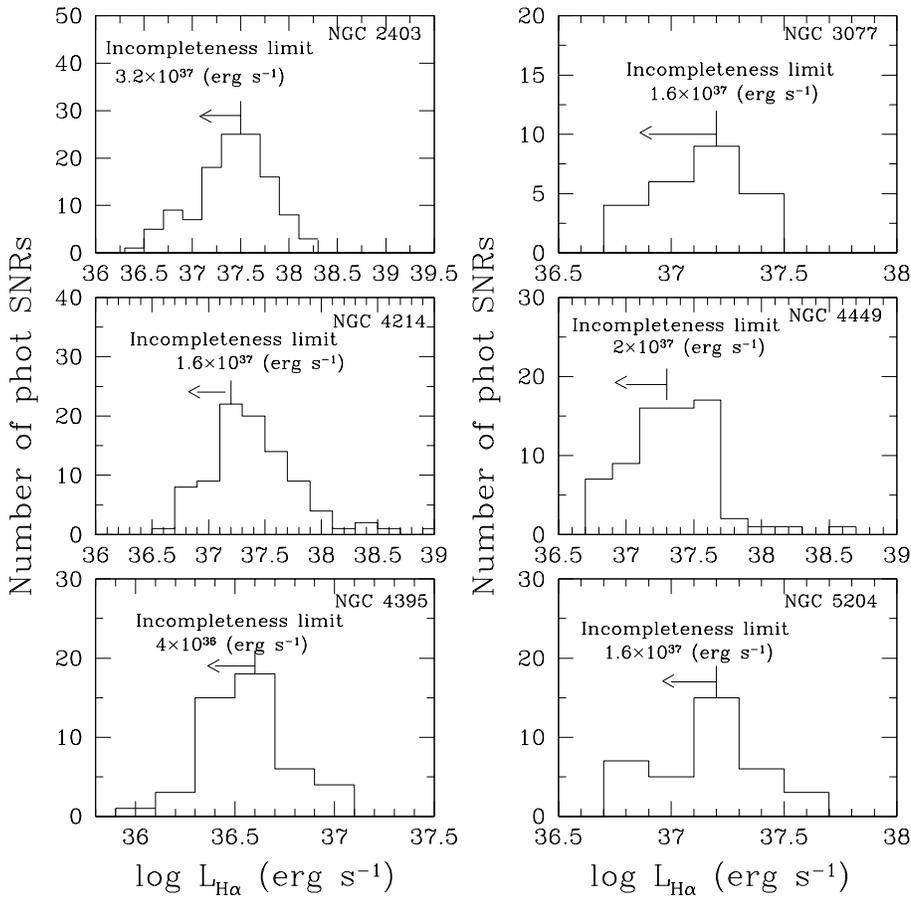}
\centering
 \caption{Histograms of the SNR \Ha\ luminosities in each galaxy of our sample. The peak at each histogram denotes the completeness limit of each galaxy.}
\end{figure}

\clearpage

\begin{figure}
\includegraphics[width=4in]{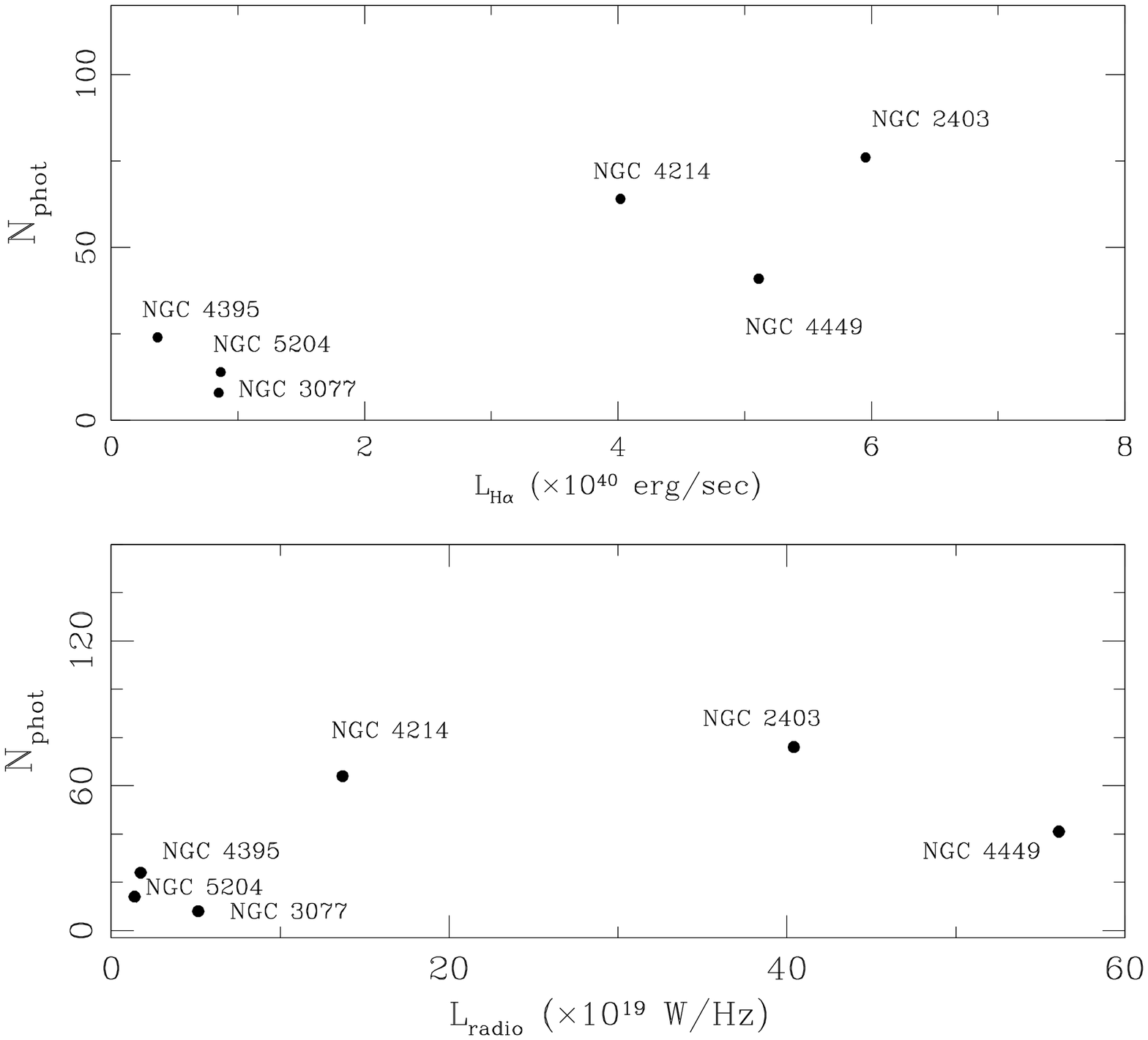}
\centering
 \caption{{\it Top}: Number of photometric SNRs above the completeness limit of each galaxy against the integrated \Ha\ luminosity of the host galaxy. {\it Bottom}: Number of photometric SNRs above the completeness limit of each galaxy against the radio luminosity.}
\end{figure}




\begin{thebibliography}{}
\bibitem[\protect\citeauthoryear{Allen et al.}{2008}]{Allen08} Allen, M. G., Groves, B. A., Dopita, M. A., Sutherland, R. S., Kewley, L. J. 2008, ApJS, 178, 20
\bibitem[\protect\citeauthoryear{Annibali et al.}{2008}]{Annibali08} Annibali, F., Aloisi, A., Mack. J., Tosi, M., van der Marel, R.P., Angeretti, L., Leitherer, C., Sirianni, M. 2008, AJ, 135, 1900
\bibitem[\protect\citeauthoryear{Asaoka \& Koyama}{1990}]{Asaoka90} Asaoka, I. \& Koyama, K. 1990, PASJ, 42, 625
\bibitem[\protect\citeauthoryear{Baldwin et al.}{1981}]{Baldwin81} Baldwin, J. A., Phillips, M. M., Terlevich, R. 1981, PASP, 93, 5
\bibitem[\protect\citeauthoryear{Bertin \& Arnouts}{1996}]{Bertin96} Bertin, E., Arnouts, S. 1996, A\&AS, 117, 393
\bibitem[\protect\citeauthoryear{Blair et al.}{1983}]{Blair83} Blair, W. P., Kirshner, R. P., \& Winkler, P. F. 1983, ApJ, 272, 84
\bibitem[\protect\citeauthoryear{Blair \& Long}{2004}]{Blair04} Blair, W. P., \& Long, K. S. 2004, ApJS, 155, 101
\bibitem[\protect\citeauthoryear{Boumis et al.}{2002}]{Boumis02} Boumis, P., Mavromatakis, F., Paleologou, E.V. 2002, A\&A, 385, 1042
\bibitem[\protect\citeauthoryear{Boumis et al.}{2005}]{Boumis05} Boumis, P., Mavromatakis, F., Xilouris, E.M., Alikakos, J., Redman, M.P., Goudis, C.D. 2005, A\&A, 443, 175
\bibitem[\protect\citeauthoryear{Boumis et al.}{2007}]{Boumis07} Boumis, P., Meaburn, J., Alikakos, J., Redman, M. P., Akras, S., Mavromatakis, F., Lopez, J. A., Caulet, A., Goudis, C. D. 2007, MNRAS, 381, 308 
\bibitem[\protect\citeauthoryear{Boumis et al.}{2009}]{Boumis09} Boumis, P., Xilouris, E. M., Alikakos, J., Christopoulou, P. E., Mavromatakis, F., Katsiyannis, A. C., Goudis, C. D. 2009, A\&A, 499, 789
\bibitem[\protect\citeauthoryear{Charles \& Seward}{1995}]{CS95} Charles, P.A.,\& Seward, F. D. 1995, Exploring the X-ray Universe (Cambridge: Cambridge Univ. Press)
\bibitem[\protect\citeauthoryear{Chen et al.}{1999}]{Chen99} Chen, C.-H. Rosie, Chu, You-Hua, Points, S. D. 1999, AAS, 194, 7207
\bibitem[\protect\citeauthoryear{Chen et al.}{2000}]{Chen00} Chen, C.-H. Rosie, Chu, You-Hua, Gruendl, R. A., Points, S. D. 2000, AJ, 119, 131
\bibitem[\protect\citeauthoryear{Chomiuk \& Wilcots}{2009}]{Chomiuk09} Chomiuk, L., \& Wilcots, E. M. 2009, AJ, 137, 3869
\bibitem[\protect\citeauthoryear{Chu \& Mac Low}{1990}]{Chu90} Chu, You-Hua, Mac Low, Mordecai-Mark 1990, ApJ, 365, 510
\bibitem[\protect\citeauthoryear{Condon}{1987}]{Condon87} Condon, J. J. 1987, ApJS, 65, 485
\bibitem[\protect\citeauthoryear{Condon \& Yin}{1990}]{Condon90} Condon, J. J., \& Yin, Q. F. 1990, ApJ, 357, 97
\bibitem[\protect\citeauthoryear{Cox \& Raymond}{1985}]{Cox85} Cox, D. P., Raymond, J. C. 1985, ApJ, 298, 651
\bibitem[\protect\citeauthoryear{de Vaucouleurs et al.}{1995}]{de Vaucouleurs95} de Vaucouleurs, G., de Vaucouleurs, A., Corwin, H. G., Buta, R. J., Paturel, G., \& Fouque, P. 1995, Third Reference Catalog of Bright Galaxies (New York: Springer) (RC3)
\bibitem[\protect\citeauthoryear{Dickel}{1999}]{Dickel99} Dickel, J. R. 1999, in IAU Symp. 190, New Views of the Magellanic Clouds, ed. Y. H. Chu, N. B. Suntzeff, J. E. Hesser, \& D. A. Bohlender (San Francisco, CA: ASP), 139
\bibitem[\protect\citeauthoryear{Dopita et al.}{2010}]{Dopita10} Dopita, M.A., et al. 2010, Astrophys Space Science, 330, 123
\bibitem[\protect\citeauthoryear{Eck et al.}{2002}] {Eck02} Eck, C. R., Cowan, J. J., \& Branch, D. 2002, ApJ, 573, 306
\bibitem[\protect\citeauthoryear{Fesen \& Milisavljevic}{2010}]{Fesen10} Fesen, R. A., Milisavljevic, D. 2010, AJ, 140, 1163
\bibitem[\protect\citeauthoryear{Franchetti et al.}{2012}]{Franchetti12} Franchetti, N. A., Gruendl, R. A., Chu, You-Hua, Dunne, B. C., Pannuti, T. G., Kuntz, K. D., Chen, C.-H. Rosie, Grimes, C. K., Aldridge, T. M. 2012, AJ, 143, 85
\bibitem[\protect\citeauthoryear{Freedman \& Madore}{1988}]{Freedman88} Freedman, W.L., \& Madore, B.F. 1988, ApJ, 332, L63
\bibitem[\protect\citeauthoryear{Freedman et al.}{1994}] {Freedman94} Freedman, W.L., et al. 1994, ApJ, 427, 628
\bibitem[\protect\citeauthoryear{Gaensler \& Slane}{2006}]{Gaensler06} Gaensler, B.M., Slane, P.O. 2006, ARA\&A, 44, 17
\bibitem[\protect\citeauthoryear{Garcia et al.}{1991}] {Garcia91}Garcia-Lario, P., Manchado, A., Riera, A., Mampaso, A. \& Pottasch, S. R., 1991, A\&A, 249, 223
\bibitem[\protect\citeauthoryear{Garnett}{2002}]{Garnett02} Garnett, D.R. 2002, ApJ, 581, 1019
\bibitem[\protect\citeauthoryear{Ghavamian et al.}{2005}]{Ghavamian05} Ghavamian, P., Blair, W. P., Long, K. S., Sasaki, M., Gaetz, T. J., \& Plucinsky, P. 2005, AJ, 130, 539
\bibitem[\protect\citeauthoryear{Green}{2009}]{Green09} Green, D.A. 2009, Bulletin of the Astronomical Society of India, 37, 45. (See: arxiv:0905.3699)
\bibitem[\protect\citeauthoryear{Hamuy et al.}{1992}]{Hamuy92} Hamuy, M., Walker, A.R., Suntzeff, N.B., et al. 1992, PASP, 104, 533
\bibitem[\protect\citeauthoryear{Hartigan et al.}{1987}]{Hartigan87} Hartigan, P., Raymond, J., Hartmann, L. 1987, ApJ, 316, 323
\bibitem[\protect\citeauthoryear{Kennicutt et al.}{2008}]{Kennicutt08} Kennicutt, R.C., Jr., Janice C. Lee, José G. Funes, S. J., Shoko Sakai, Sanae Akiyama 2008, ApJS, 178, 247
\bibitem[\protect\citeauthoryear{Lasker}{1977}]{Lasker77} Lasker, B.M. 1977, ApJ, 212, 390
\bibitem[\protect\citeauthoryear{Leonidaki et al.}{2010}]{Leonidaki10} Leonidaki, I., Zezas, A. \& Boumis, P. 2010, ApJ, 725, 842
\bibitem[\protect\citeauthoryear{Lequeux et al.}{1979}] {Lequeux79} Lequeux, J., Peimbert, M., Rayo, J. F., Serrano, A., Torres-Peimbert, S. 1979, A\&A, 80, 155 
\bibitem[\protect\citeauthoryear{Long et al.}{2010}]{Long10} Long, K. S., Blair, W. P., Winkler, P. F., Becker, R. H., Gaetz, T. J., Ghavamian, P., Helfand, D. J., Hughes, J. P., Kirshner, R. P., Kuntz, K. D., McNeil, E. K., Pannuti, T. G., Plucinsky, P. P., Saul, D., Tüllmann, R., Williams, B. 2010, ApJS, 187, 495
\bibitem[\protect\citeauthoryear{Martin}{1997}]{Martin97} Martin, C.L. 1997, ApJ, 491, 561
\bibitem[\protect\citeauthoryear{Mathewson \& Clarke}{1973}]{MC73} Mathewson, D. S. \& Clarke, J.N. 1973, ApJ, 180, 725
\bibitem[\protect\citeauthoryear{Mathewson et al.}{1983}]{Mathewson83} Mathewson, D. S., Ford, V. L., Dopita, M. A., Tuohy, I. R., Long, K. S., Helfand, D. J. 1983, ApJS, 51, 345
\bibitem[\protect\citeauthoryear{Matonick \& Fesen}{1997}]{MF97} Matonick, D. M., Fesen, R. A., 1997, ApJS, 112, 49 
\bibitem[\protect\citeauthoryear{Matonick et al.}{1997}]{MFBL97} Matonick, D. M., et al., 1997,  ApJS, 113, 333
\bibitem[\protect\citeauthoryear{Meaburn et al.}{2010}]{Meaburn10} Meaburn, J., Redman, M. P., Boumis P. \& Harvey, E., 2010, MNRAS, 408, 1249
\bibitem[\protect\citeauthoryear{Osterbrock \& Ferland}{2006}]{Osterbrock06} Osterbrock, D. E., \& Ferland, G. J. 2006, Astrophysics of gaseous nebulae and AGN, 2nd edn (Sausalito, CA: University Science Books)
\bibitem[\protect\citeauthoryear{Ott et al.}{2003}]{Ott03} Ott, J., Martin, C. L., \& Walter, F. 2003, ApJ, 594, 776
\bibitem[\protect\citeauthoryear{Pagel \& Endmunds}{1981}]{Pagel81} Pagel, B. E. J., \& Endmunds, M. G. 1981, ARA\&A, 19, 77
\bibitem[\protect\citeauthoryear{Pannuti et al.}{2007}]{Pannuti07} Pannuti, T. G., Schlegel, E. M., \& Lacey, C. K. 2007, AJ, 133, 1361
\bibitem[\protect\citeauthoryear{Pilyugin et al.}{2004}]{Pilyugin04} Pilyugin, L. S., Vílchez, J. M., Contini, T. 2004, A\&A, 425, 849
\bibitem[\protect\citeauthoryear{Raymond et al.}{1988}]{Raymond88} Raymond, J. C., Hester, J. J., Cox, D., Blair, W. P., Fesen, R. A., Gull, T. R. 1988, ApJ, 324, 869 
\bibitem[\protect\citeauthoryear{Reach et al.}{2006}]{Reach06} Reach, W. T., Rho, J., Tappe, A., Pannuti, T. G., Brogan, C. L., Churchwell, E. B., Meade, M. R., Babler, B., Indebetouw, R., Whitney, B. A. 2006, AJ, 131, 1479
\bibitem[\protect\citeauthoryear{Reynolds et al.}{2009}]{Reynolds09} Reynolds, S. P., Borkowski, K. J., Green, D. A., Hwang, U., Harrus, I., Petre, R. 2009, ApJ, 695, 149
\bibitem[\protect\citeauthoryear{Richer \& McCall}{1995}]{Richer95} Richer, M. G., \& McCall, M. L. 1995, ApJ, 445, 642
\bibitem[\protect\citeauthoryear{Rosa-Gonzalez}{2005}]{Rosa05} Rosa-Gonzalez, D. 2005, MNRAS, 364, 1304
\bibitem[\protect\citeauthoryear{Rosado et al.}{1983}]{Rosado83} Rosado M., Georgelin Y. M., Laval A., Monnet G., 1983, Proc. IAU Symp. 101, Supernova Remnants and Their X-ray Emission, ed. P. Gorenstein and I.J. Danziger, Reidel, Dordrecht, p. 567
\bibitem[\protect\citeauthoryear{Russell \& Dopita}{1992}]{Russell92} Russell, S. \& Dopita, M. A., 1992, ApJ, 384, 508
\bibitem[\protect\citeauthoryear{Sabbadin et al.}{1977}]{Sabbadin77} Sabbadin, F., Minello, S. \& Bianchini, A. 1977, A\&A, 60, 147
\bibitem[\protect\citeauthoryear{Sabbadin et al.}{1984}]{Sabbadin84} Sabbadin, F., Ortolani, S., Bianchini, A. 1984,  A\&A, 131, 1S 
\bibitem[\protect\citeauthoryear{Safi-Harb \& Petre}{1999}]{Safi99} Safi-Harb, S., \& Petre, R. 1999, ApJ, 512, 784
\bibitem[\protect\citeauthoryear{Safi-Harb et al.}{2001}]{Safi-Harb01} Safi-Harb, S., Harrus, I. M., Petre, R., Pavlov, G. G., Koptsevich, A. B., \& Samwal, D. 2001, ApJ, 561, 308
\bibitem[\protect\citeauthoryear{Saha et al.}{1994}]{Saha94} Saha, A., Labhardt, L., Schwengeler, H., Macchetto, F.D., Panagia, N., Sandage, A. \& Tammann, G.A. 1994, ApJ, 425, 14
\bibitem[\protect\citeauthoryear{Slane et al.}{2002}]{Slane02} Slane, P., Smith, R. K., Hughes, J. P., Petre, R. 2002, ApJ, 564, 284
\bibitem[\protect\citeauthoryear{Storchi-Bergmann et al.}{1994}]{Storchi-Bergmann94} Storchi-Bergmann, T., Calzetti, D./ Kinney, A.L. 1994, ApJ, 429, 572
\bibitem[\protect\citeauthoryear{Stupar \& Parker}{2009}]{Stupar09} Stupar, M., \& Parker, Q. A. 2009, MNRAS, 394, 1791
\bibitem[\protect\citeauthoryear{Summers et al.}{2003}]{Summers03} Summers, L.K., Stevens, I.R., Strickland, D.K. \& Heckman, T.M. 2003, MNRAS, 342, 690
\bibitem[\protect\citeauthoryear{Tully}{1988}]{Tully88} Tully, R. 1988, Nearby galaxies Catalog (Cambridge: Cambridge University Press)
\bibitem[\protect\citeauthoryear{Turner \& Ho}{1994}]{Turner94} Turner, J. L., \& Ho, P. T. P. 1994, ApJ, 421, 122
\bibitem[\protect\citeauthoryear{Viironen et al.}{2007}]{Viironen07} Viironen, K., Delgado-Inglada, G., Mampaso, A., Magrini, L., \& Corradi, R.L.M. 2007, MNRAS, 381, 1719 
\bibitem[\protect\citeauthoryear{Vukotic et al.}{2005}]{Vukotic05} Vukotic, B., Bojiicic, I., Pannuti, T. G., \& Urosevic, D. 2005, Serb. Astron. J., 170, 101
\bibitem[\protect\citeauthoryear{Williams et al.}{1999}]{Williams99} Williams, R. M., Chu, You-Hua, Dickel, J. R., Petre, R., Smith, R. C., Tavarez, M. 1999, ApJS, 123, 467
\end{thebibliography}
\end{document}